\documentclass[]{raa}
\usepackage{graphicx, times}
\usepackage{natbib}

\providecommand{\bm}[1]{\mbox{\boldmath$#1$\unboldmath}}

\def\ptb{180pt}
\def\ptm{140pt}
\def\pts{120pt}

\newcommand{\path}{./figures}

\begin{document}

\title{Detailed Analysis of Fan-Shaped jets in Three Dimensional Numerical Simulation}

\volnopage{ {\bf 20xx} Vol.\ {\bf 9} No. {\bf XX}, 000--000} \setcounter{page}{1}

\author{R. L. Jiang \inst{1,2,4} \and K. Shibata\inst{2} \and H. Isobe\inst{3} \and C. Fang\inst{1,4}}

\email{rljiang@nju.edu.cn}

\institute{Department of Astronomy, Nanjing University, Nanjing 210093, China; {\it rljiang@nju.edu.cn}
\and Kwasan and Hida Observatories, Kyoto University, Yamashina, Kyoto 607-8471, Japan
\and Unit of synergetic Studies for Space, Kyoto University, Yamashina, Kyoto 607-8471, Japan
\and Key Laboratory of Modern Astronomy and Astrophysics (Nanjing University), Ministry of Education, China
\vs \no
{\small Received [year] [month] [day]; accepted [year] [month] [day] }}

\abstract{
We performed three dimensional resistive magnetohydrodynamic simulations to study the magnetic reconnection using an initially shearing magnetic field configuration (force free field with a current sheet in the middle of the computational box). It is shown that there are two types of reconnection jets: the ordinary reconnection jets and fan-shaped jets, which are formed along the guide magnetic field. The fan-shaped jets are much different from the ordinary reconnection jets which are ejected by magnetic tension force. There are two driving forces for accelerating the fan-shaped jets. The one is the Lorentz force which dominates the motion of fluid elements at first and then the gas pressure gradient force accelerates the fluid elements in the later stage. The dependence on magnetic reconnection angle and resistivity value has also been studied. The formation and evolution of these jets provide a new understanding of dynamic magnetohydrodynamic jets.
}

\authorrunning{R. L. Jiang, K. Shibata, H. Isobe \& C. Fang}
\titlerunning{Fan-Shaped jets}
\maketitle

\section{Introduction}

Magnetic reconnection plays a very important role in solar flares, coronal mass ejections, and other solar activities. The problem of flare energy release has been puzzling people for many years before the development of magnetic reconnection mechanism. The first pioneers in this field of magnetic reconnection are \cite{Sweet1958} and \cite{Parker1957}, who developed a theory well known as Sweet-Parker mechanism. However, it can not explain the solar flares. The biggest problem in their theory is that the time scale of energy release is much longer than the realistic one of flares. Several years later, \cite{Petschek1964} improved this work by considering a pair of slow mode shocks outside a very small diffusion region, which greatly increases the reconnection rate. The fast reconnection model is successful, though some people do not believe that the large energy release process has such a small diffusion region. The famous CSHKP model (\citealt{Carmichael1964, Sturrock1966, Hirayama1974, Kopp1976}) based on magnetic reconnection for flares has been developed since 1960s.

Besides observations, magnetohydrodynamic (MHD) simulations play a key role in studying the magnetic reconnection (\citealt{Ugai1977, Forbes1983, Forbes1990, Forbes1991, Yokoyama1994, Magara1996, Ugai1996, Chen1999}). With the development of computer and observation technology, the CSHKP flare model has been extended. More and more simulations show that the plasmoid ejection may greatly increase the magnetic reconnection rate (\citealt{Shibata1995, Shibata1996, Shibata1997, Magara1997, Nishida2009}). The plasmoids can form in an anti-parallel magnetic field. Before ejection magnetic energy is restored in the diffusion region, and simultaneously the plasmoids merge with each other. Finally, the restored magnetic energy is released in a very short time scale after the ejection of the big plasmoid.

Recently, the three dimensional (3D) simulation is one of the hottest points in astrophysics. In this case, the reconnecting process is much more complicated than that in two dimensional (2D) case. The 3D simulations can provide us very complex and realistic structures (\citealt{Yokoyama1995, Isobe2005, Ugai2005, Isobe2006, Shimizu2009}). It is also a good tool to explain the mechanism of many small scale solar activities (\citealt{Cirtain2007, Shibata2007, Katsukawa2007}) which have been observed by Hinode (\citealt{Kosugi2007}). Some of these observed jetlike features are believed to be a result of magnetic reconnection in the so-called interlocking-comb like magnetic configuration in sunsport penumbra (i.e., penumbral jets). The motivation of this paper is to study the effect of the guide field using 3D numerical simulations. In our new simulation, we found that some parts of the reconnection jets which are ejected from the diffusion region can move along the magnetic guide field lines in the interlocking-comb like magnetic configuration. Moreover, the reconnection jets which move along the guide field lines are almost perpendicular to the ambient magnetic field. In order to distinguish the jets moving in different directions, we call the jets moving along the ambient magnetic field as ordinary reconnection jets and the jets moving along the guide field as fan-shaped jets (we will see the shape of the jets is similar to a fan in the next section). Besides our simulations, two very recent paper by~\cite{Pontin2005} and~\cite{Ugai2010} also discussed the effect of guide field. Their simulations showed that the guide field can distort the propagation of 3D plasmoids and the plasma is almost ejected along the guide field. It seems that the guide field plays a very important role for forming these jets in the 3D magnetic reconnection. In our first paper (\citealt{Jiang2011}) we briefly reported the main results of our 3D simulation, while in this paper the detailed analysis and discussion of fan-shaped jets are presented.

This paper is organized as follows. A detailed description of the magnetohydrodynamics (MHD) equations and the initial condition are given in the next sections. In Section~\ref{typical} we show the 3D, 2D and 1D structures of our typical simulation results. Moreover, the analysis of the Lagrangian fluid elements, reconnection rate and slow mode shock are also given. Section \ref{dependence} shows dependence on different reconnection angles (defined in the initial conditon) and different values of resistivity in the diffusion region. Finally, discussion and summary are given in the last Section.

\section{NUMERICAL METHOD}
\label{method}

We solve the three-dimensional, nonlinear, resistive, compressible MHD equations numerically. The role of gravity, thermal conduction, radiation or some other effects are not so important in our simulations, because what we are really interested in is the dynamics of the magnetic reconnection. The additional consideration may affect the final results but not essential. The simplified MHD equations are given in cgs units as follows:

\begin{equation}
\frac{\partial \rho}{\partial t} + \mathbf{v} \cdot \nabla \rho = - \rho \nabla \cdot \mathbf{v} , %
\label{mhd_1}
\end{equation}
\begin{equation}
\frac{\partial \mathbf{v}}{\partial t} + \mathbf{v} \cdot \nabla
\mathbf{v} = - \frac{\nabla p}{\rho} - \frac{1}{8 \pi \rho} \nabla
B^2
+ \frac{1}{4 \pi \rho}(\mathbf{B} \cdot \nabla) \mathbf{B} , %
\label{mhd_2}
\end{equation}
\begin{equation}
\frac{\partial p}{\partial t} + \mathbf{v} \cdot \nabla p = - \gamma p \nabla \cdot \mathbf{v} + (\gamma-1) \eta j^2 , %
\label{mhd_3}
\end{equation}
\begin{equation}
\frac{\partial \mathbf{B}}{\partial t} = - c \nabla \times \mathbf{E} , %
\label{mhd_4}
\end{equation}
\begin{equation}
\mathbf{E}=\eta \mathbf{j} - \frac{1}{c} \mathbf{v} \times \mathbf{B} , %
\label{mhd_5}
\end{equation}
\begin{equation}
\mathbf{j}=\frac{c}{4 \pi} \nabla \times \mathbf{B}  , %
\label{mhd_6}
\end{equation}
\begin{equation}
p = \frac{\kappa_B}{m} \rho T ,  %
\label{mhd_7}
\end{equation}

\noindent where the eight independent variables are the velocity ($\upsilon_{x}, \upsilon_{y}, \upsilon_{z}$), density ($\rho$), magnetic field ($B_{x}, B_{y}, B_{z}$), and pressure ($p$). $\eta$ is the resistivity, $\mathbf{j}$ the current density, $T$ the temperature, and $\mathbf{E}$ the electric field. The specific heat ratio of monoatomic perfect gas is $\gamma=5/3$. The normalized units of the variables are listed in Table~\ref{tab_units}. Hereafter, all the values used in our analysis and discussion are in non dimensional form for simplicity.

\begin{table}[htbp]
\small
\centering
\begin{minipage}[]{110mm}
\caption[]{Normalization Units.\label{tab_units}}
\end{minipage}
\tabcolsep 6mm
\begin{tabular}{ccc}
\noalign{\smallskip}\hline
\noalign{\smallskip}\hline
Variable & Quantity & Unit\\
\noalign{\smallskip}\hline
$x,y,z$           & Length                   & $L_0$                                            \\
$T$               & Temperature              & $T_0$                                            \\
$\rho$            & Density                  & $\rho_0$                                         \\
$p$               & Pressure                 & $p_0=\rho_0 T_0 \kappa_B / m$                    \\
$t$               & Time                     & $t_0=L_0 / v_0 $                                 \\
$\eta$            & Resistivity              & $\eta_0=(1/c^2)L_0 v_0 $                         \\
$\mathbf{v}$      & Velocity                 & $v_0=(p_0/\rho_0)^{1/2}$                         \\
$\mathbf{B}$      & Magnetic field           & $B_0=p_0^{1/2}$                                  \\
$\mathbf{E}$      & Electric field           & $E_0=(1/c)v_0 B_0 $                              \\
$\mathbf{j}$      & Current density          & $j_0=(c)B_0/L_0 $                                \\
$\mathbf{F}$      & Force per unit mass      & $F_0=p_0 / L_0 $                                 \\
\noalign{\smallskip}\hline
$\mathbf{v}_A$    & Alfv\'en velocity        & $v_{A0}=B_0 /(\rho_0)^{1/2}$                        \\
$\tau_A$          & Alfv\'en time scale      & $\tau_{A0}=L_0 / v_A$                               \\
$\mathbf{E}_A$    & Alfv\'en electric field  & $E_{A0}=(1/c)v_A B_0$                               \\
\noalign{\smallskip}\hline
\end{tabular}
\tablecomments{0.86\textwidth}{In this table, $m$, $\kappa_B$ and c mean molecular mass, Boltzmann constant and light speed, respectively.}
\end{table}

\begin{figure}[htbp]
   \centering
   \includegraphics[width=400pt]{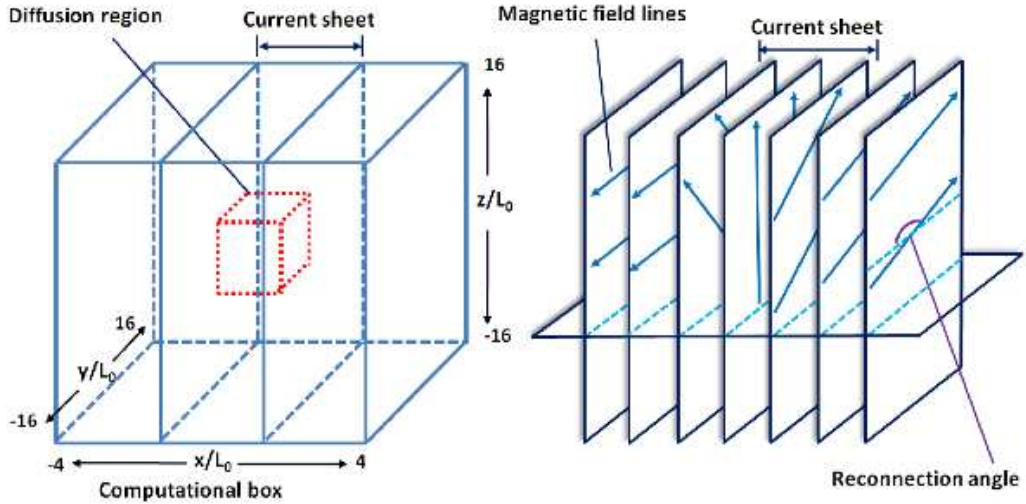}
   \caption{Initial condition, and the box size is $-4<x<4$, $-16<y<16$, $-16<z<16$ in non dimensional unit.}
   \label{fig01}
\end{figure}

The computational box is shown in Figure~\ref{fig01}. The domain of this box is $-4<x<4$, $-16<y<16$ and $-16<z<16$ in all of our cases (Note that the images of simulation results only show the central part of the computational box and the scale for $x$, $y$ and $z$ coordinates is different because of the thin current sheet in $x$ direction). The most important condition is the configuration of the force free magnetic field which is shown in the right panel of Figure~\ref{fig01}. The mathematical form of this field is

\begin{equation}
  B_x=0 , %
\end{equation}

\begin{equation}
  B_y=\left\{
  \begin{array}{ll}
    -B_{ini}                                    & \ \ \textrm{for} \ \   x < -\triangle h_x  \\
    B_{ini}\sin{\{[\theta x - (\pi - \theta)\triangle h_x]/2\triangle h_x\}}  & \ \ \textrm{for} \ \   |x| < \triangle h_x \\
    B_{ini}\sin{[(2 \theta - \pi)/2]}           & \ \ \textrm{for} \ \   x > \triangle h_x  \\
  \end{array}
 , %
  \right.
\end{equation}

\begin{equation}
  B_z=\left\{
  \begin{array}{ll}
    0                                     & \ \ \textrm{for} \ \   x < -\triangle h_x  \\
    B_{ini}\cos{\{[\theta x - (\pi - \theta)\triangle h_x]/2\triangle h_x\}}   & \ \ \textrm{for} \ \   |x| < \triangle h_x \\
    B_{ini}\cos{[(2 \theta - \pi)/2]}     & \ \ \textrm{for} \ \   x > \triangle h_x  \\
  \end{array}
 , %
  \right.
\end{equation}

\noindent where $B_{ini}$ is a constant, which means the magnetic field has an initial uniform strength. The direction of this field is shown in Figure~\ref{fig01}. The $\triangle h_x$ is the half width of the current sheet. $\theta$ is the reconnection angle shown in Figure~\ref{fig01}, and it has been taken to be one of the four values in our simulation cases, i.e., $\pi$, $3\pi/4$, $\pi/2$ and $\pi/3$, as listed in Table~\ref{cases}.

\begin{table}[htbp]
\small
\centering
\begin{minipage}[]{110mm}
\caption{Computational Cases.\label{cases}}
\end{minipage}
\begin{tabular}{cccccc}
\hline\noalign{\smallskip}
\hline\noalign{\smallskip}
Case          & $\beta$ & $\eta_{ini}$  & Reynolds number ($v_A L_x/\eta_{ini}$)            & Reconnection angle ($\theta$) & Total grid points \\
\hline\noalign{\smallskip}
$1$           & 1.0     & 0.025         & 226                                               & $\pi$            & 300 $\times$ 240 $\times$ 240  \\
\hline\noalign{\smallskip}
$2$           & 0.8     & 0.050         & 126                                               & $\pi$         & 240 $\times$ 480 $\times$ 480  \\
$3$           & 0.8     & 0.050         & 126                                               & $3\pi/4$      & 240 $\times$ 480 $\times$ 480  \\
$4$           & 0.8     & 0.050         & 126                                               & $\pi/2$       & 240 $\times$ 480 $\times$ 480  \\
$5$           & 0.8     & 0.050         & 126                                               & $\pi/3$       & 240 $\times$ 480 $\times$ 480  \\
\hline\noalign{\smallskip}
$6$           & 1.0     & 0.025         & 226                                               & $\pi$         & 240 $\times$ 480 $\times$ 480  \\
$7$           & 1.0     & 0.050         & 113                                               & $\pi$         & 240 $\times$ 480 $\times$ 480  \\
$8$           & 1.0     & 0.075         & 75                                                & $\pi$         & 240 $\times$ 480 $\times$ 480  \\
\noalign{\smallskip}\hline
\end{tabular}
\tablecomments{0.86\textwidth}{In this table, $L_x$ is the half length of the computational box size in $x$ direction.}
\end{table}

In order to get fast reconnection process, we put a localized diffusion region at the center of the current sheet (\citealt{Ugai1986, Yokoyama1994}). The adopted functional form for the anomalous resistivity is as follows:

\begin{equation}
\eta = \eta_{ini} \cos[\frac{x \pi}{2 \triangle h_x}] \cos[\frac{y
\pi}{2 \triangle h_y}] \cos[\frac{z \pi}{2 \triangle h_z}] \ \
\textrm{for} \ \  |x| < \triangle h_x,\ |y| < \triangle h_y,\ |z| <
\triangle h_z , %
\end{equation}

\noindent where $\triangle h_x = 0.4$ (it is the same value as the width of the current sheet), $\triangle h_y = 1.6$, $\triangle h_z = 1.6$ is the half width of the diffusion region in $x$, $y$ and $z$ direction, respectively. $\eta_{ini}$ is the maximum of the resistivity in the diffusion region and is taken as a constant.

In the numerical procedures, we solve the MHD equations using CIP-MOCCT scheme (\citealt{Yabe1991a, Yabe1991b, Kudoh1999}) with an artificial viscosity~(\citealt{Stone1992}) and libraries of CANS (Coordinated Astronomical Numerical Softwares). The calculation domain is discretized into $240 \times 480 \times 480$ nonuniform grids in most of cases. The minimum grid size is $\triangle x=0.00667$, $\triangle y = 0.0667$ and $\triangle z = 0.0667$. Since there is no gravity in our simulations, the free boundary condition is simply applied in all of six boundaries (namely, an equivalent extrapolation was applied for all quantities). The non-divergence condition of the magnetic field is guaranteed by the constraint transport (CT) algorithm of~\cite{Evans1988} in CIP-MOCCT scheme. For initial condition, the density and pressure are uniform ($\rho=1$, $p=1$) in the computational box and the other parameters are shown in Table~\ref{cases}.

\section{Typical Case}
\label{typical}

\begin{figure}[htbp]
\centering
\includegraphics[width=\ptb]{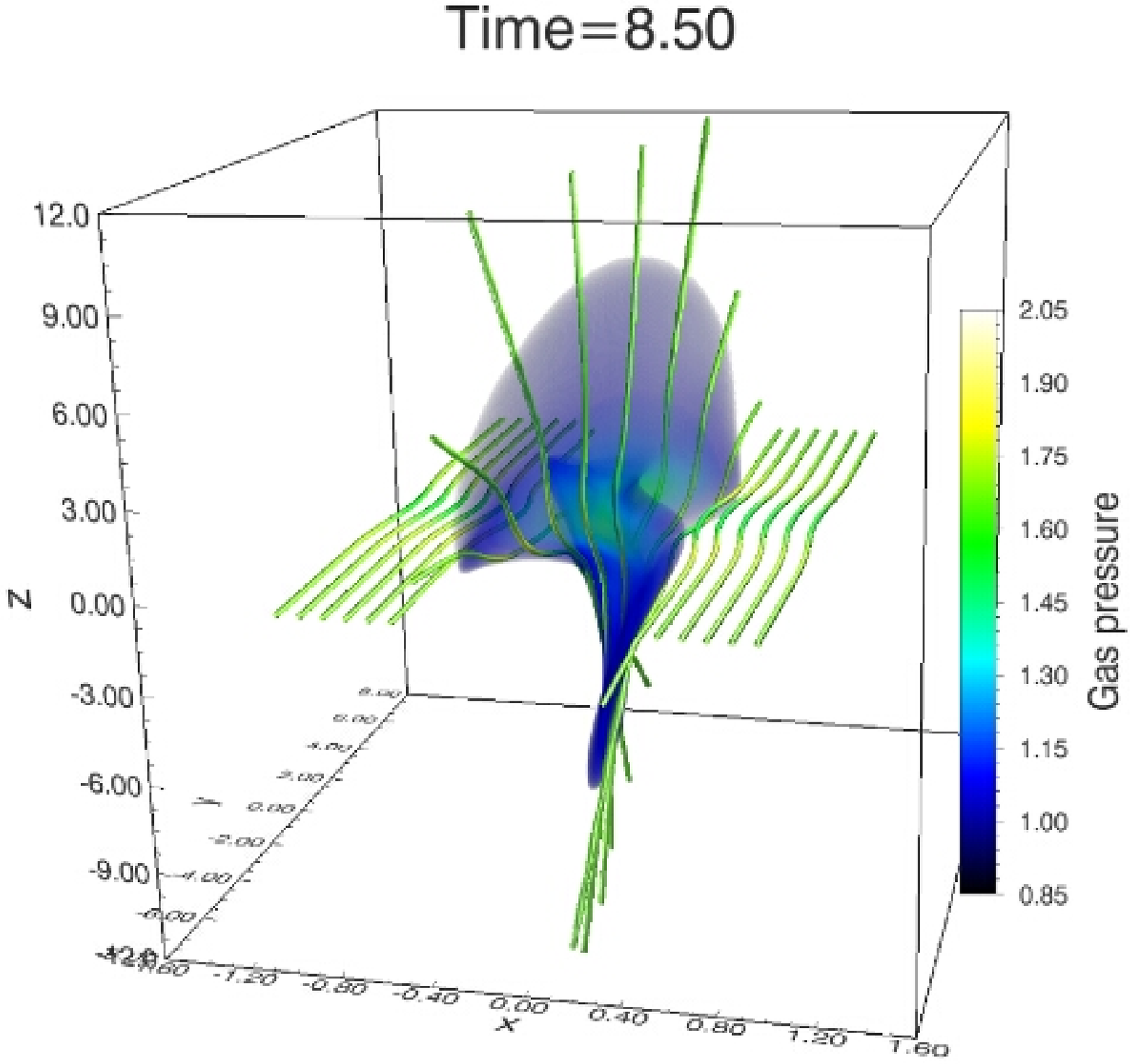}
\includegraphics[width=\ptb]{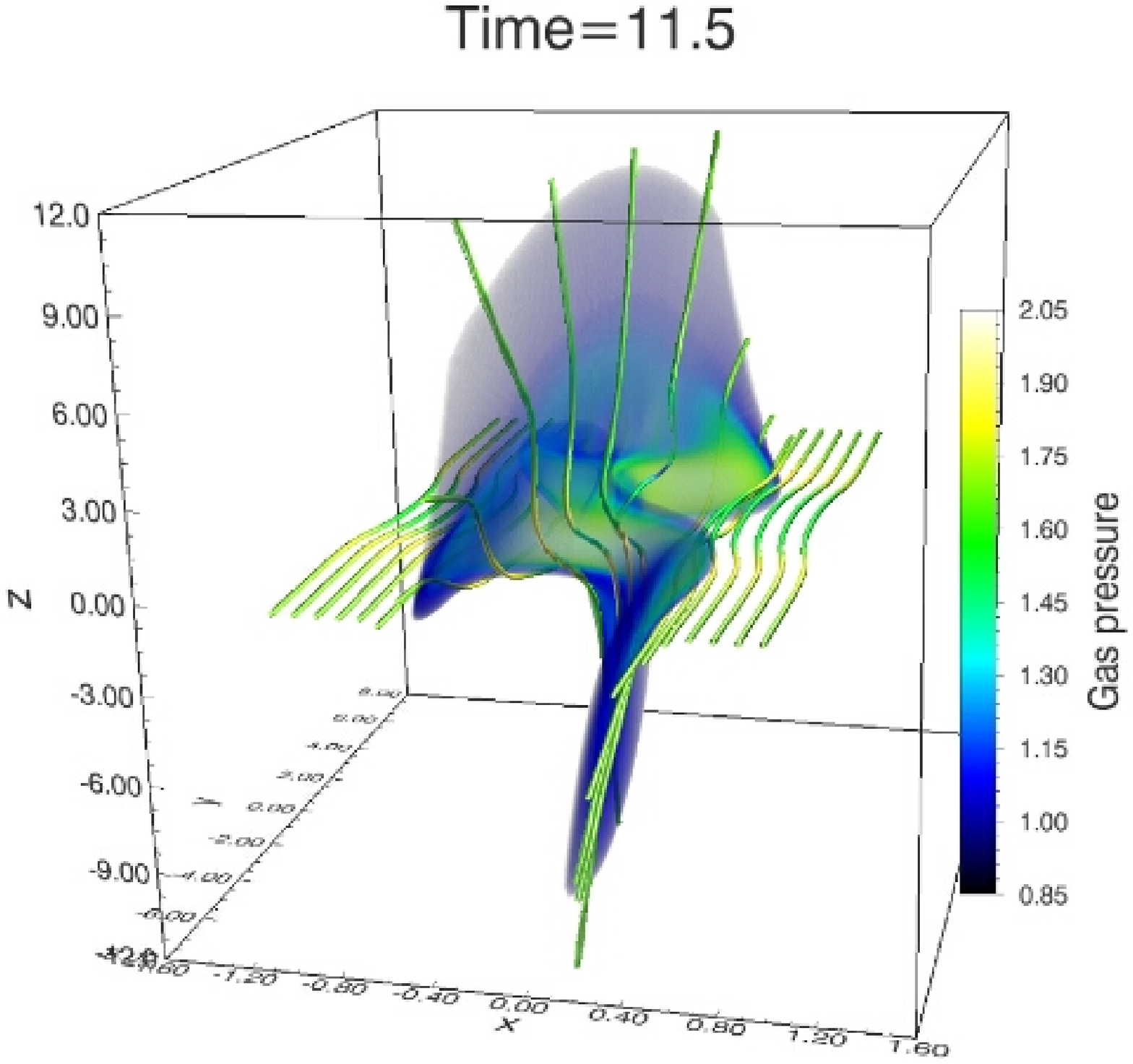}
\includegraphics[width=\ptb]{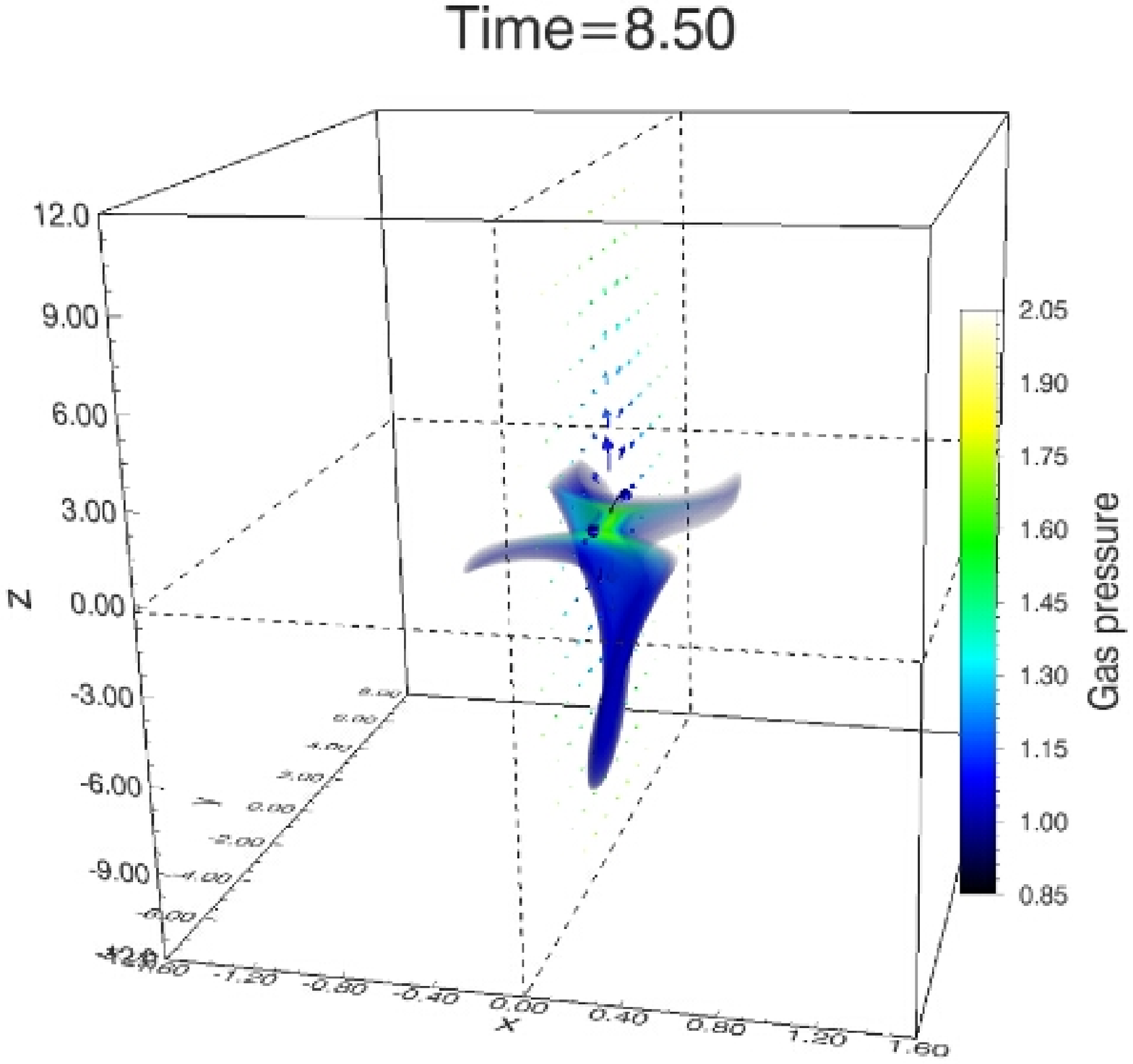}
\includegraphics[width=\ptb]{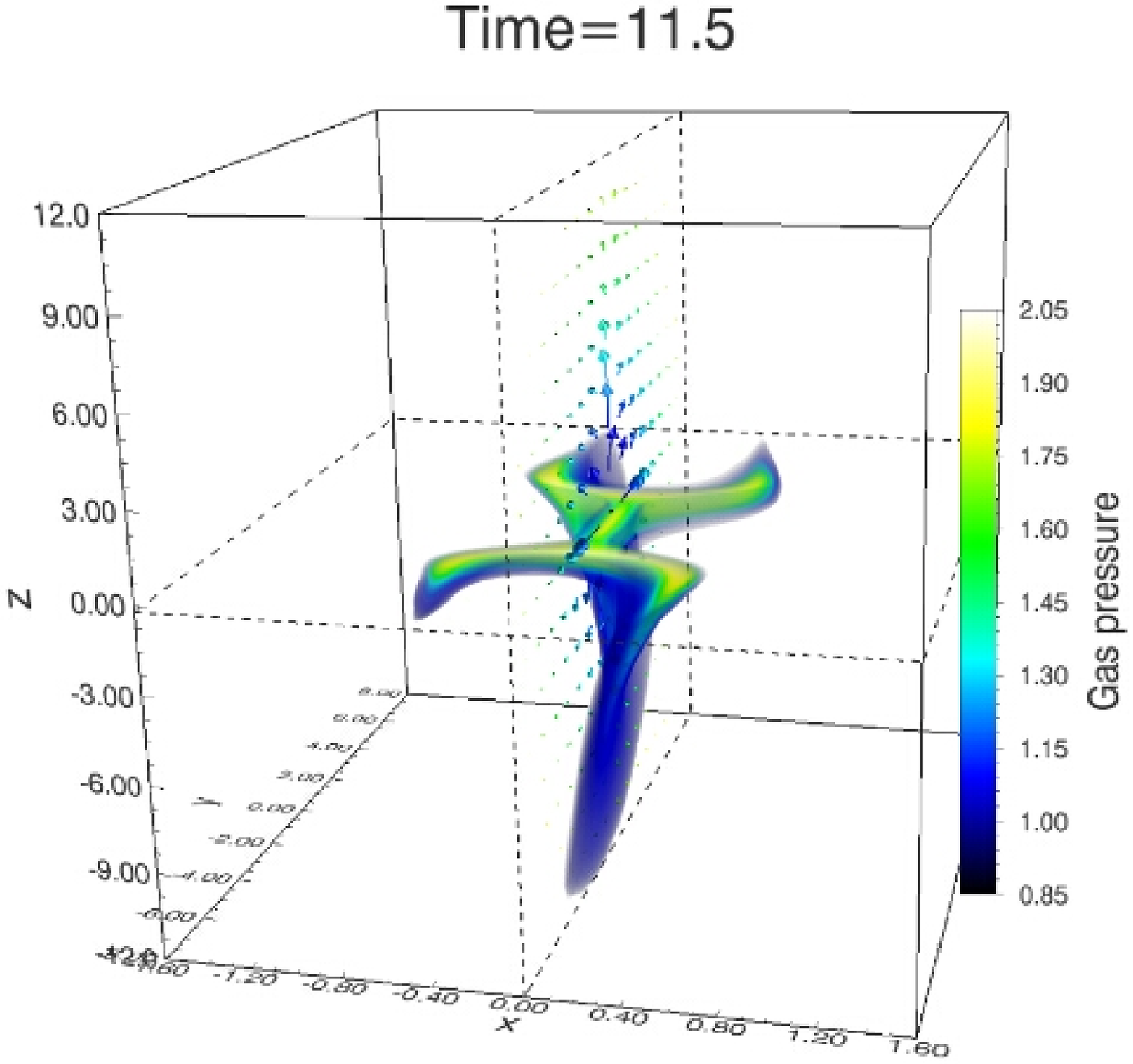}
\caption{3D visualization of the gas pressure distributions at different times (typical case). The bright colors and high opacity stand for the high gas pressure and the low value is transparent. Solid tubes are magnetic field lines in the upper panels and arrows are velocity in the lower panels. Since the current sheet is very thin, a different scale in $x$ direction is used in these panels.} \label{fig02}
\end{figure}

\begin{figure}[htbp]
\centering
\includegraphics[width=\ptb]{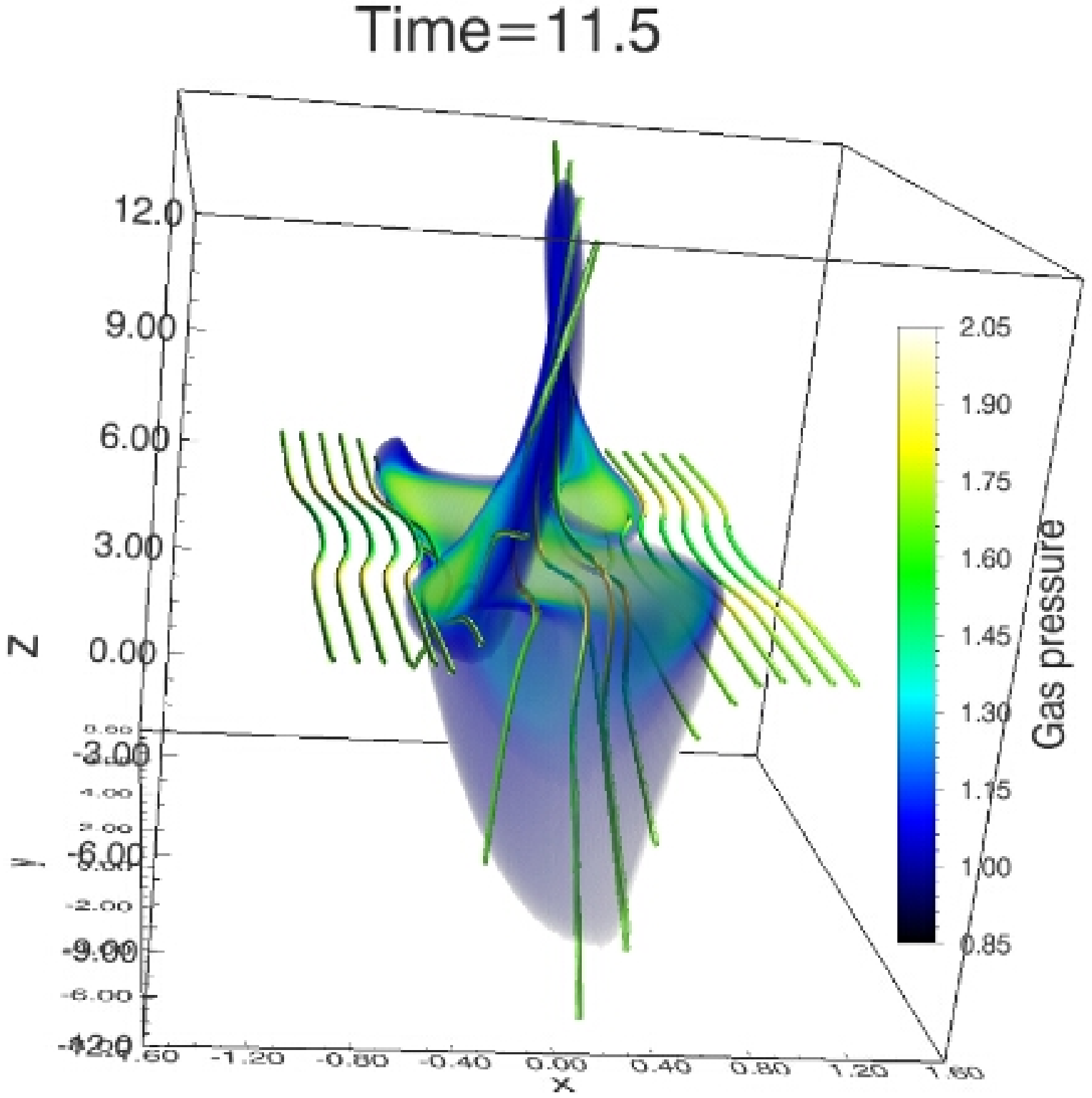}
\includegraphics[width=\ptb]{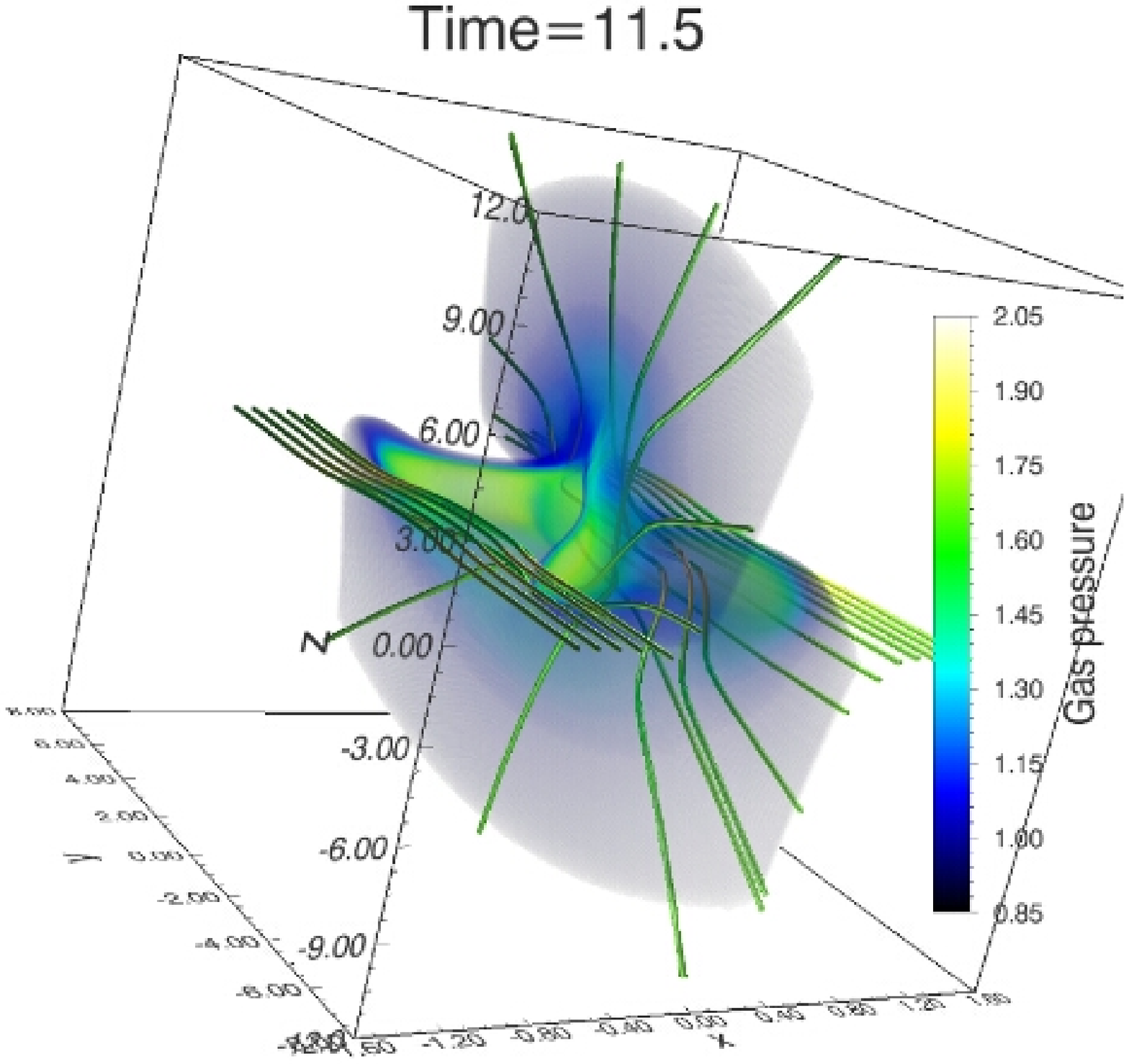}
\includegraphics[width=\ptb]{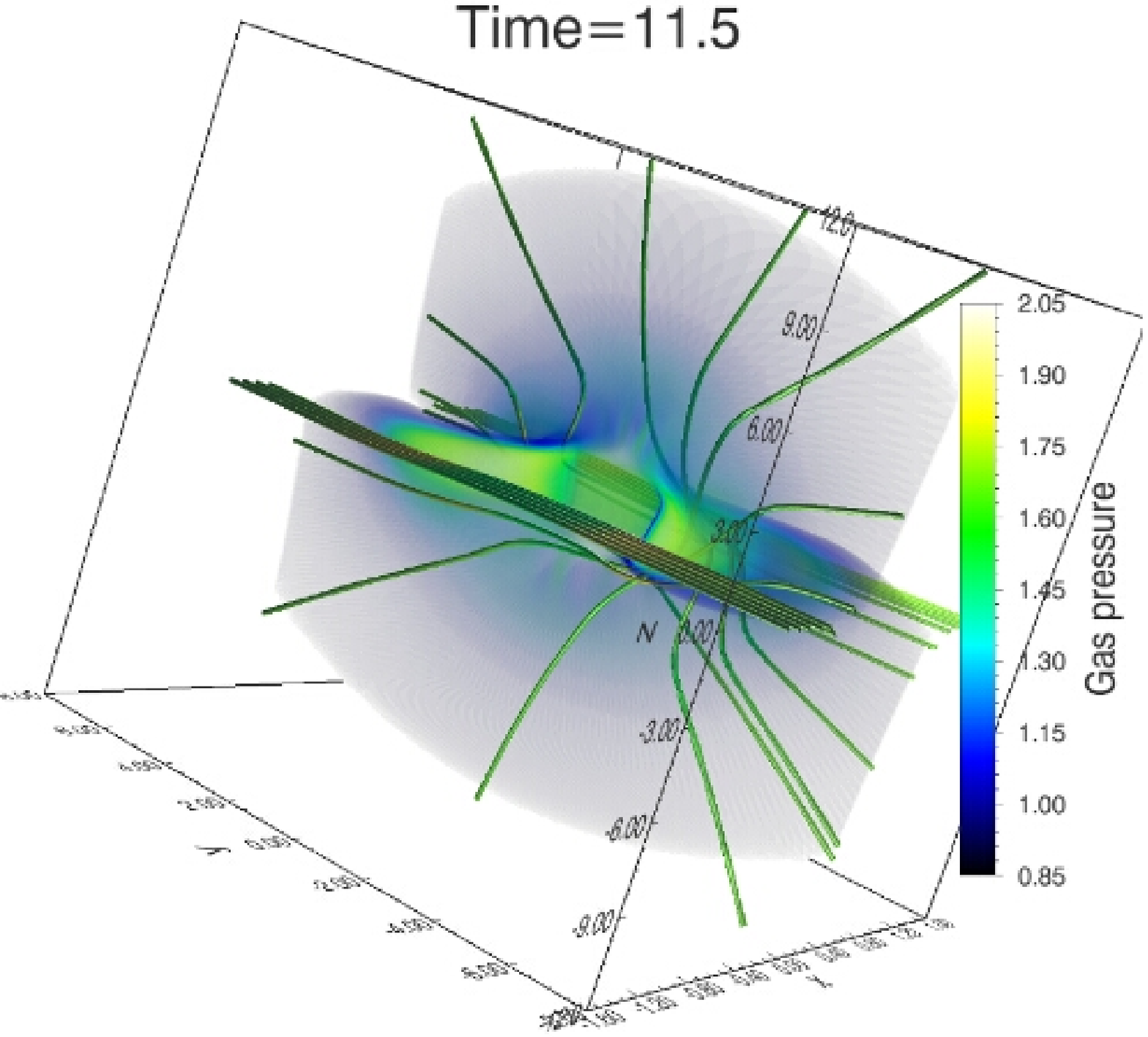}
\includegraphics[width=\ptb]{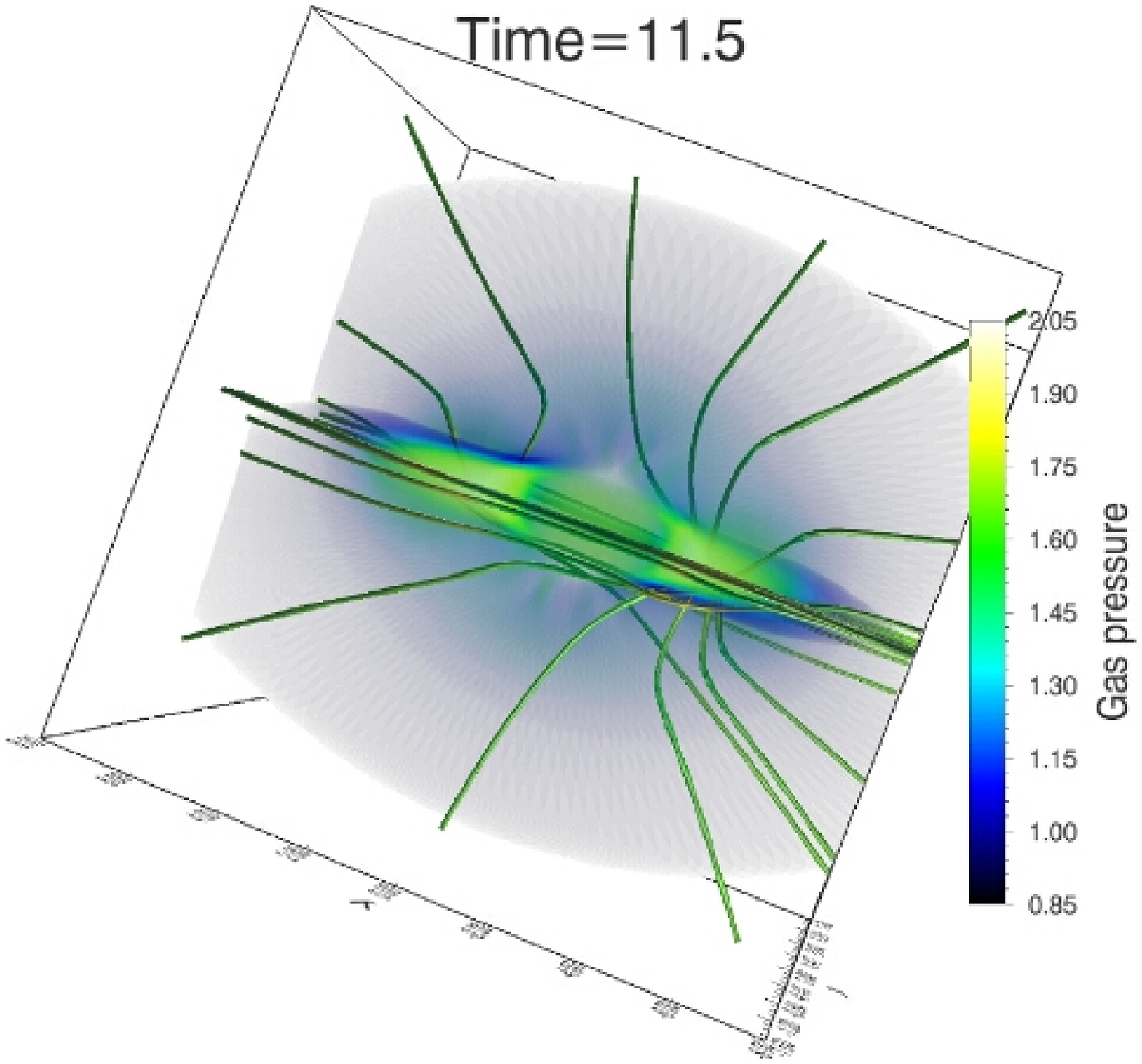}
\caption{3D visualization of the gas pressure distributions at different view points (typical case). The bright colors and high opacity stand for the high gas pressure and the low value is transparent. Solid tubes are magnetic field lines.} \label{fig03}
\end{figure}

Our typical case is the one in which the reconnection angle is $\pi$ with a plasma beta $\beta = 0.8$ and $\eta_{ini}=0.050$ (case 2 in Table~\ref{cases}). Figure~\ref{fig02} shows the 3D visualization of gas pressure distribution at different times. The bright colors and high opacity stand for the high values. The volume rendering in the 3D visualization does not show the low gas pressure volumes which exist everywhere in the computational box. Magnetic field lines are drawn from Lagrangian fluid elements located outside of the diffusion region, and the same field lines are shown in the upper panels of Figure~\ref{fig02}. We can see a fan-shaped structure in the current sheet. For the lower panels, no magnetic field lines are plotted but we add a cross section near $z=0$ and a velocity field plane at $x=0$. The velocity arrows show the fan-shaped structure almost along the guide field lines (there is no legend for the velocity field, the scale for the velocity is useless in these 3D images because of the projection effect). The cross section located near $z=0.0$ shows the ordinary magnetic reconnection jets in this plane which is very similar to the 2D or 2.5D (two dimension with three components) simulations (\citealt{Chen2001, Yokoyama2001, Jiang2010}). We will discuss this in detail in the next paragraph. Figure~\ref {fig03} shows the 3D results at different view points at the time $=11.5$. From this figure we can see the fan-shaped structure jets more clearly.

\begin{figure}[htbp]
\centering
\includegraphics[height=\ptb,trim= 4mm 10mm 36mm 16mm,clip]{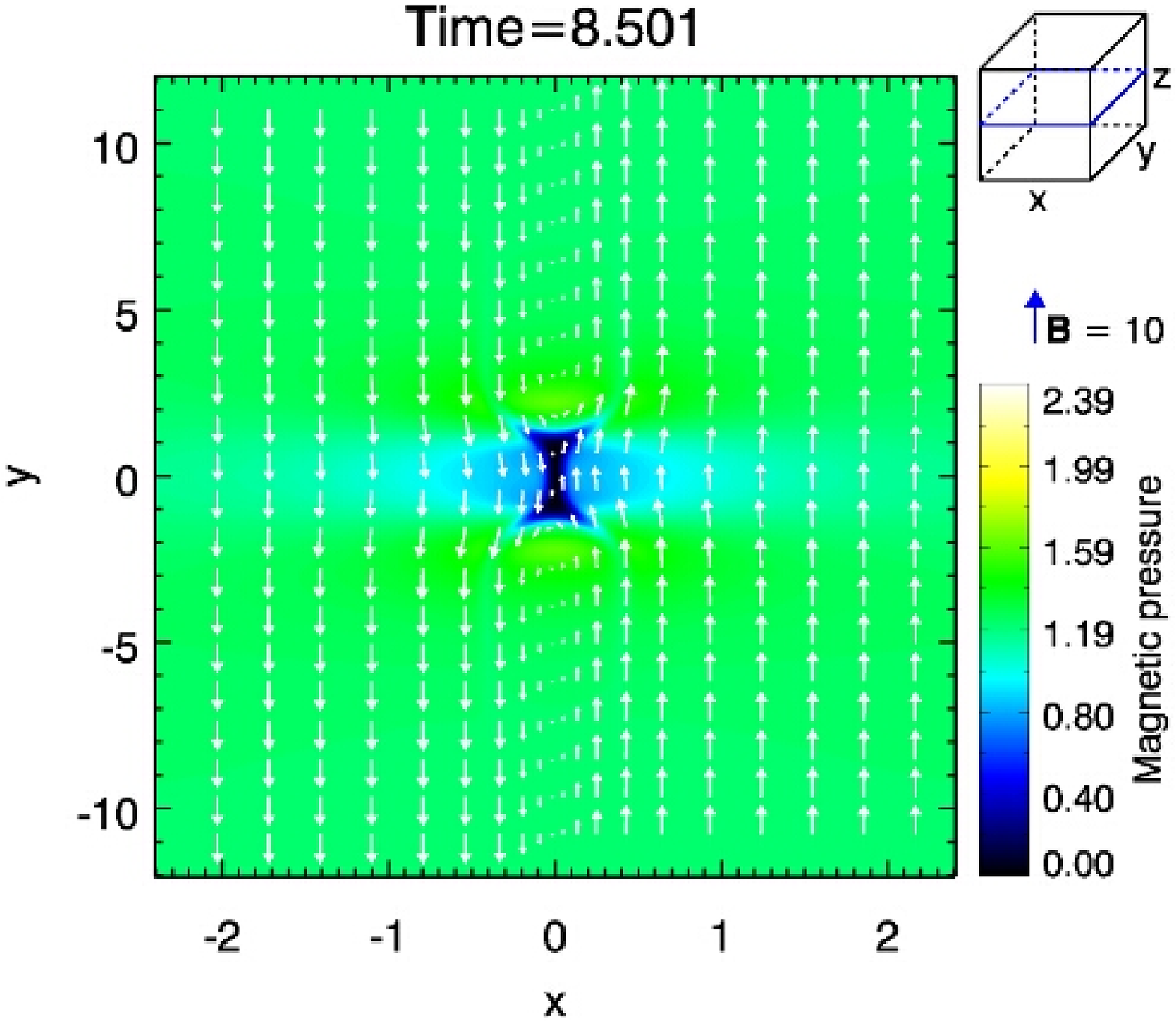}
\includegraphics[height=\ptb,trim= 4mm 10mm  4mm 16mm,clip]{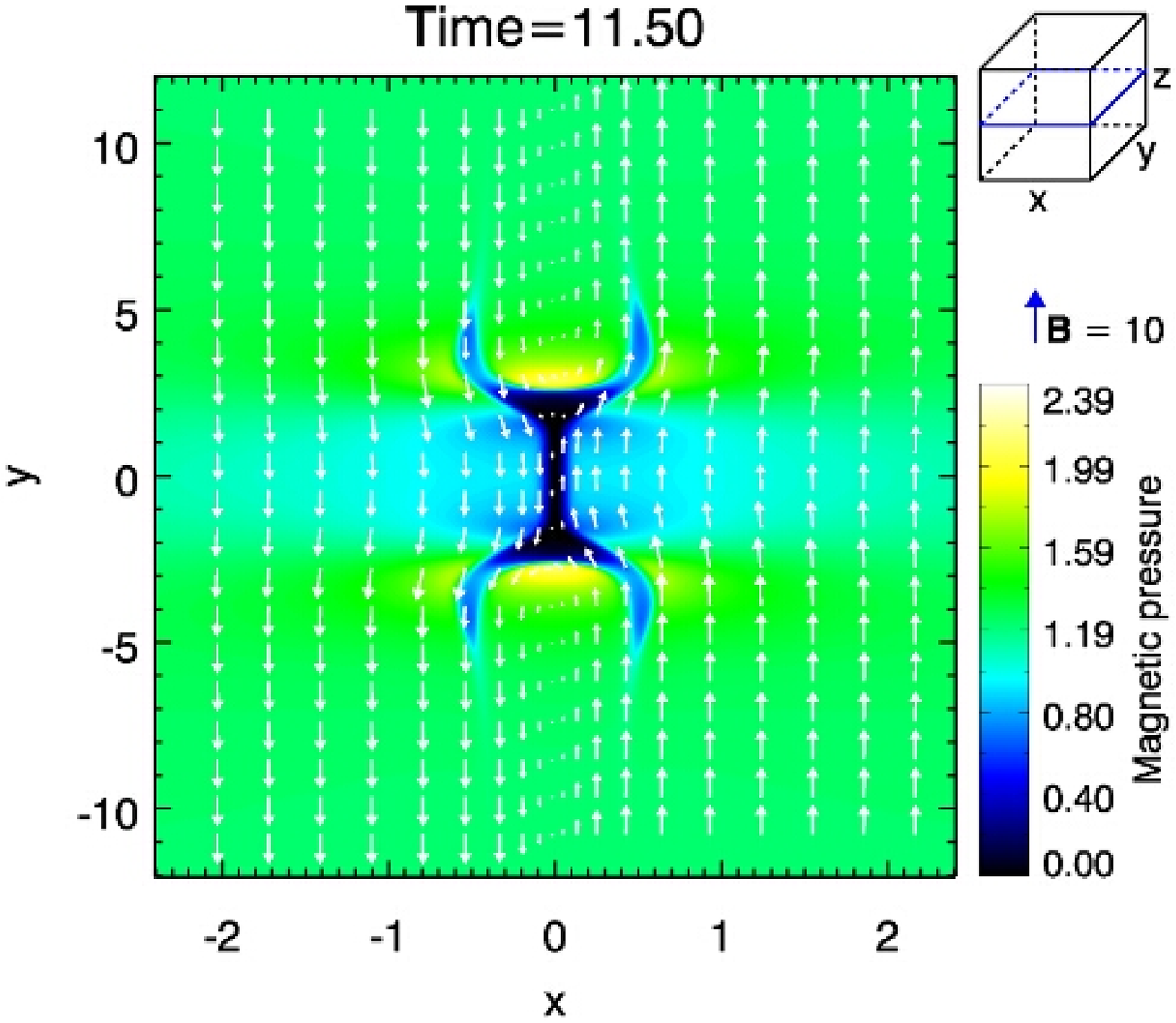}
\includegraphics[height=\ptb,trim= 4mm 10mm 36mm 16mm,clip]{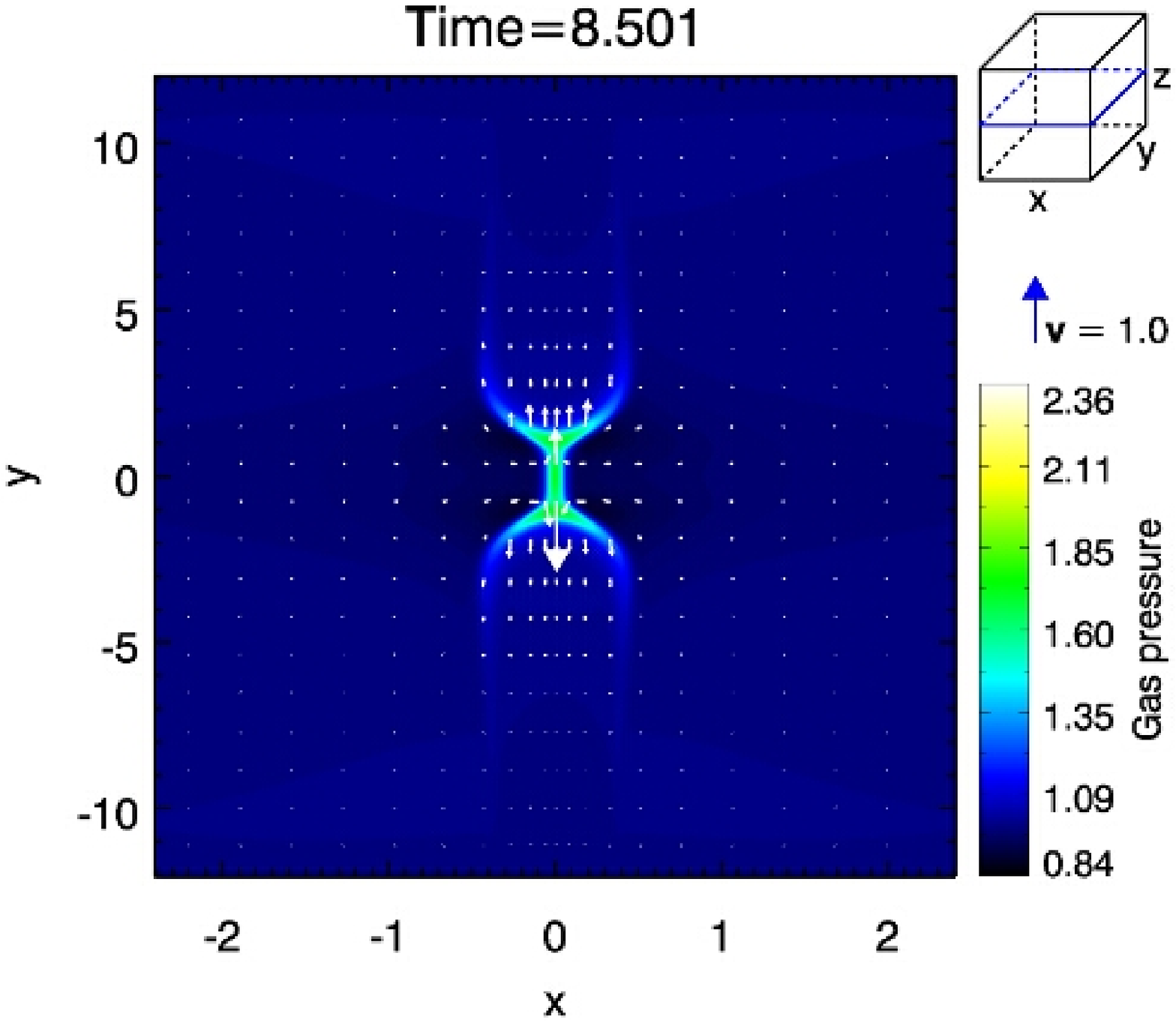}
\includegraphics[height=\ptb,trim= 4mm 10mm  4mm 16mm,clip]{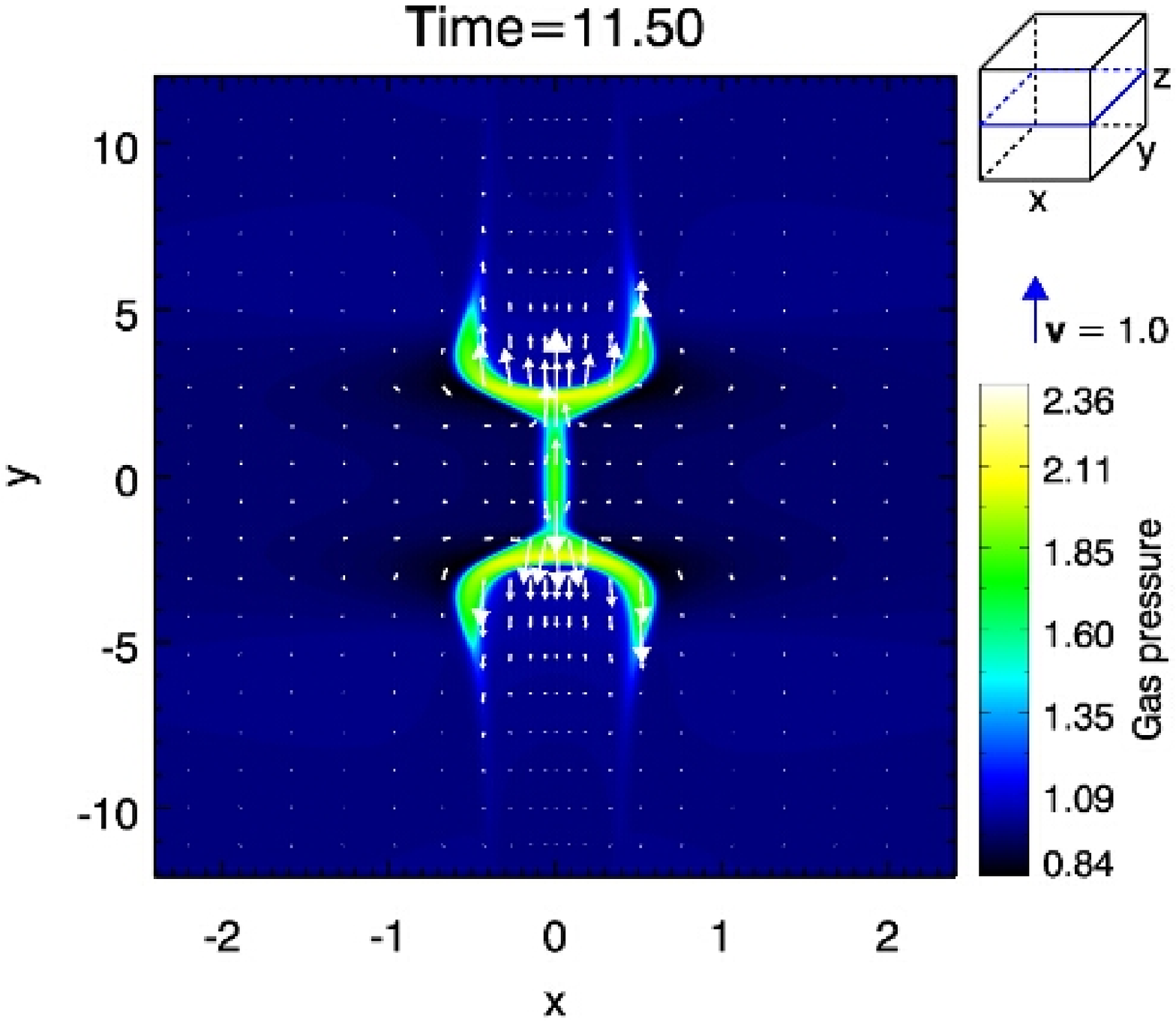}
\includegraphics[height=\ptb,trim= 4mm 10mm 36mm 16mm,clip]{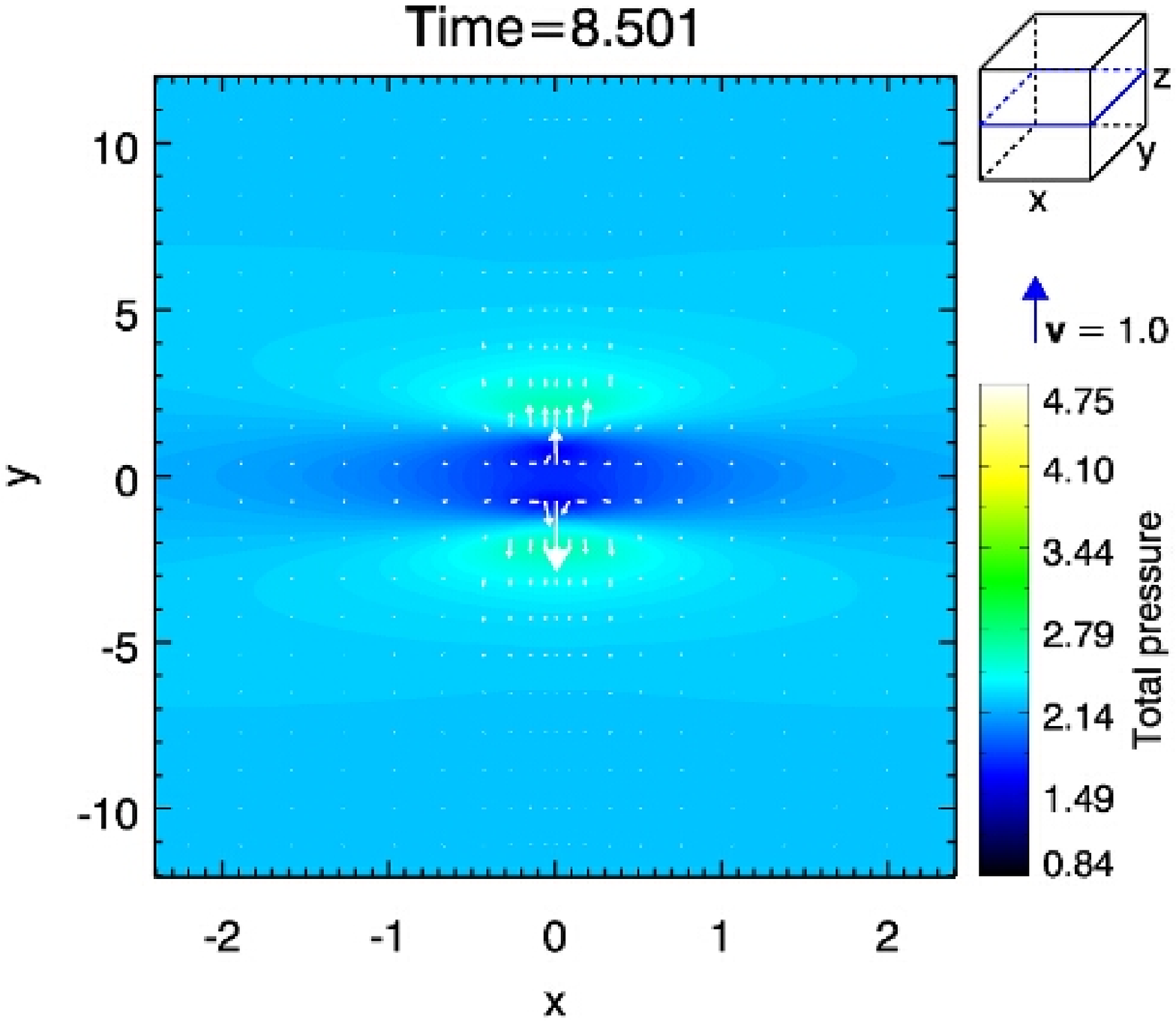}
\includegraphics[height=\ptb,trim= 4mm 10mm  4mm 16mm,clip]{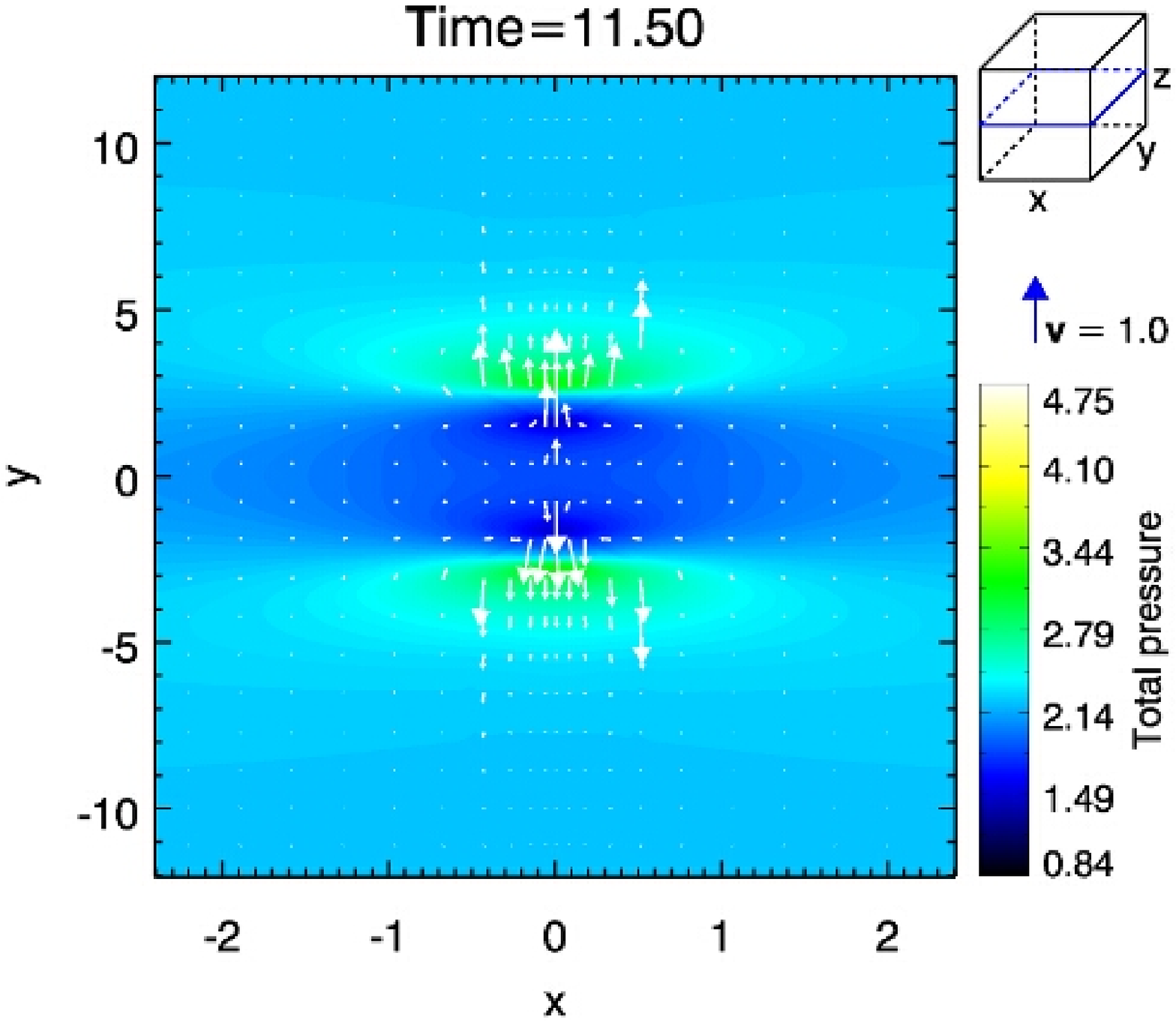}
\caption{2D distributions (typical case) of the gas pressure, magnetic pressure and total pressure in $x-y$ plane at $z=0.0$. The arrows in the upper panels indicate the magnetic field and others the velocity field.}
\label{fig04}
\end{figure}

\begin{figure}[htbp]
\centering
\includegraphics[height=\ptb,trim= 4mm 10mm 36mm 16mm,clip]{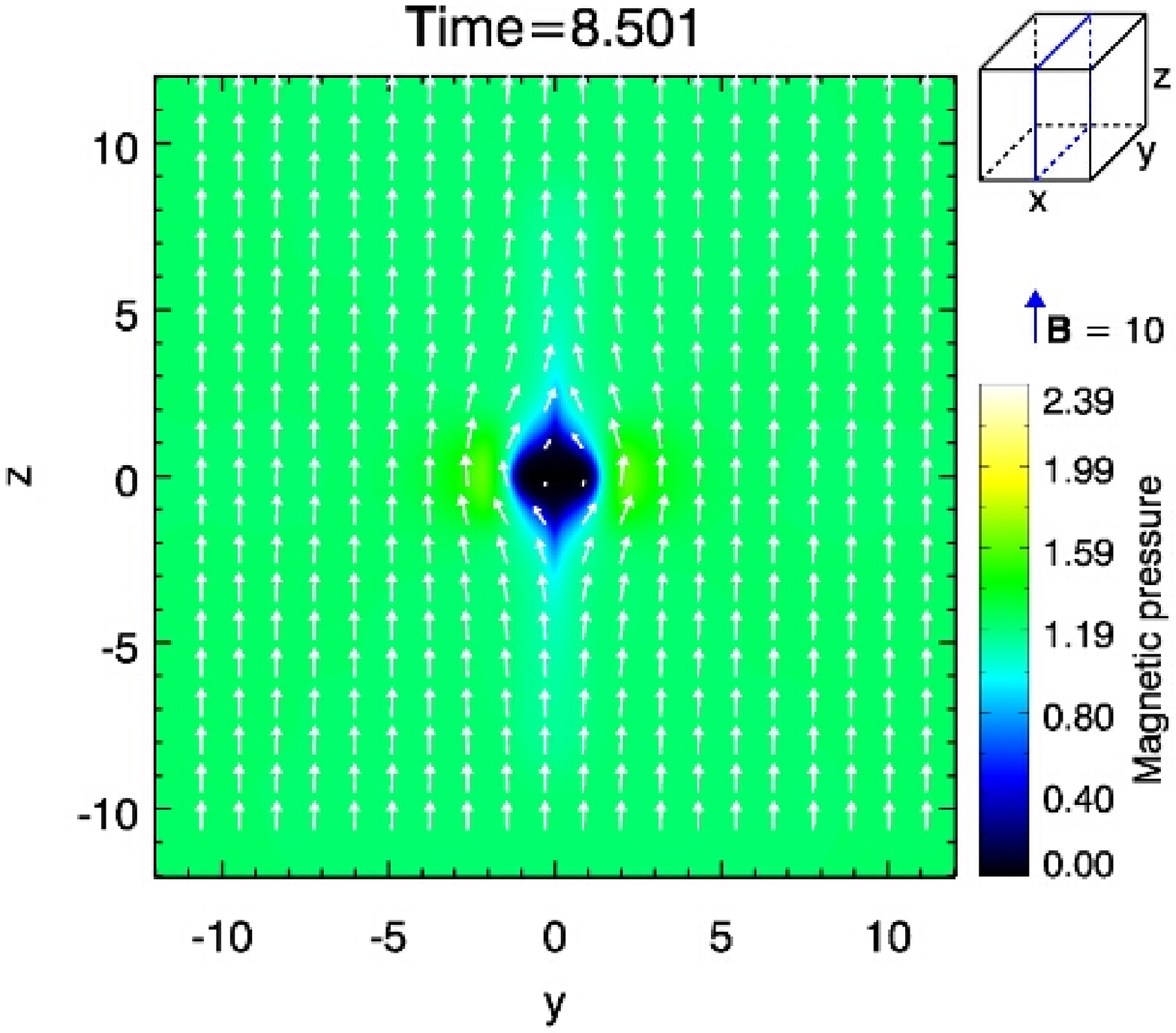}
\includegraphics[height=\ptb,trim= 4mm 10mm  4mm 16mm,clip]{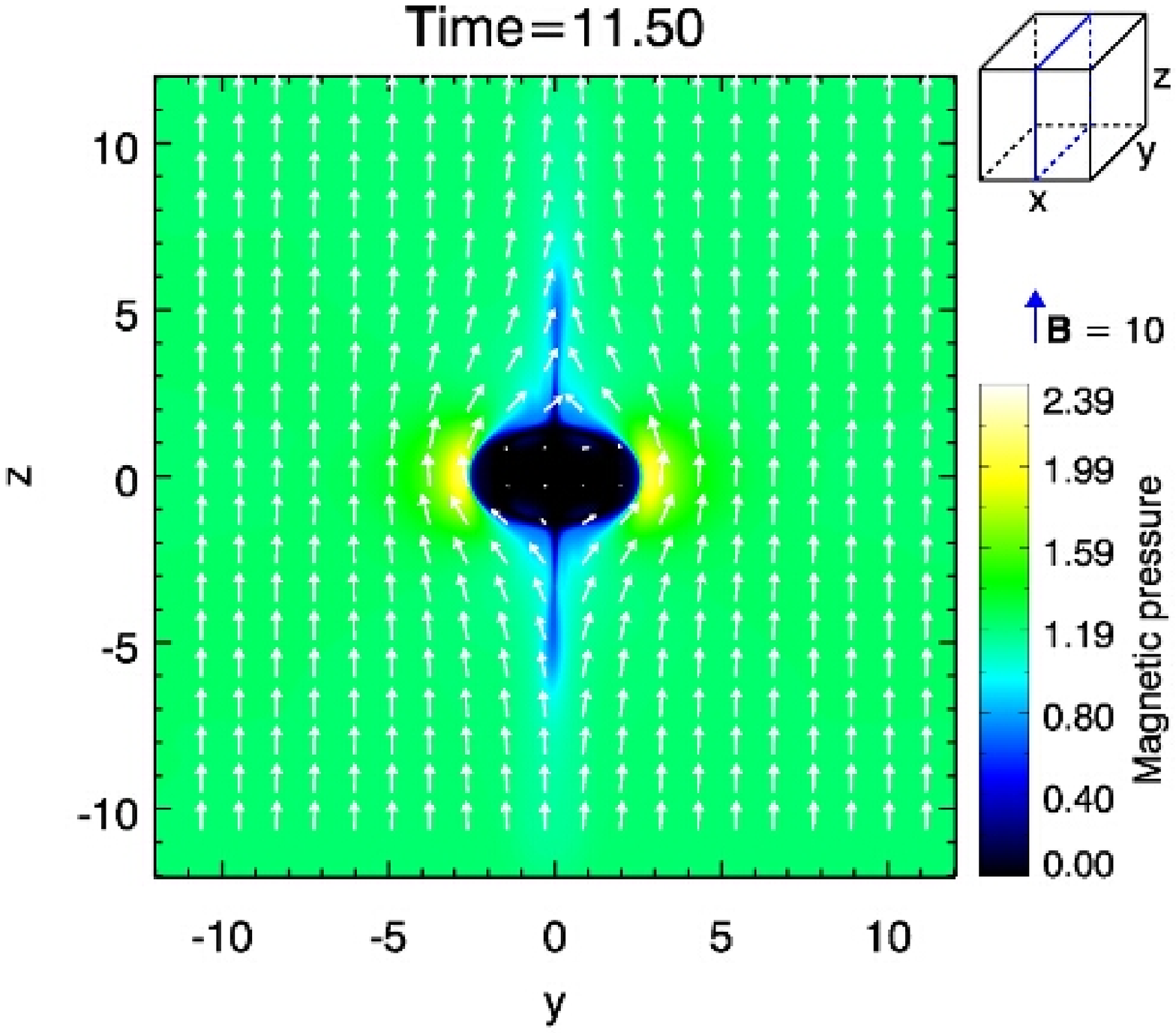}
\includegraphics[height=\ptb,trim= 4mm 10mm 36mm 16mm,clip]{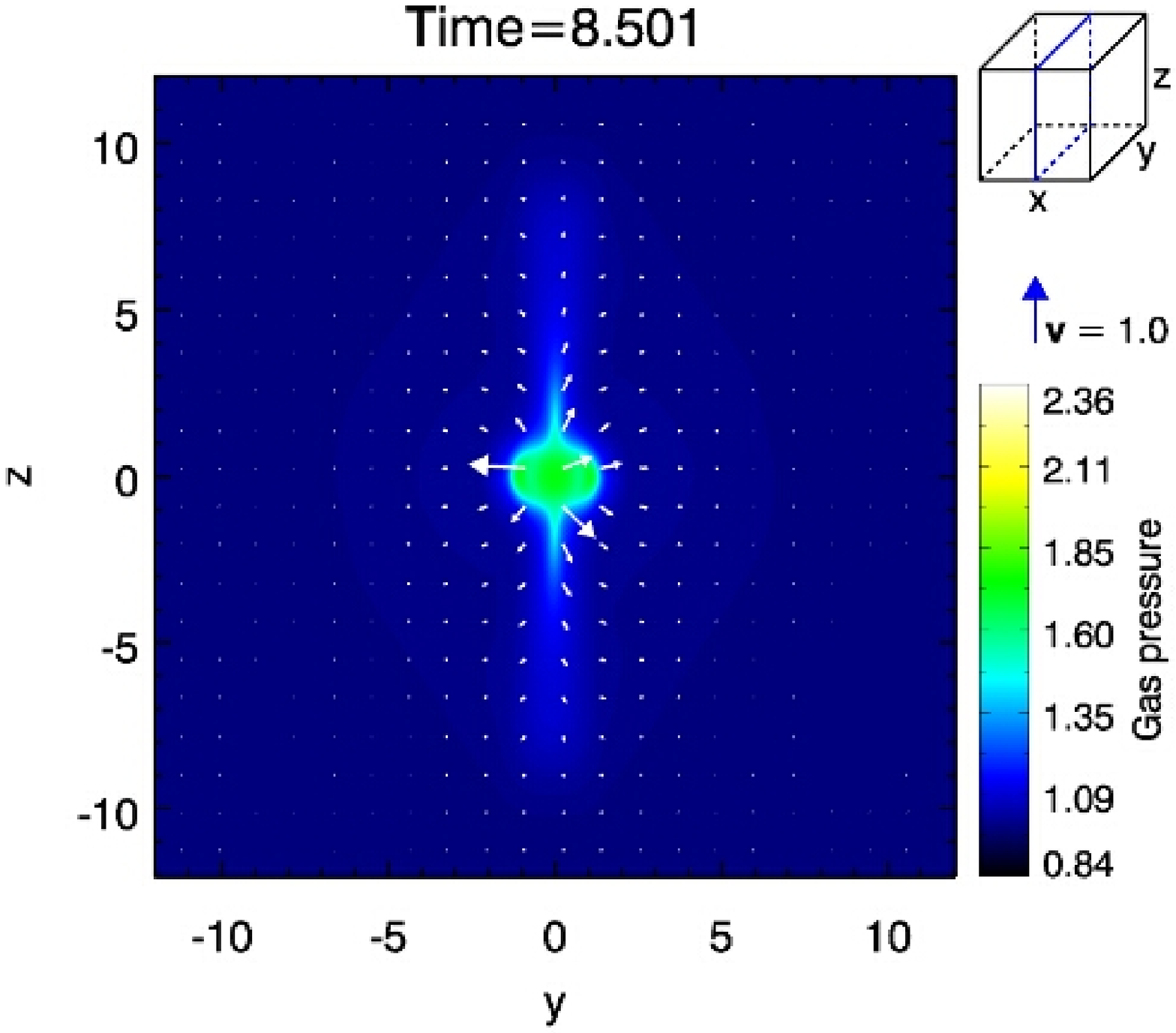}
\includegraphics[height=\ptb,trim= 4mm 10mm  4mm 16mm,clip]{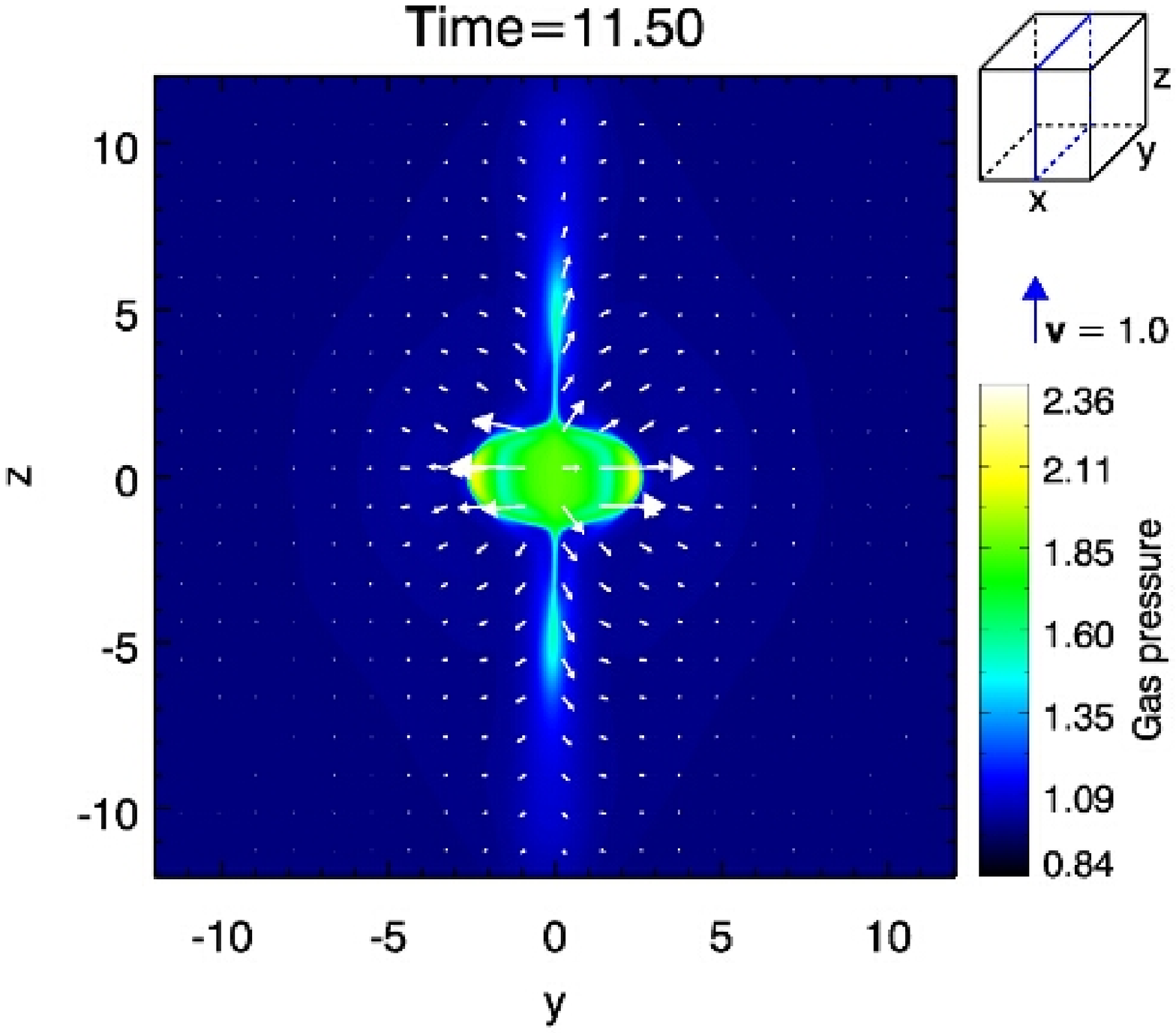}
\includegraphics[height=\ptb,trim= 4mm 10mm 36mm 16mm,clip]{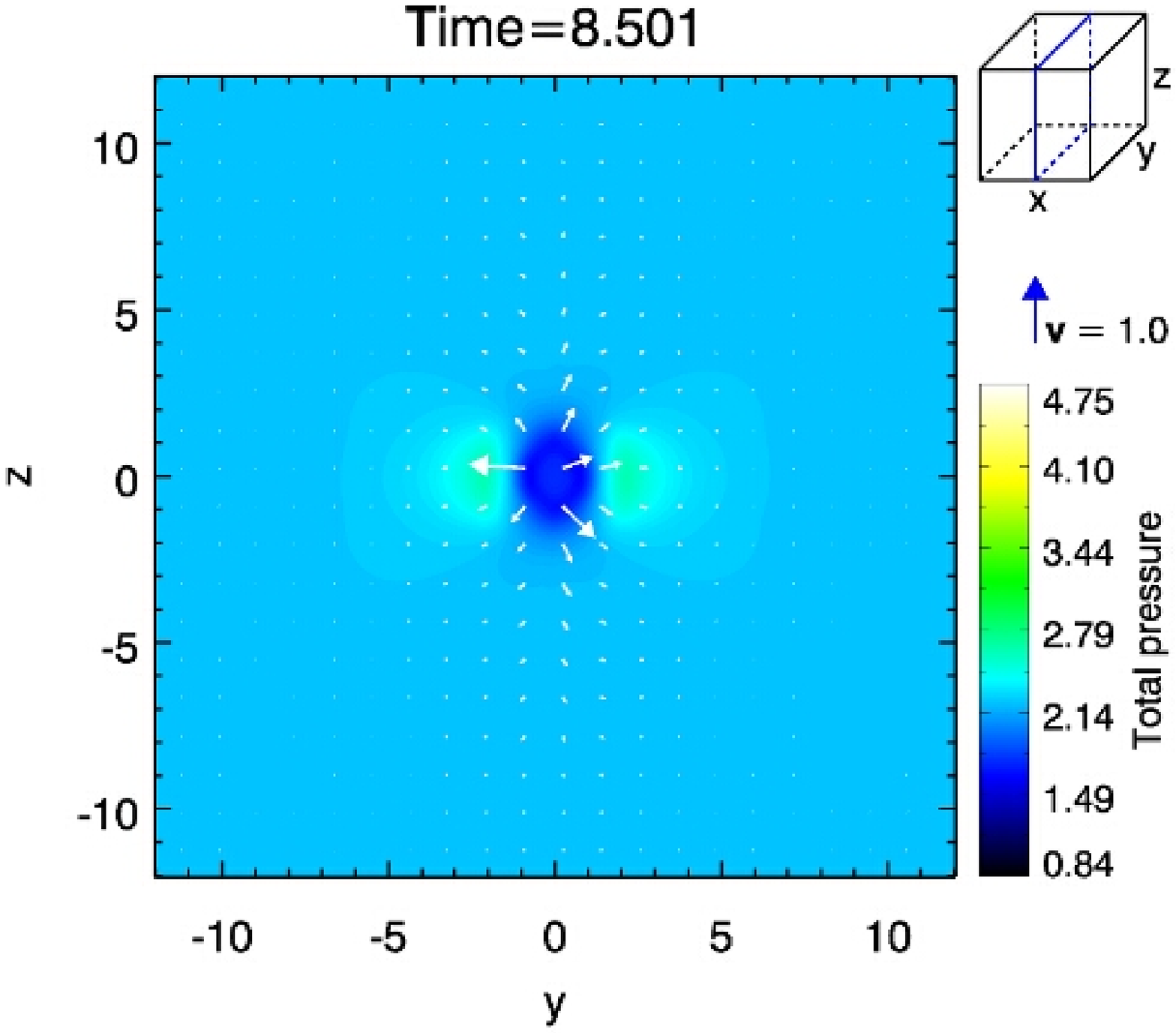}
\includegraphics[height=\ptb,trim= 4mm 10mm  4mm 16mm,clip]{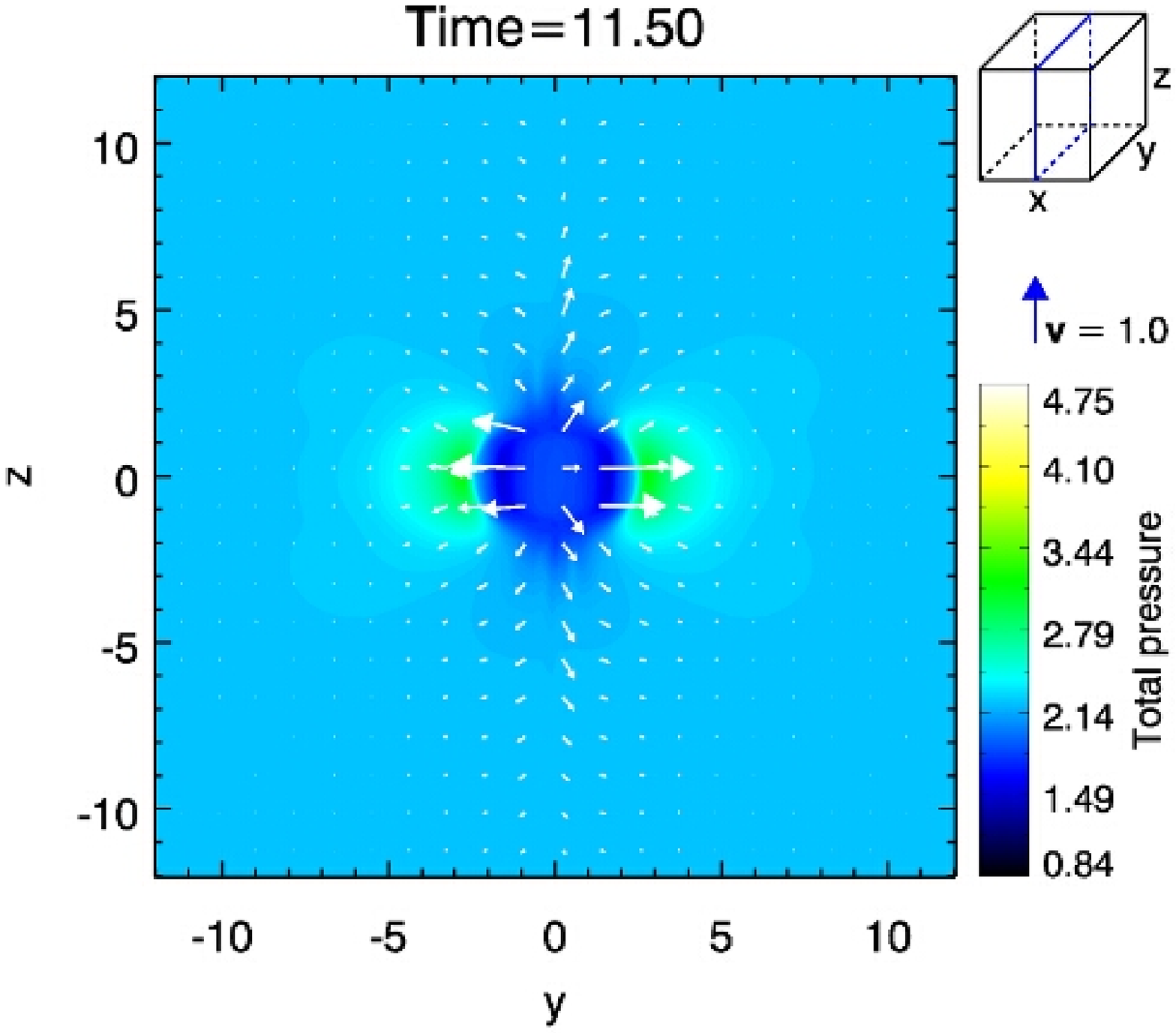}
 \caption{2D distributions (typical case) of gas pressure, magnetic pressure and total pressure in $y-z$ plane at $x=0.0$. The arrows in the upper panels indicate the magnetic field and others the velocity field.}
\label{fig05}
\end{figure}

\begin{figure}[htbp]
\centering
\includegraphics[height=\ptb,trim=4mm 10mm 36mm 16mm,clip]{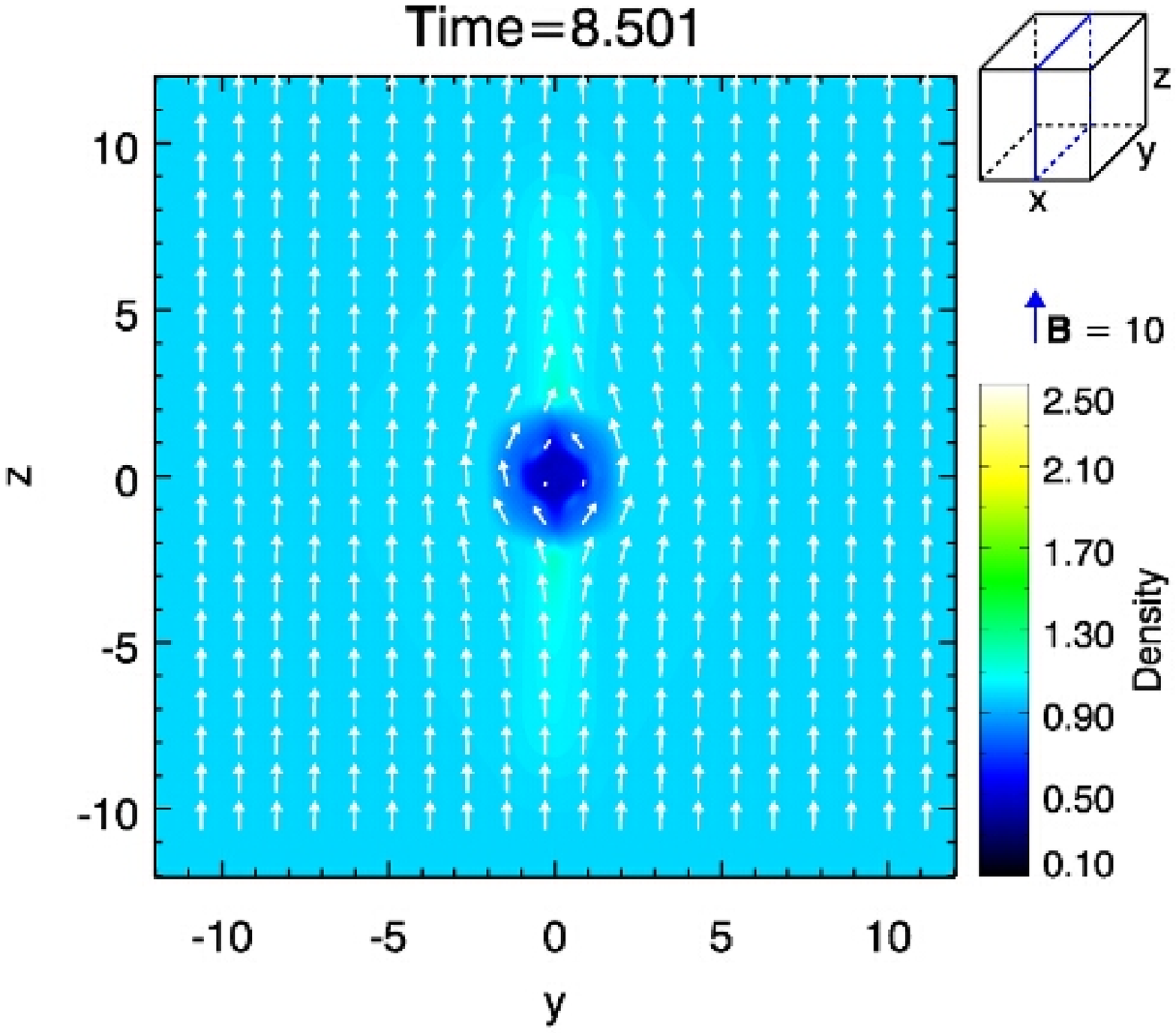}
\includegraphics[height=\ptb,trim=4mm 10mm  4mm 16mm,clip]{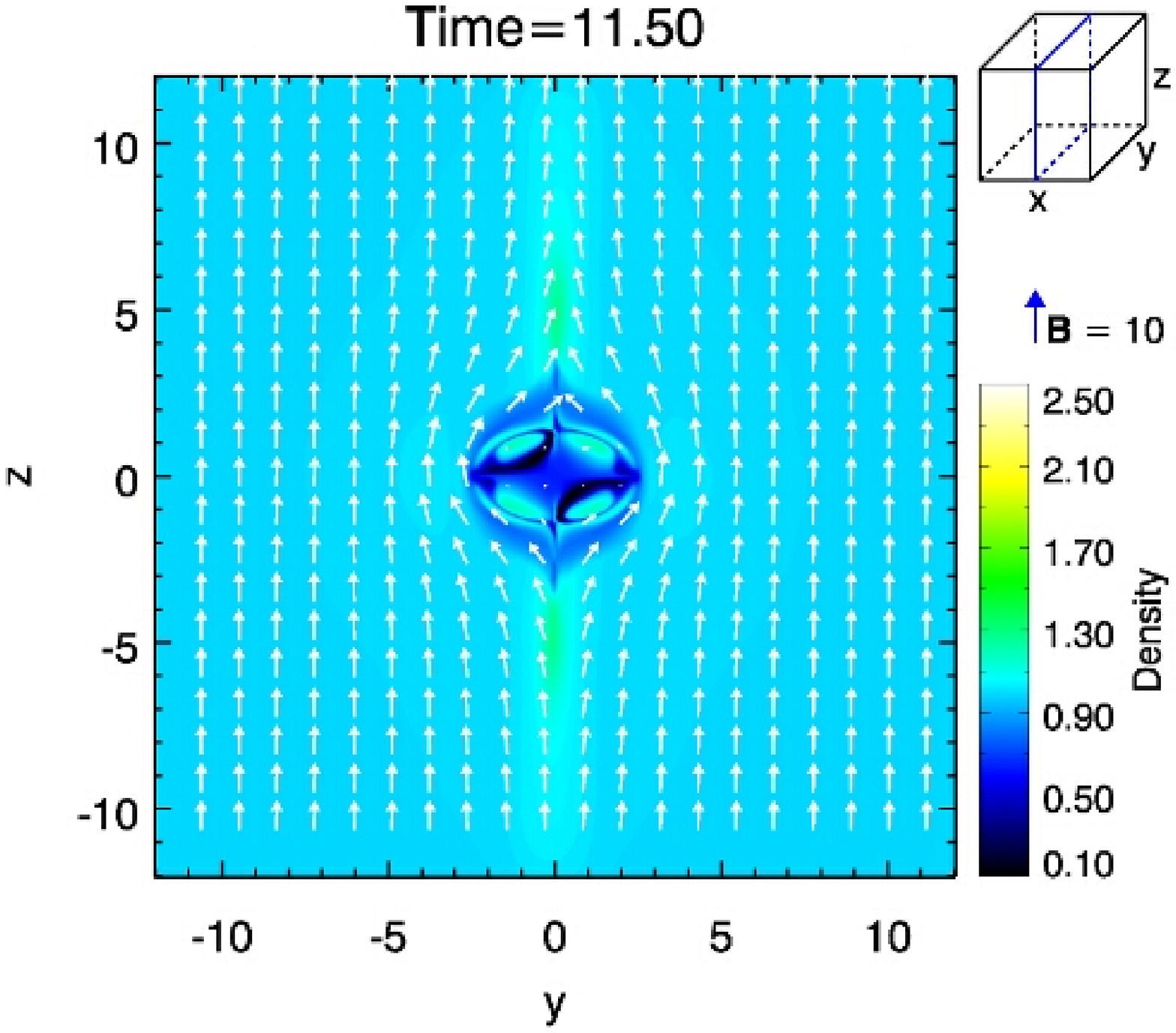}
\includegraphics[height=\ptb,trim=4mm 10mm 36mm 16mm,clip]{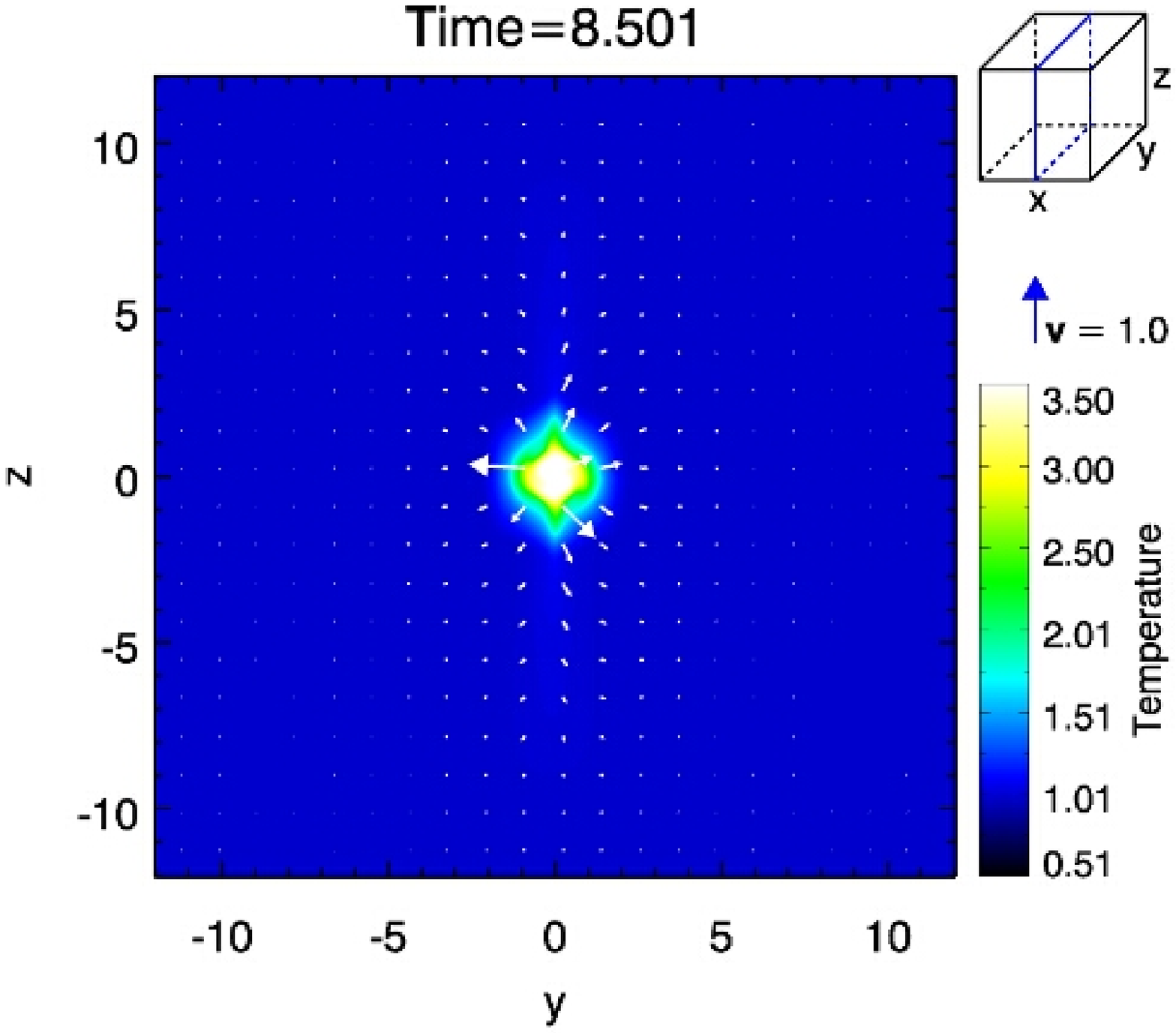}
\includegraphics[height=\ptb,trim=4mm 10mm  4mm 16mm,clip]{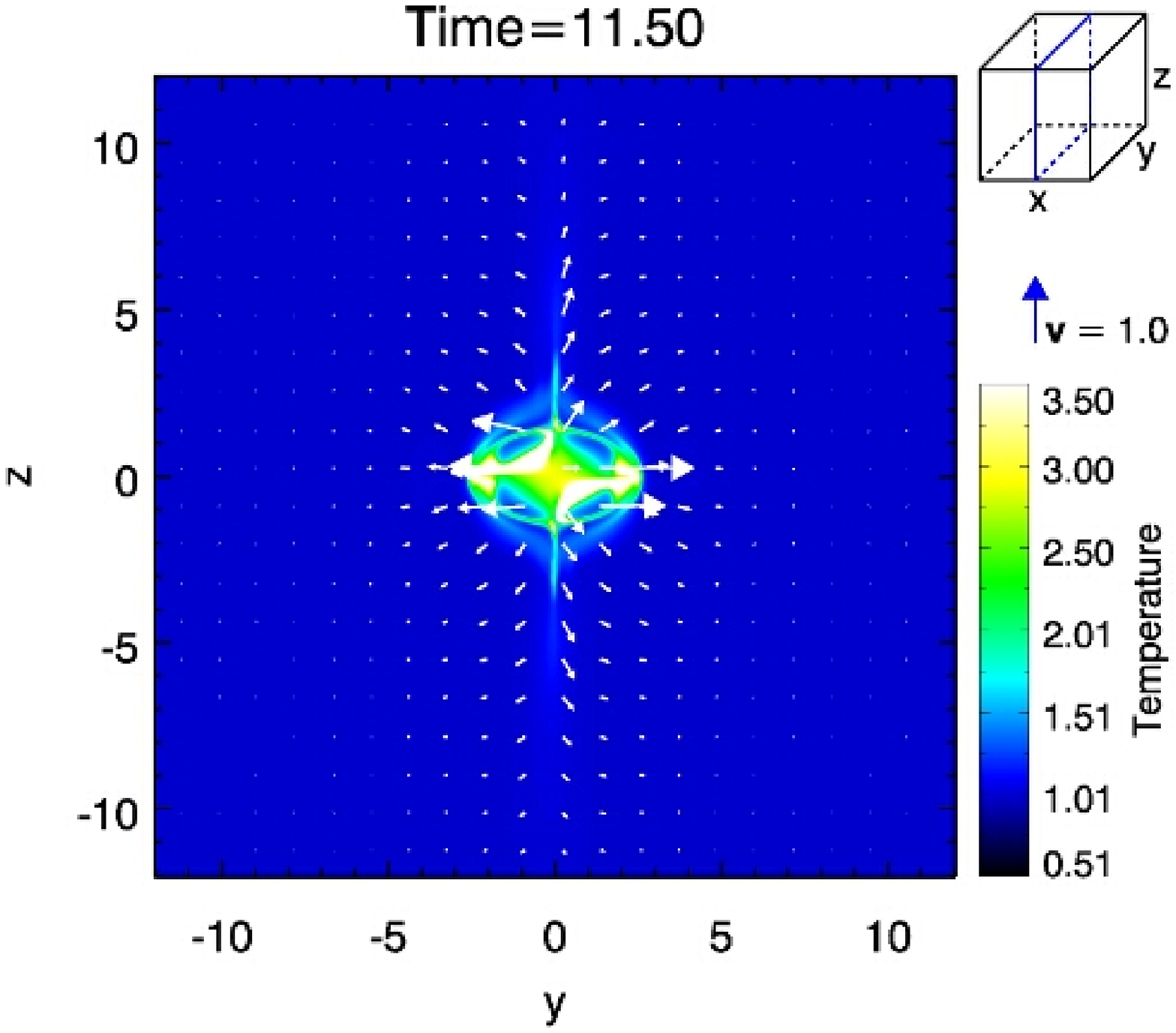}
\caption{2D distributions (typical case) of density and temperature in $y-z$ plane at $x=0.0$. The arrows in the upper panels indicate the magnetic field and others the velocity field.} \label{fig06}
\end{figure}

The 2D cross sections at different positions and different times are shown in Figures~\ref{fig04}-\ref{fig06}. Figure~\ref{fig04} depicts the gas pressure, magnetic pressure and total pressure distributions in $x-y$ plane at $z=0.0$ (we can also briefly understand the position by the small bos above the color bar in this figure). The guide field is perpendicular to this plane, which means what we present in this plane is similar to simulations in 2D or 2.5D case as we mentioned above. As reconnection goes on, the central diffusion region is heated by Joule dissipation, and a pair of ordinary reconnection jets are ejected by magnetic tension force. The central low magnetic pressure region is due to the magnetic reconnection process, which can be compared to the high gas pressure region. For a steady state, the two pressures are almost balanced with each other in the $x$ direction of the inflow region shown by the total pressure in the lower panels of Figure~\ref{fig04}. The X-shaped structures (we can find them in the panels showing the gas and magnetic pressures) almost disappear.

Figure~\ref{fig05} shows the gas pressure, magnetic pressure and total pressure distributions in $y-z$ plane at $x=0.0$. In this plane, the ordinary reconnection jets are almost perpendicular to the magnetic field lines (in other words, moving along the $y$ and $-y$ directions) and the jets almost moving along the magnetic field lines (along the $z$ and $-z$ directions) are shown only in a cross-section of the fan-shaped jets. The velocity of the fan-shaped jets is smaller than that of the ordinary reconnection jets whose velocity is approximately the Alfv\'en speed (about 1.58). High magnetic pressure regions have also been ejected as shown in the magnetic pressure distribution images in Figures~\ref{fig04} and~\ref{fig05} (high magnetic pressure parts, yellow color). The temperature and density distributions in this plane are shown in Figure~\ref{fig06}. Similar to the gas and magnetic pressure distributions, the distributions of the density and temperature are roughly opposite with each other.

\begin{figure}[htbp]
\centering
\includegraphics[height=\ptm,trim=5mm 10mm 36mm 16mm,clip]{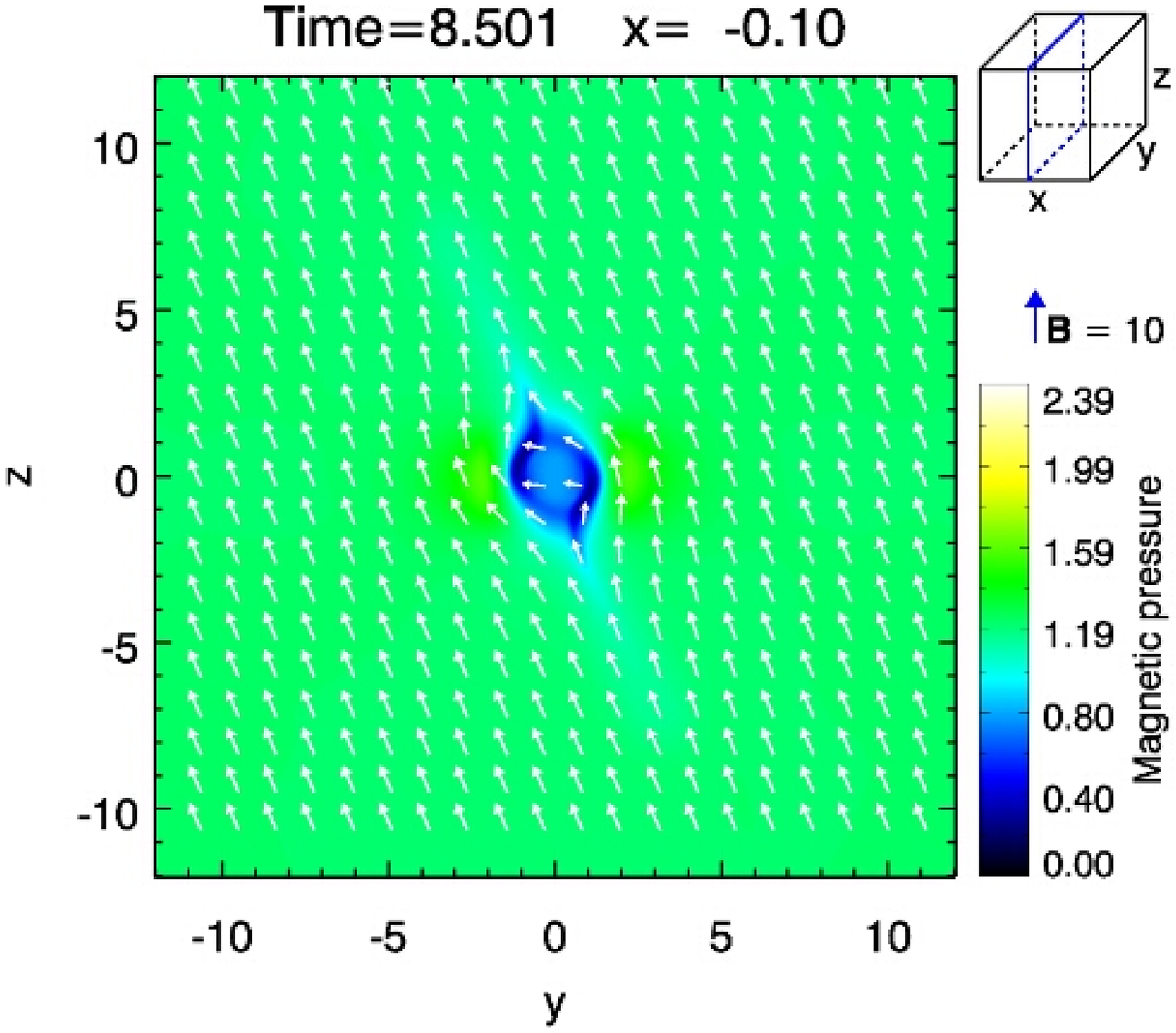}
\includegraphics[height=\ptm,trim=5mm 10mm 36mm 16mm,clip]{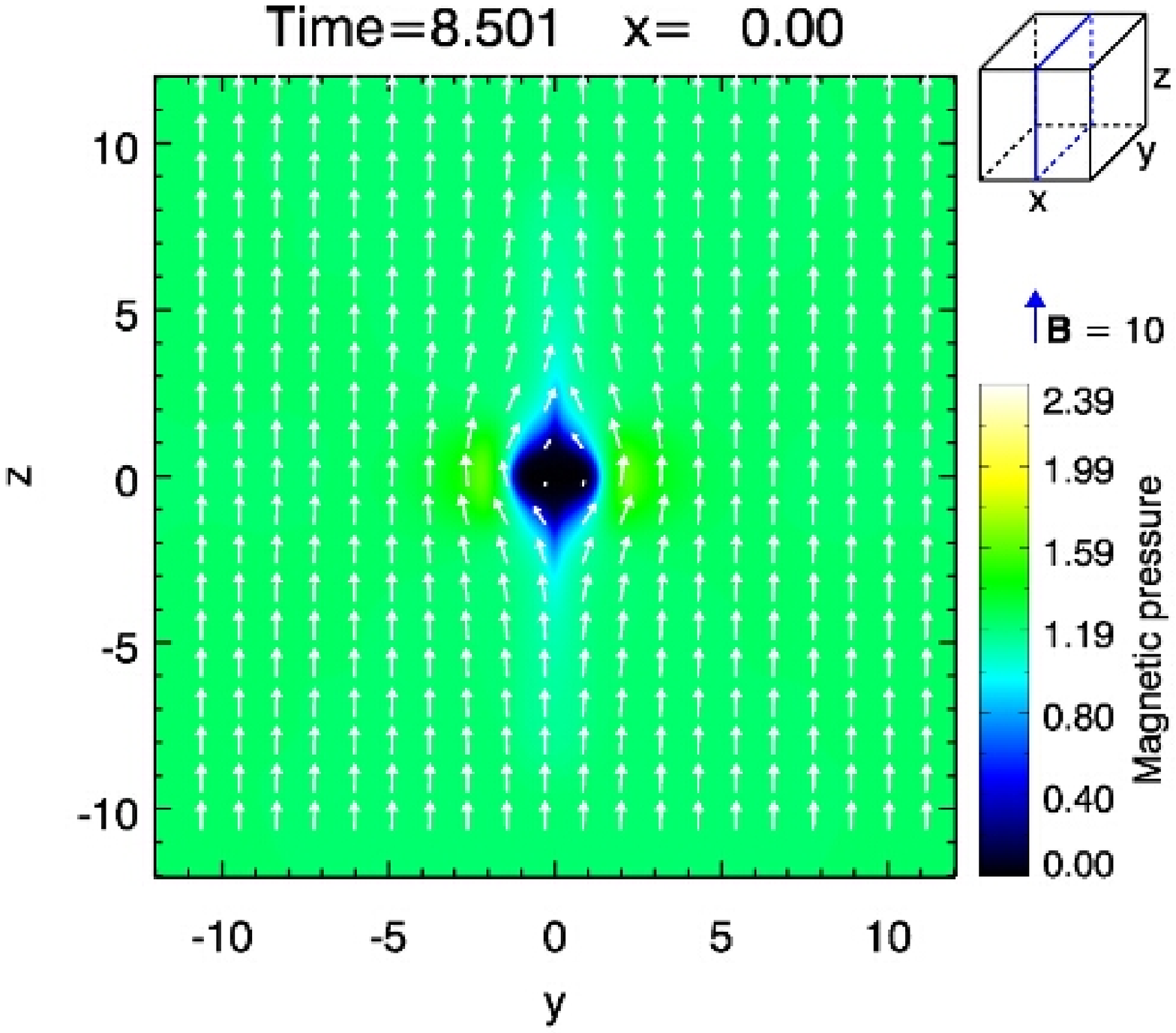}
\includegraphics[height=\ptm,trim=5mm 10mm  4mm 16mm,clip]{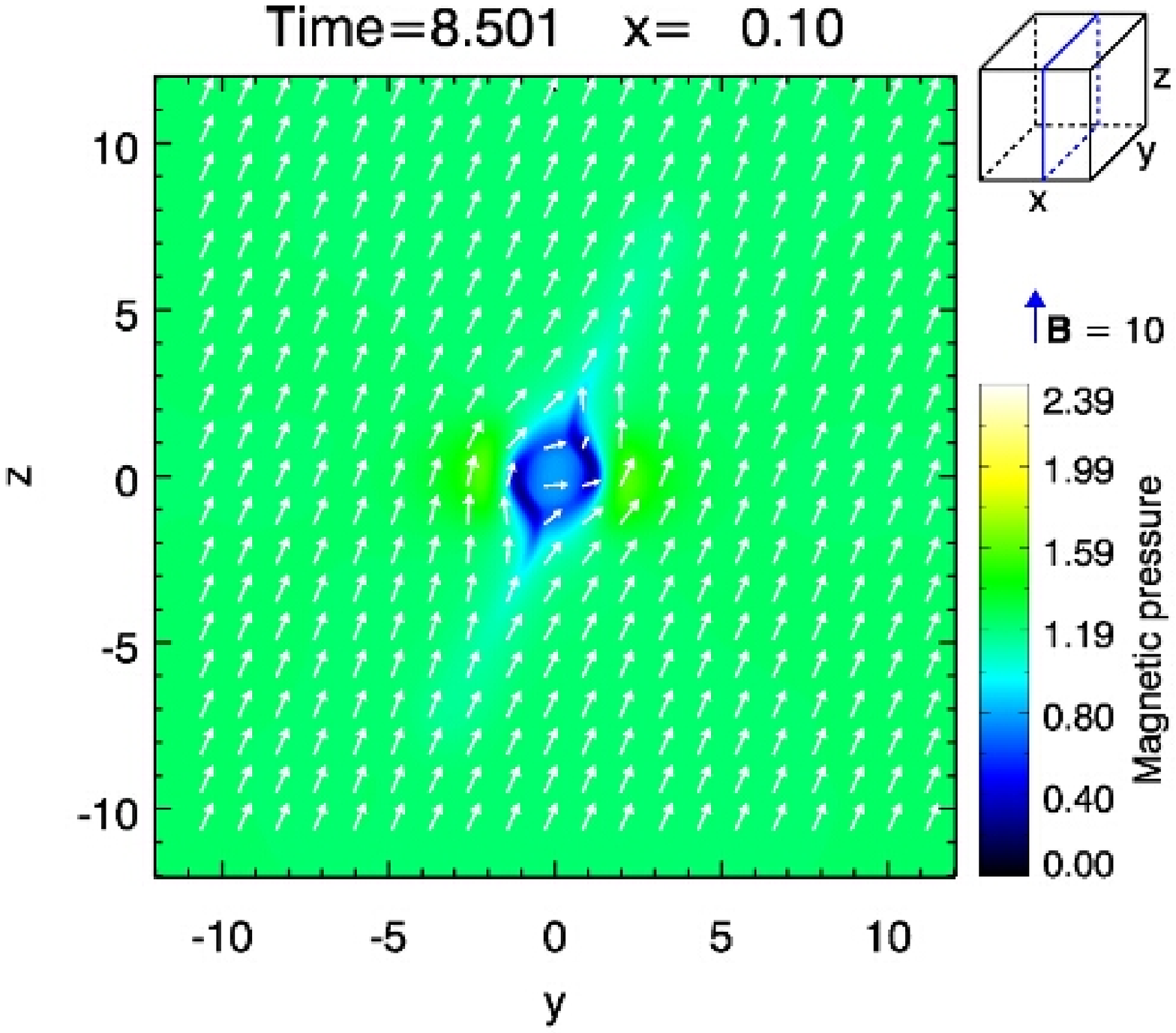}
\includegraphics[height=\ptm,trim=5mm 10mm 36mm 16mm,clip]{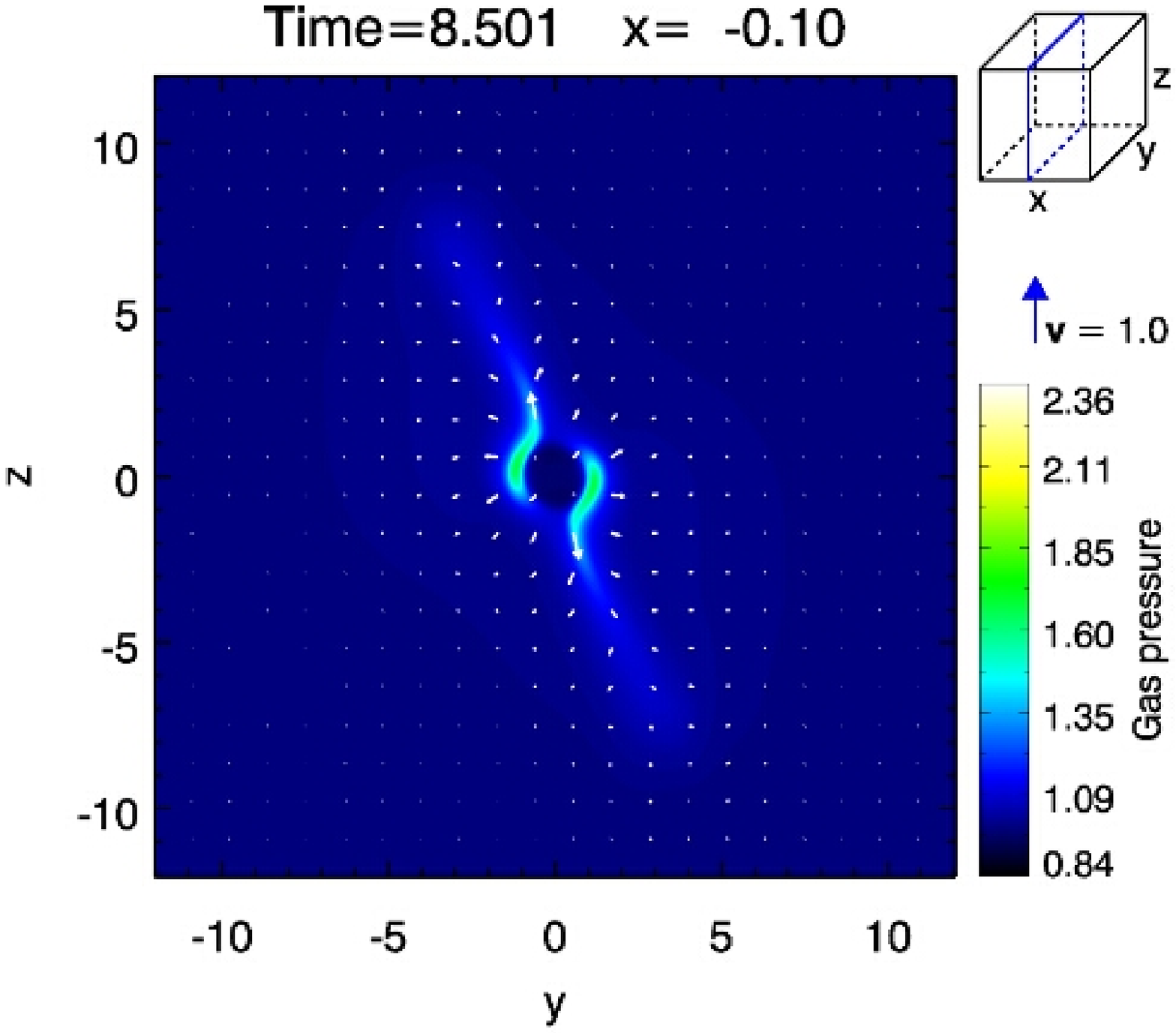}
\includegraphics[height=\ptm,trim=5mm 10mm 36mm 16mm,clip]{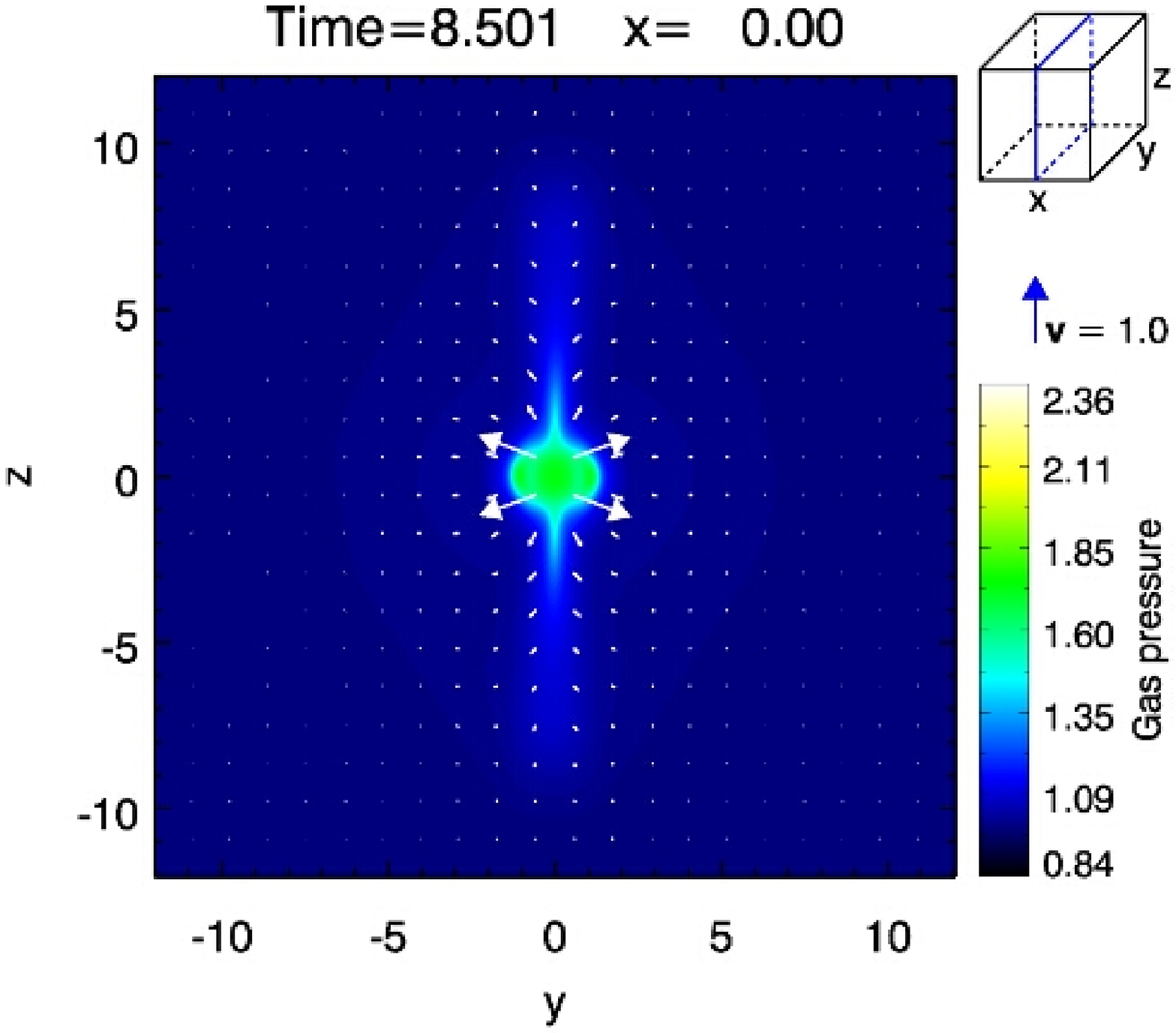}
\includegraphics[height=\ptm,trim=5mm 10mm  4mm 16mm,clip]{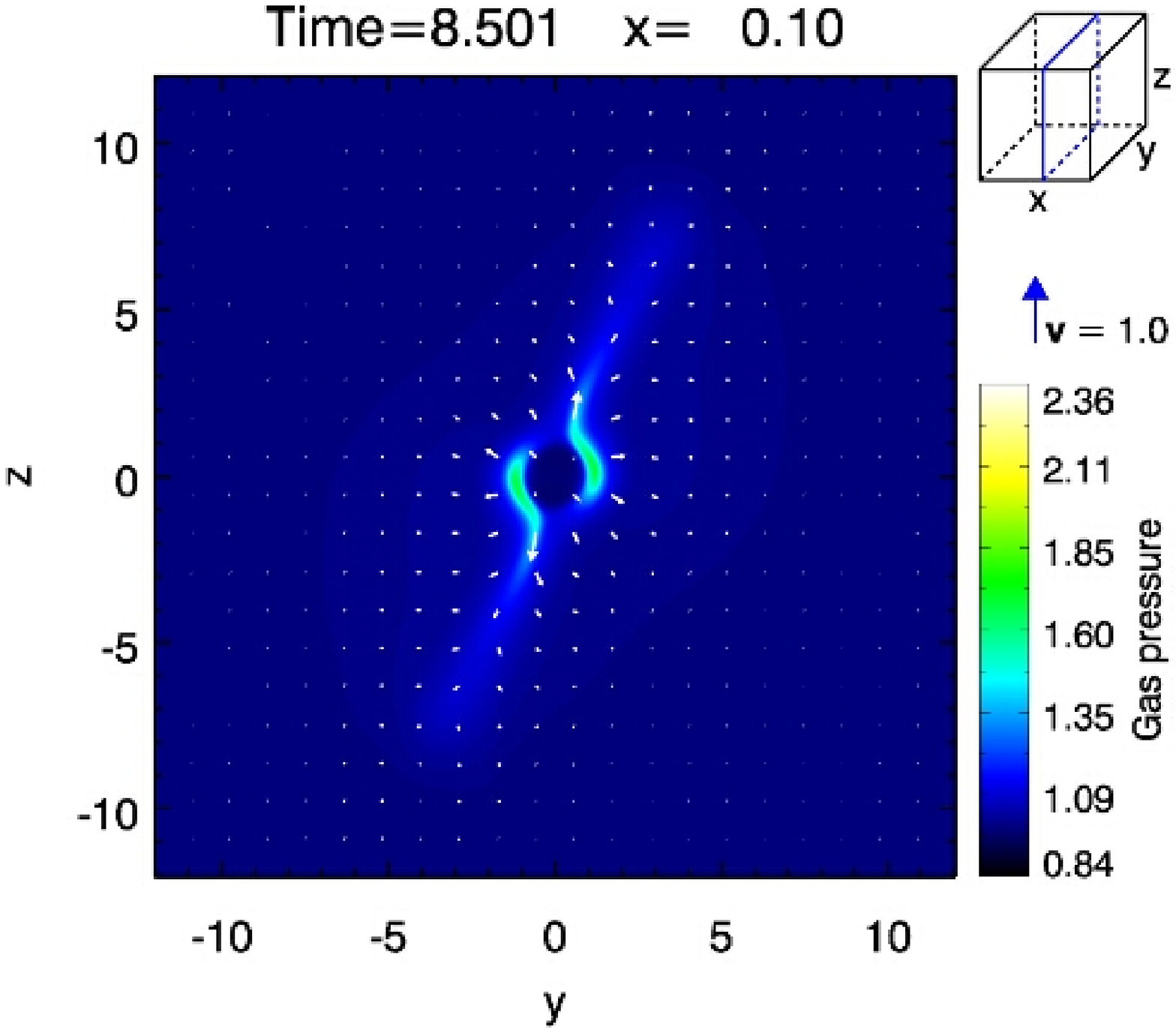}
\includegraphics[height=\ptm,trim=5mm 10mm 36mm 16mm,clip]{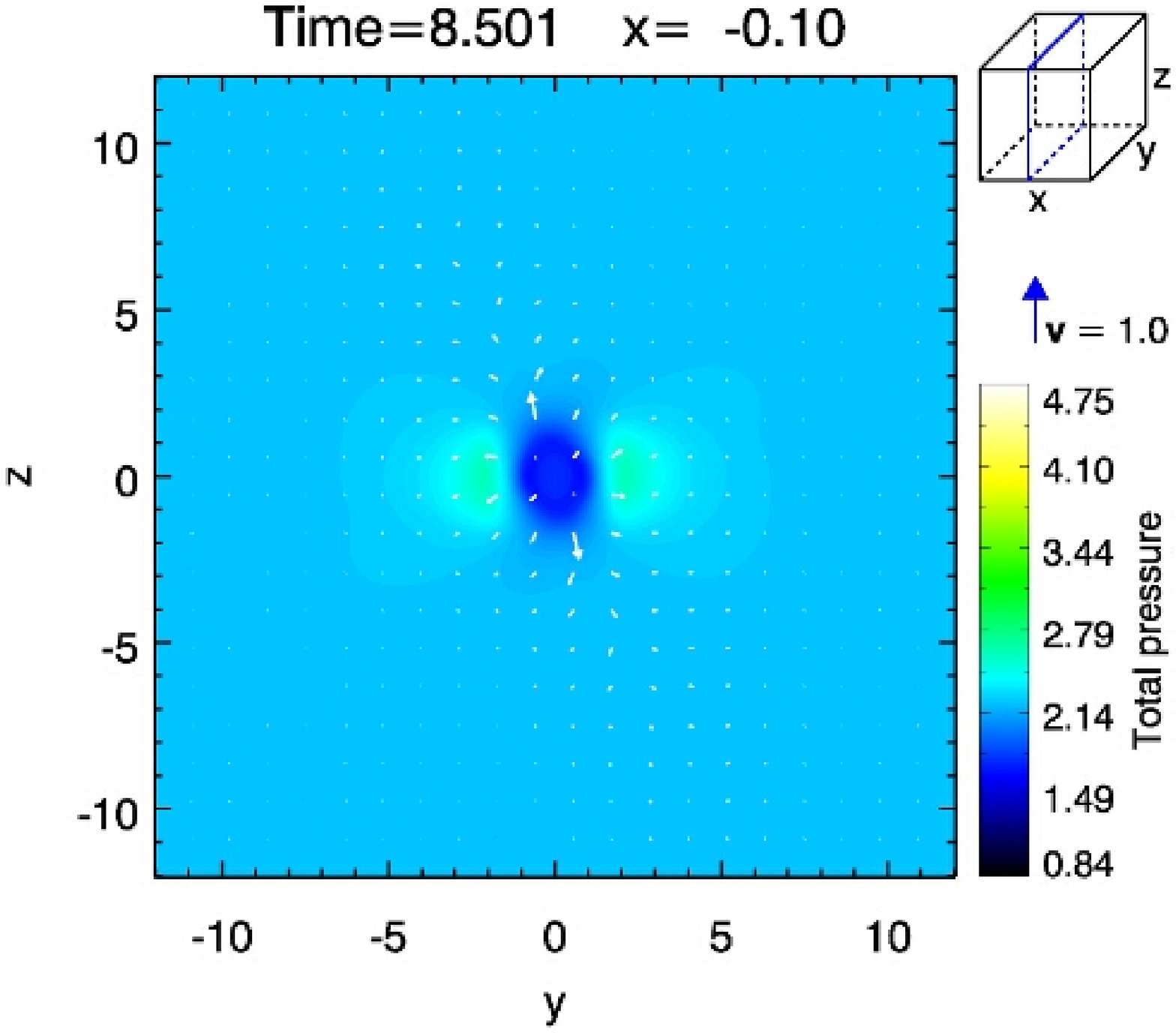}
\includegraphics[height=\ptm,trim=5mm 10mm 36mm 16mm,clip]{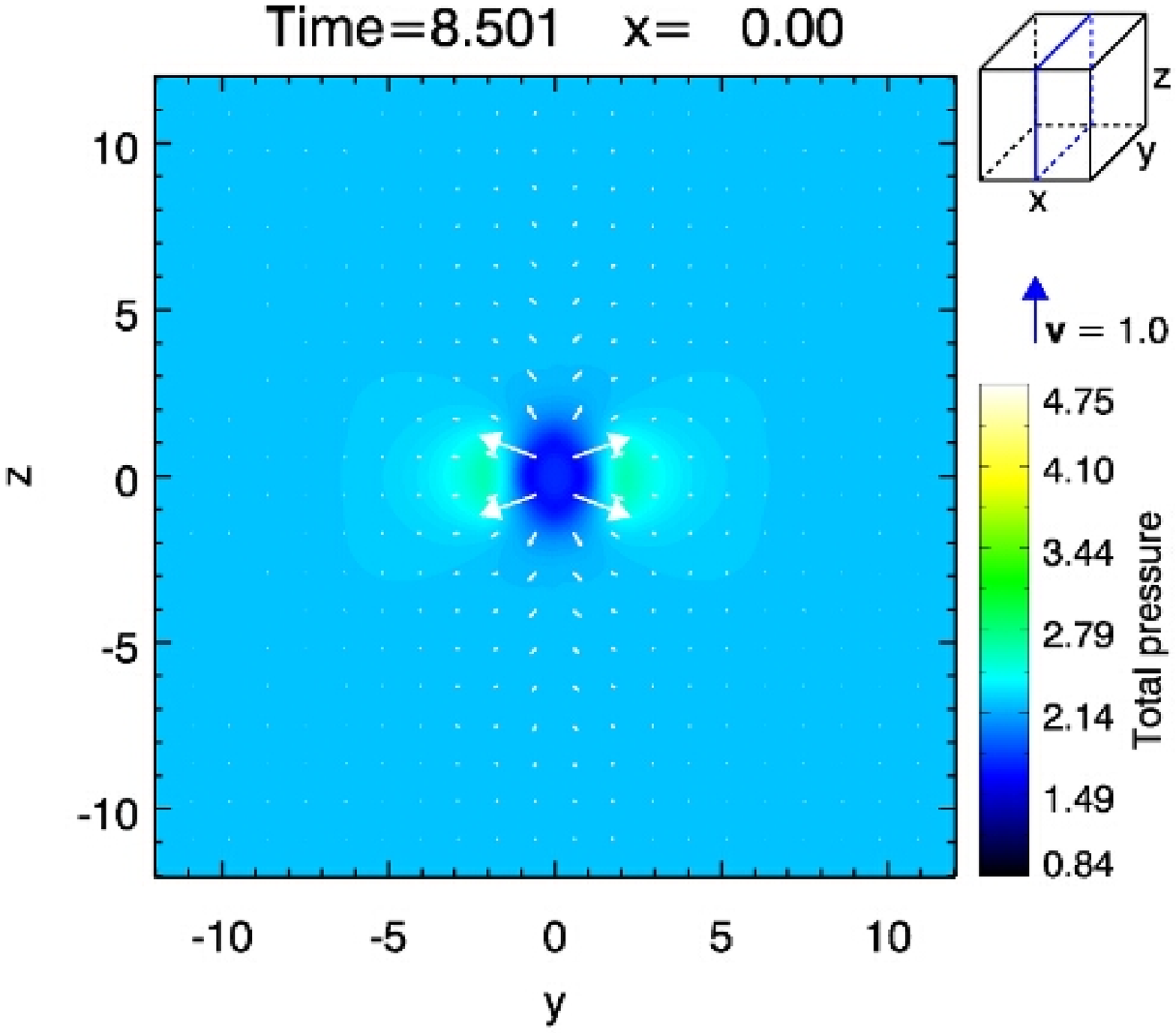}
\includegraphics[height=\ptm,trim=5mm 10mm  4mm 16mm,clip]{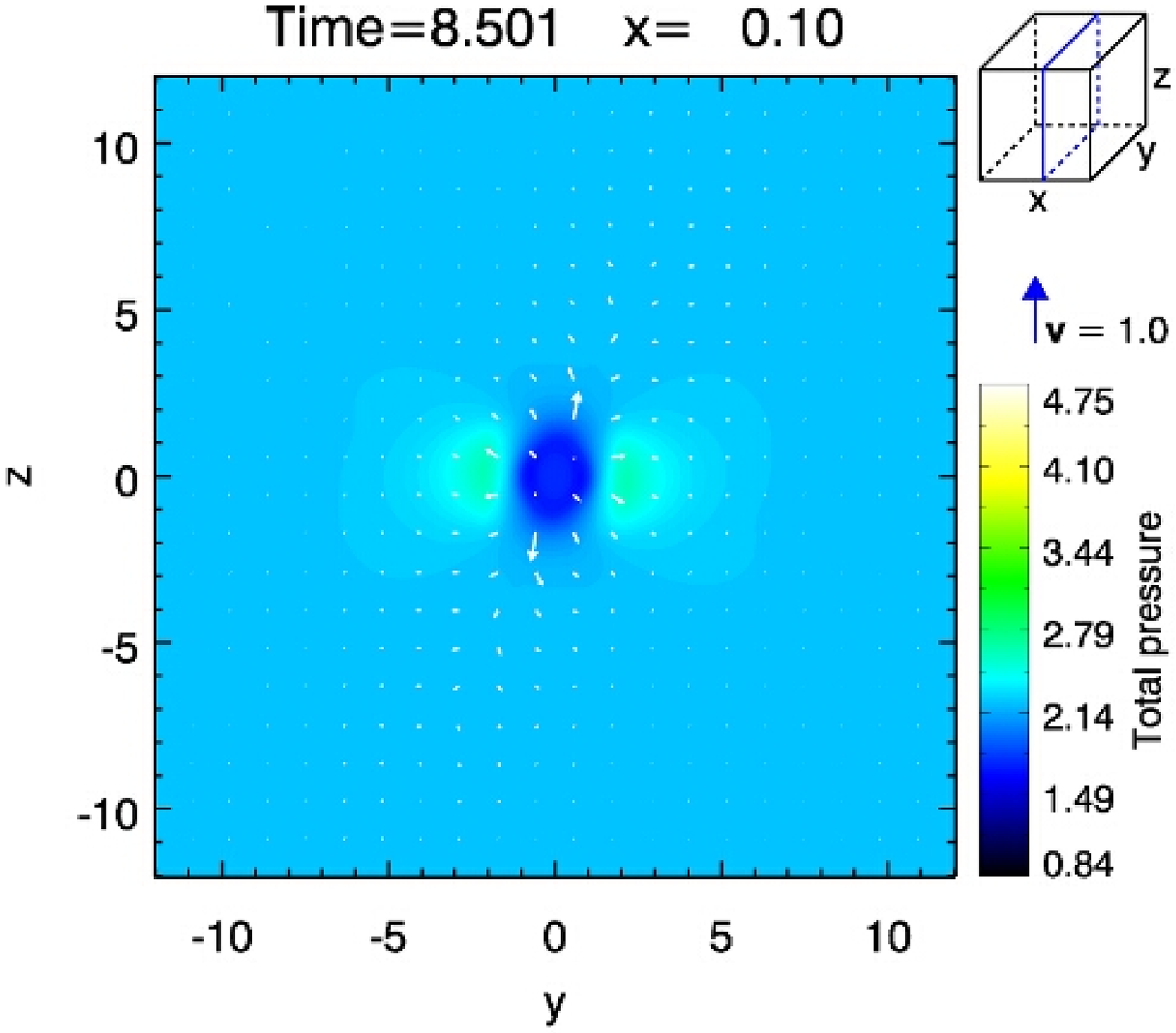}
\caption{2D scanning figures (typical case) of the gas pressure, magnetic pressure and total pressure in $y-z$ plane at $x=-0.1$, $0.0$, $0.1$ and at time $=8.501$. The arrows in the upper panels indicate the magnetic field and others the velocity field.} \label{fig07}
\end{figure}

\begin{figure}[htbp]
\centering
\includegraphics[height=\ptm,trim=5mm 10mm 36mm 16mm,clip]{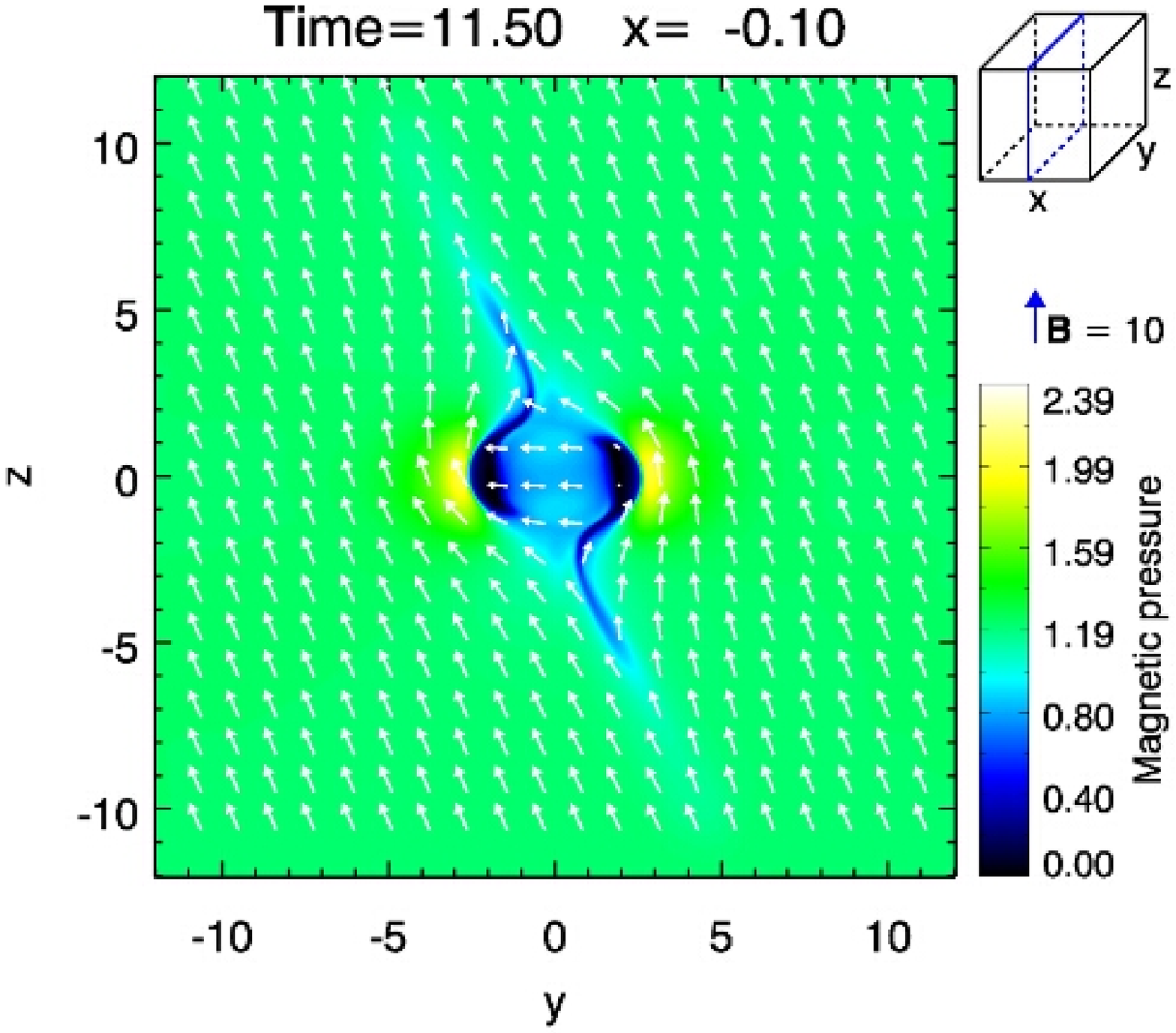}
\includegraphics[height=\ptm,trim=5mm 10mm 36mm 16mm,clip]{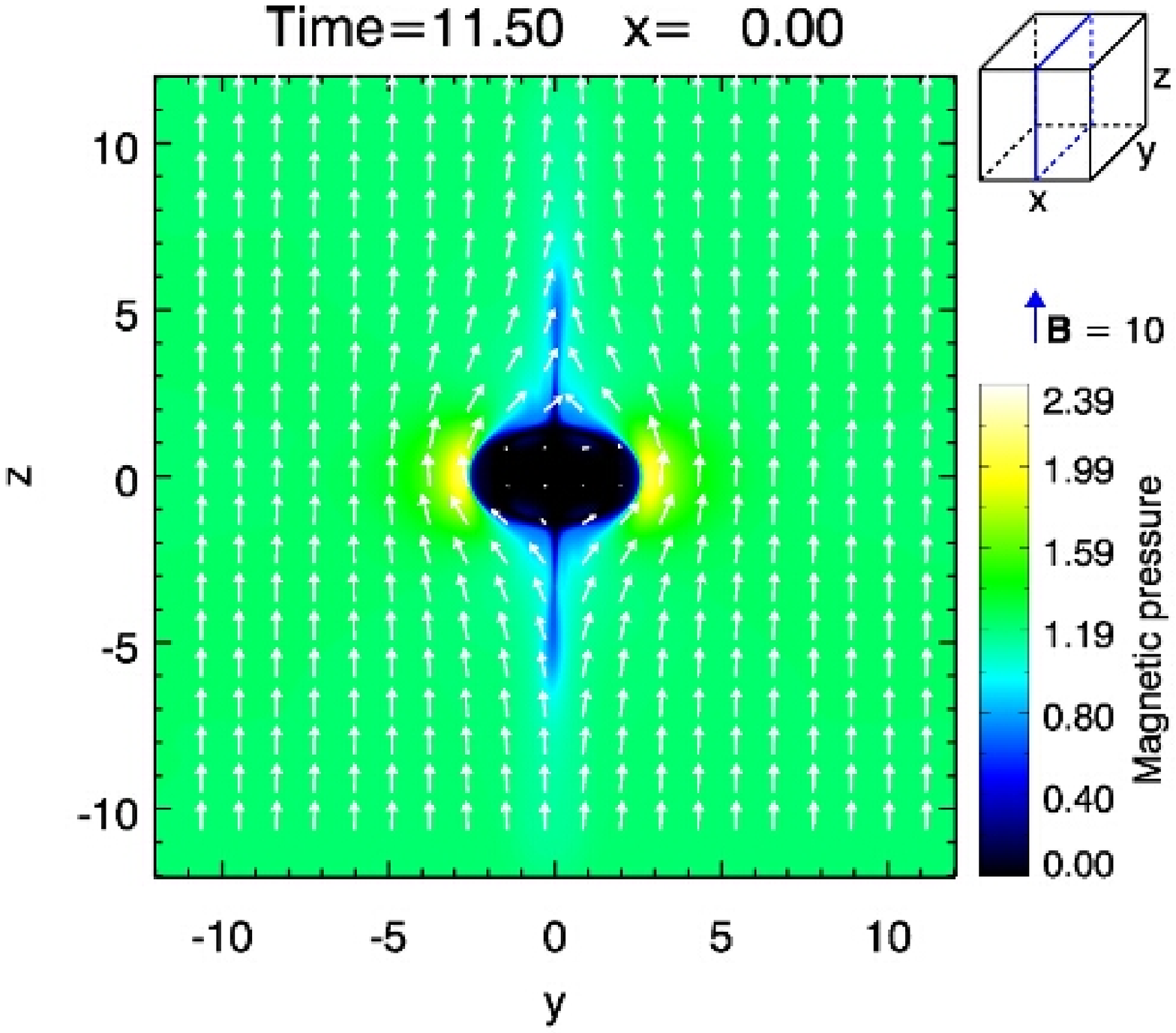}
\includegraphics[height=\ptm,trim=5mm 10mm  4mm 16mm,clip]{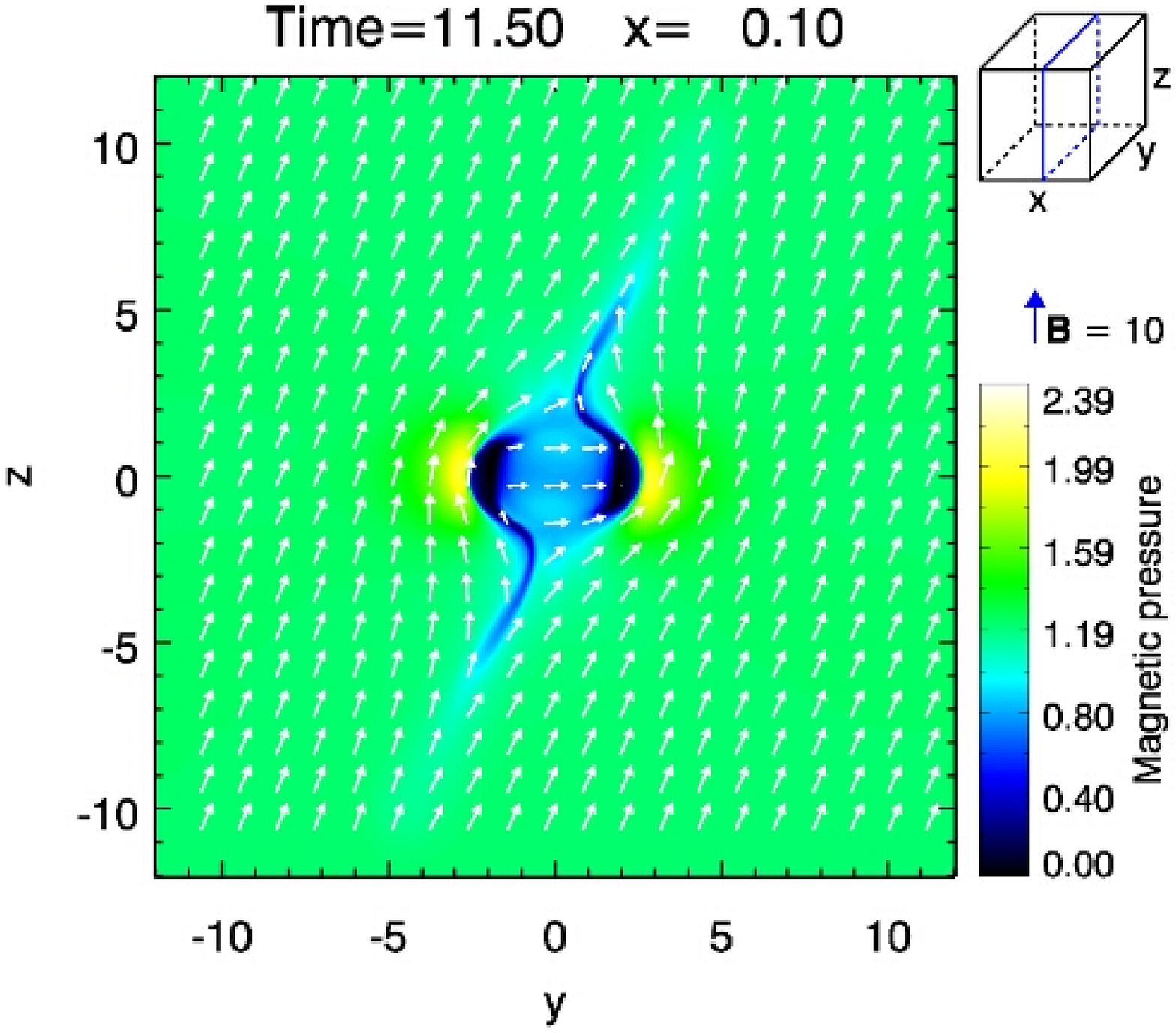}
\includegraphics[height=\ptm,trim=5mm 10mm 36mm 16mm,clip]{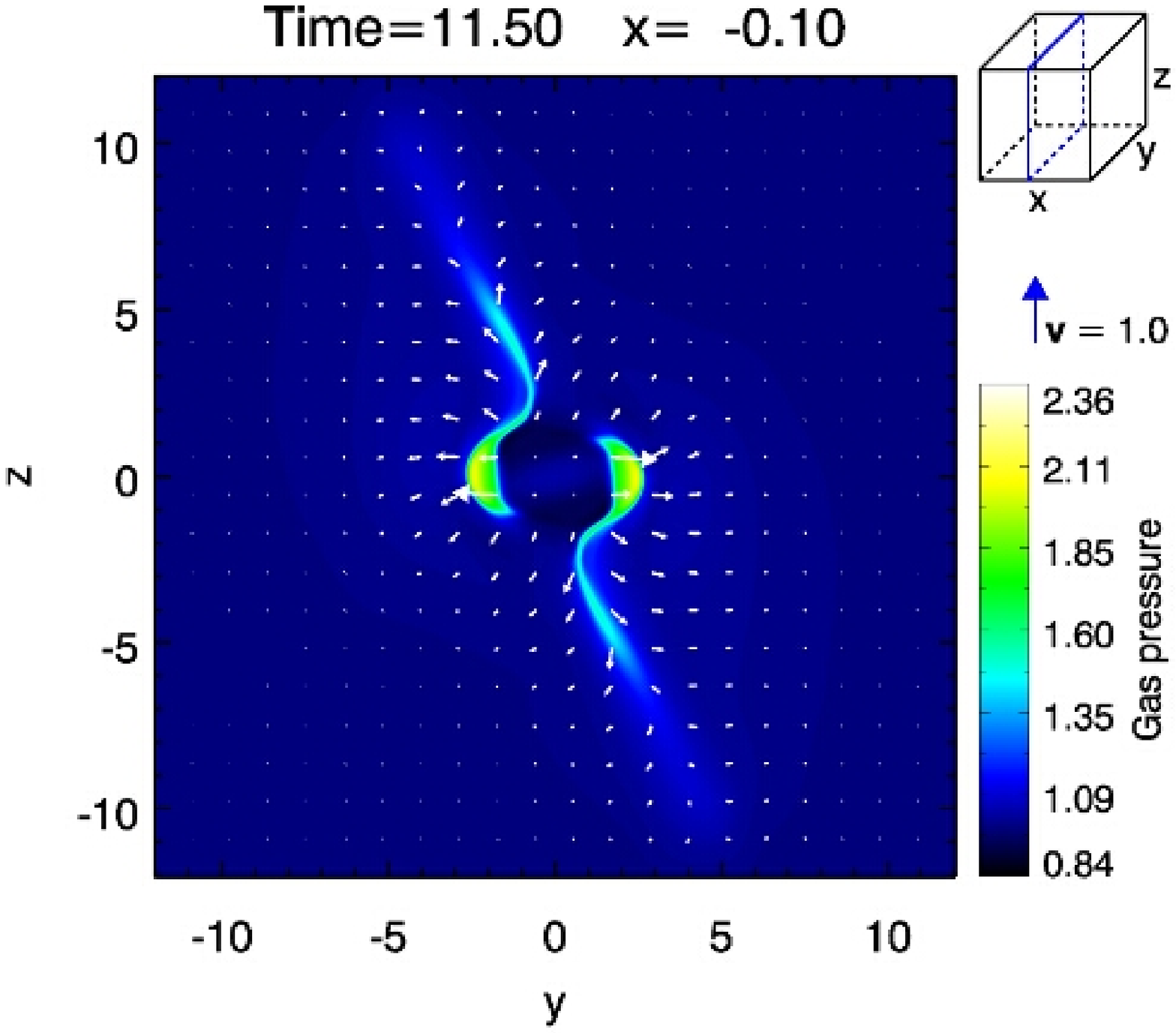}
\includegraphics[height=\ptm,trim=5mm 10mm 36mm 16mm,clip]{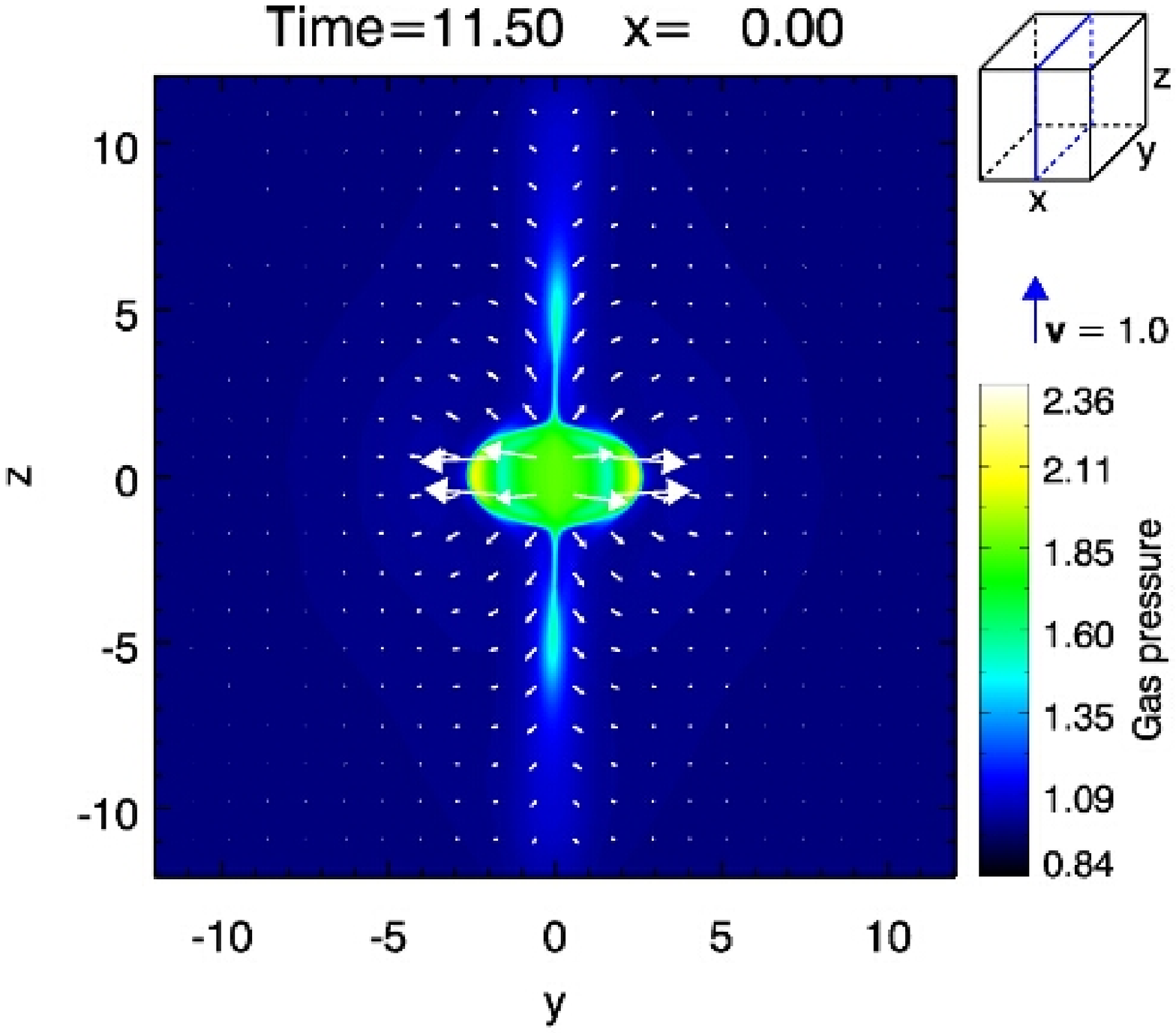}
\includegraphics[height=\ptm,trim=5mm 10mm  4mm 16mm,clip]{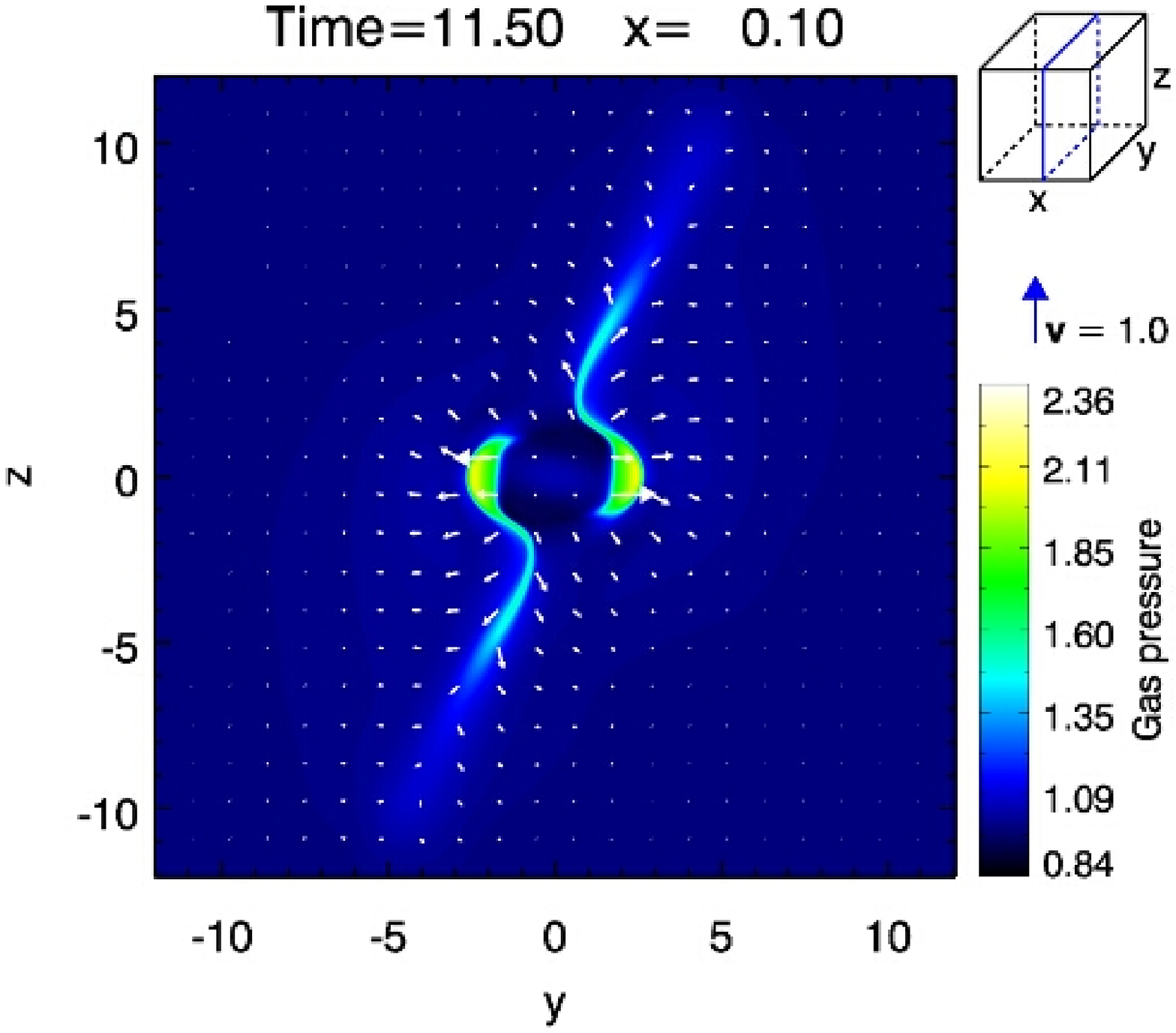}
\includegraphics[height=\ptm,trim=5mm 10mm 36mm 16mm,clip]{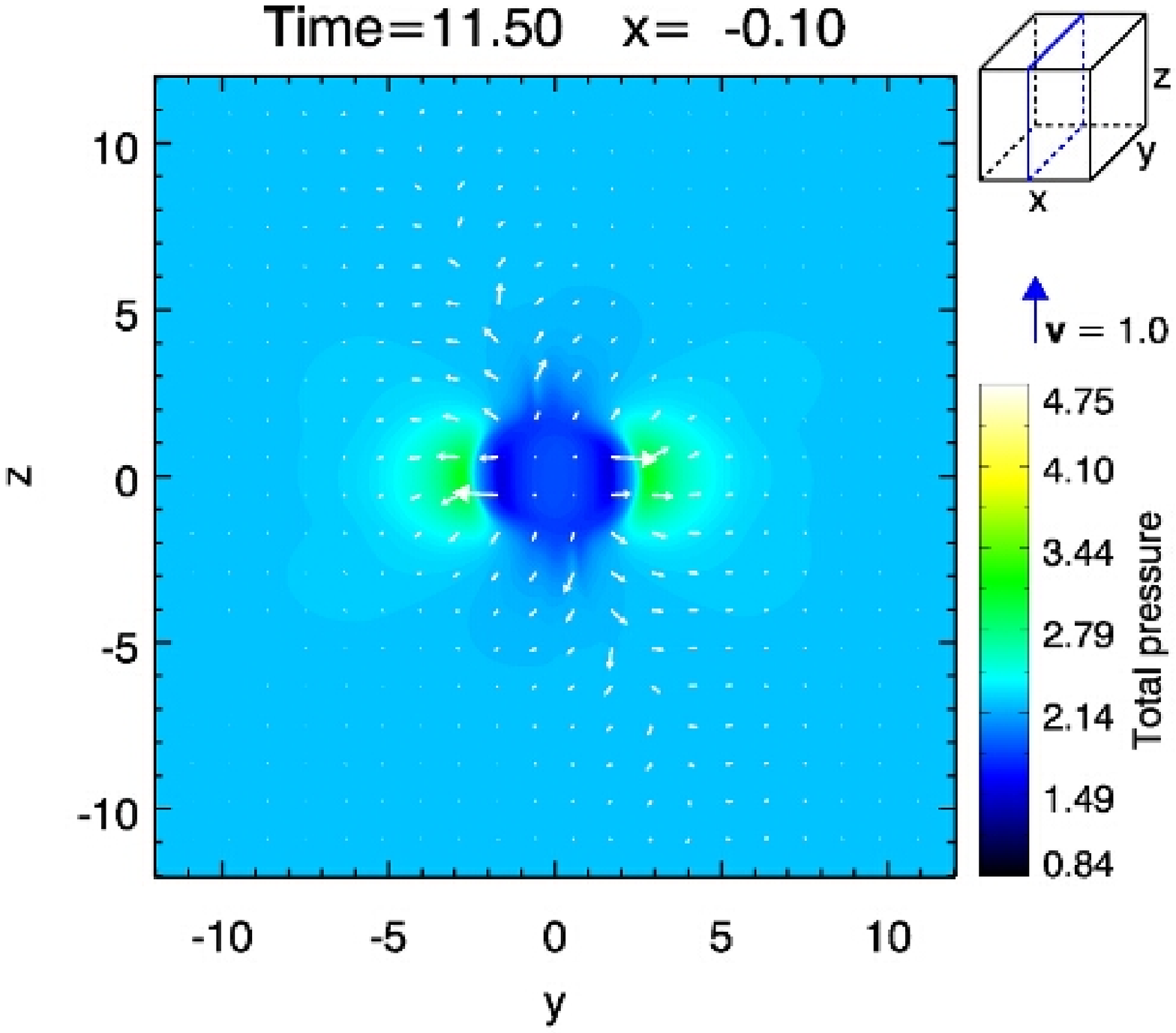}
\includegraphics[height=\ptm,trim=5mm 10mm 36mm 16mm,clip]{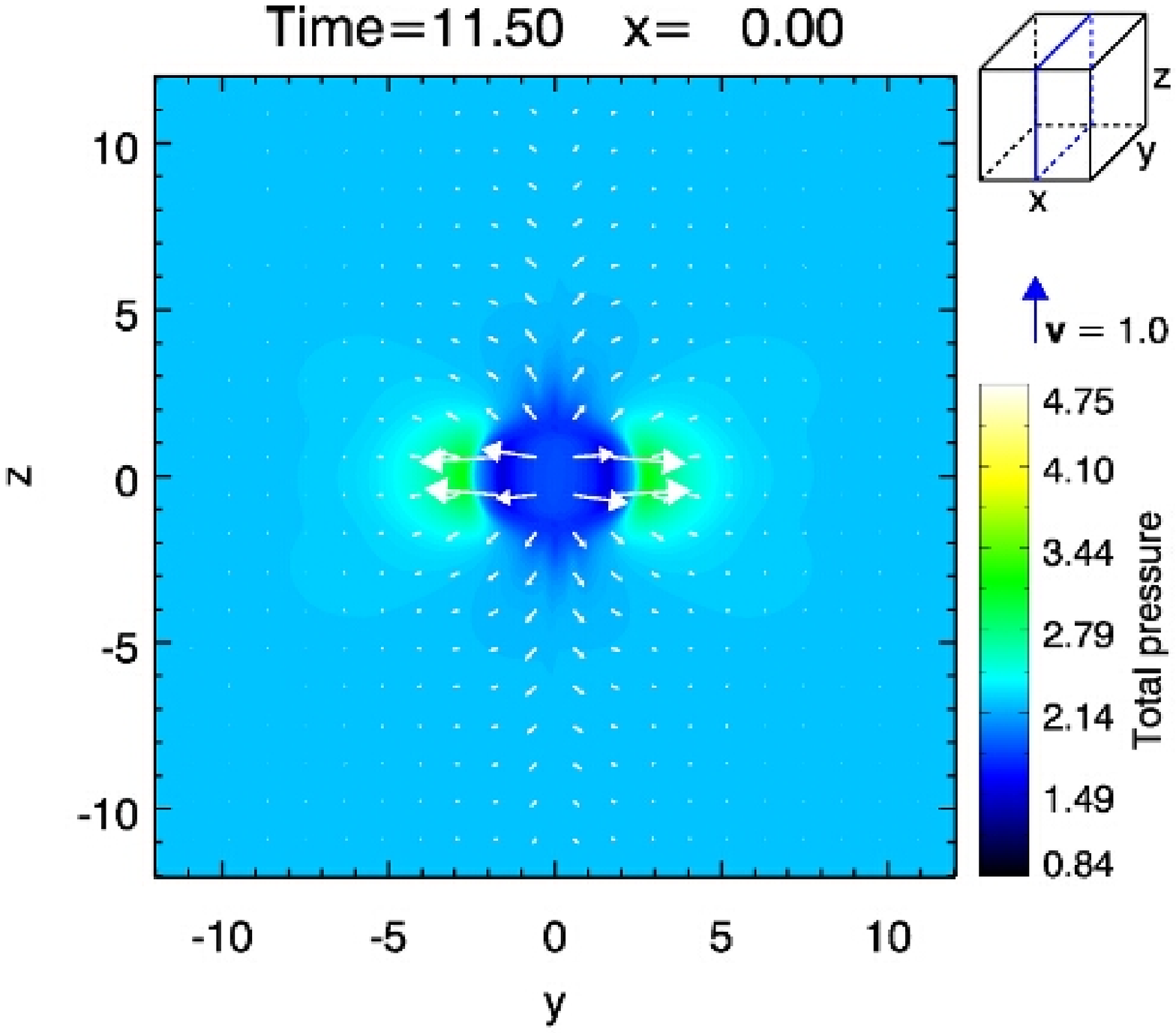}
\includegraphics[height=\ptm,trim=5mm 10mm  4mm 16mm,clip]{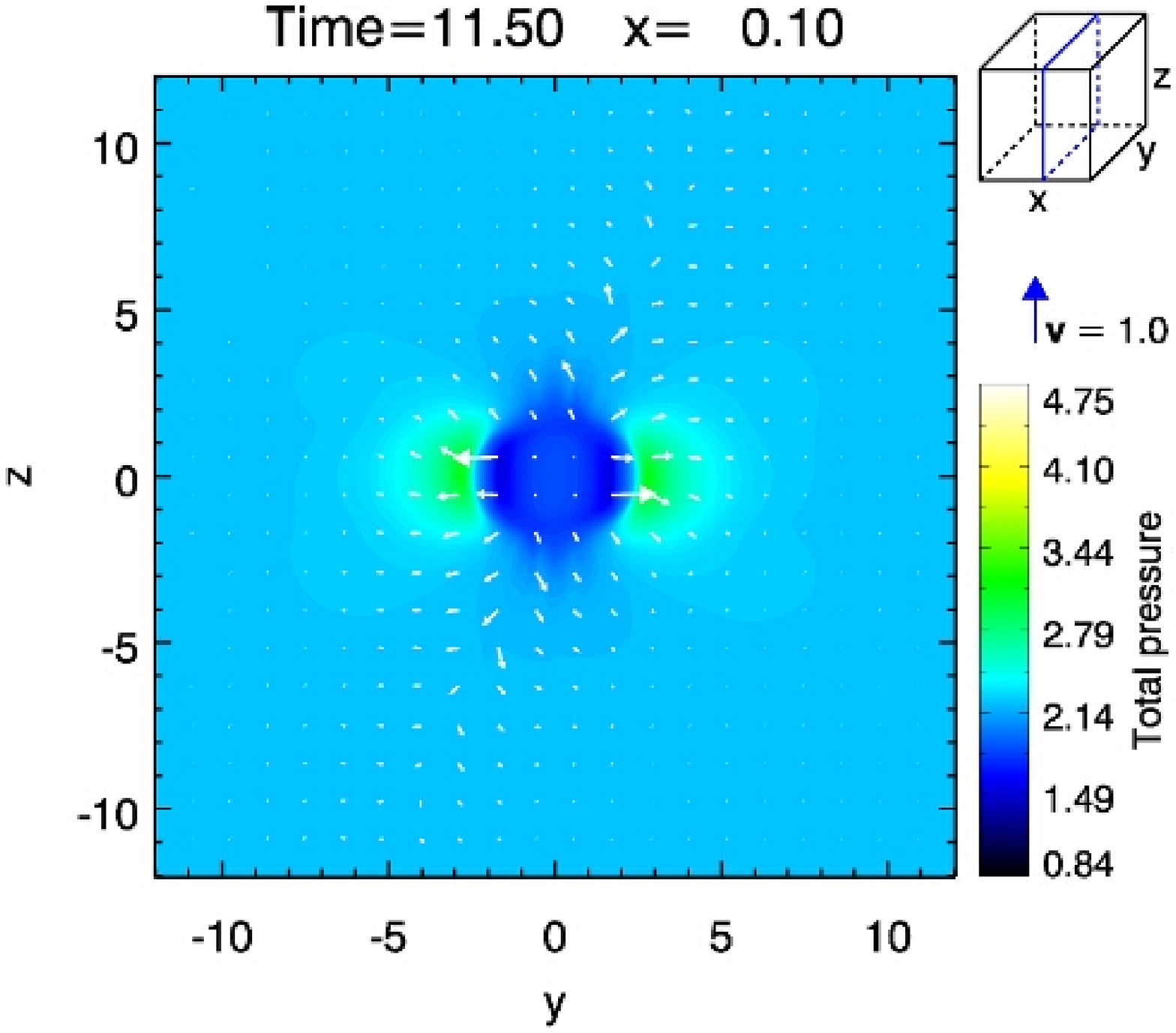}
\caption{2D scanning figures (typical case) of the gas pressure, magnetic pressure and total pressure in $y-z$ plane at $x=-0.1$, $0.0$, $0.1$ and at time $=11.50$. The arrows in the upper panels indicate the magnetic field and others the velocity field.}
 \label{fig08}
\end{figure}

\begin{figure}[htbp]
\centering
\includegraphics[width=320pt]{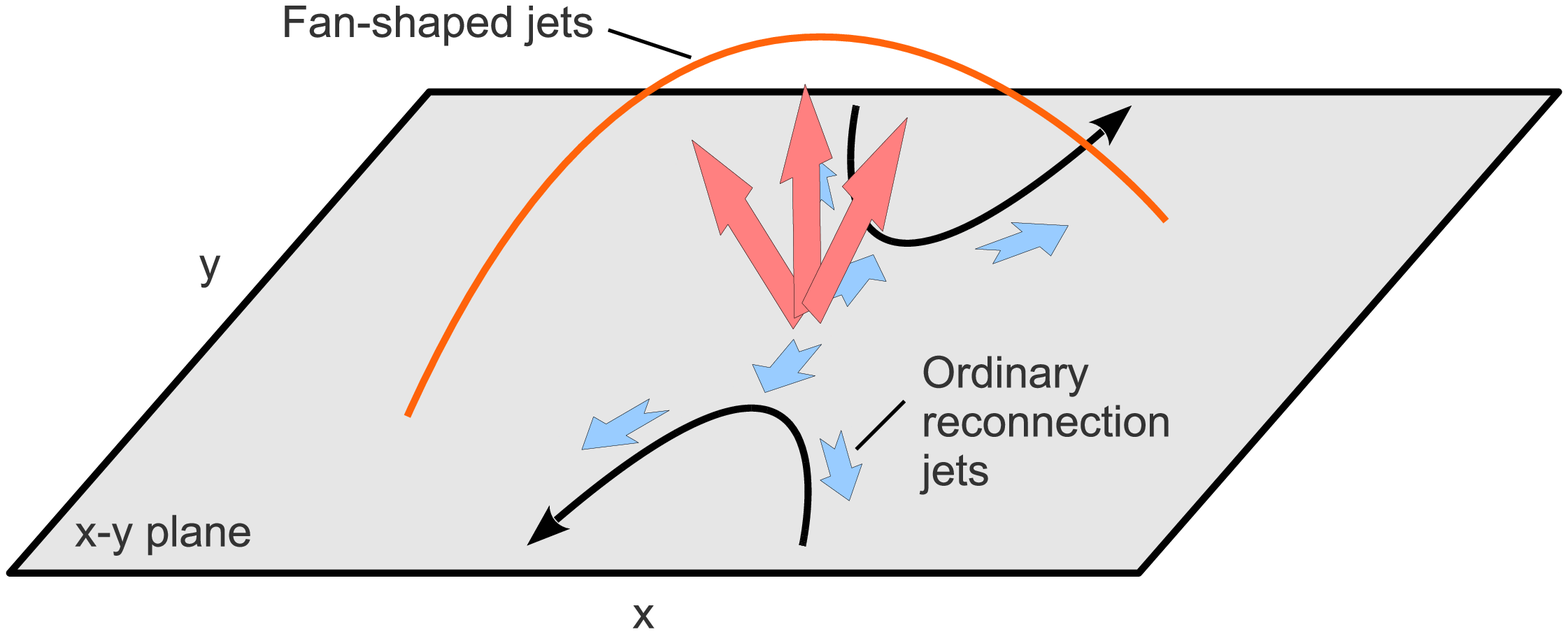}
 \caption{A cartoon for explaining the fan-shaped jets.}
 \label{fig09}
\end{figure}

Figures~\ref{fig07} and~\ref{fig08} depict pressure distributions in $y-z$ plane scanning at $x=-0.1$, $x=0.0$ and $x=0.1$ at the time $=8.5$ and $11.5$, respectively. Because the domains of these figures are in the current sheet, we can see that the magnetic field is shearing between different columns. The interesting thing is that the jets also have different directions for different $x$ positions. In other words, the direction of the jets continuously changes following the direction of the magnetid field which is shearing in the current sheet for forming fan-shaped jets as shown in the 3D visualization in Figures~\ref{fig02} and~\ref{fig03}. The cartoon in Figure~\ref{fig09} shows both ordinary reconnection jets and the fan-shaped jets. One can clearly understand the shape and the stucture of the jets by this cartoon.

\begin{figure}[htbp]
\begin{center}
\includegraphics[width=\ptb,trim=4mm 10mm 12mm 2mm,clip]{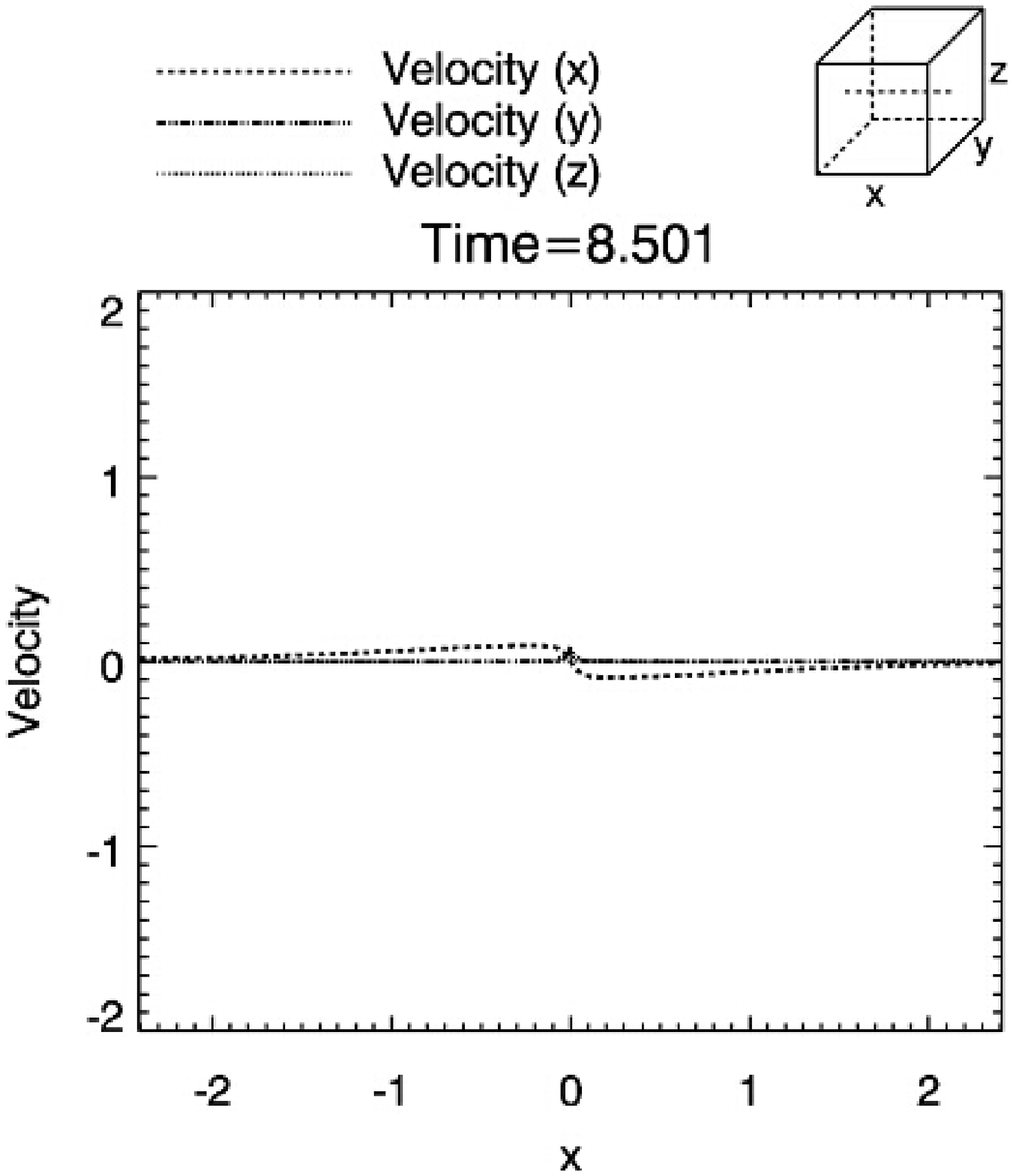}
\includegraphics[width=\ptb,trim=4mm 10mm 12mm 2mm,clip]{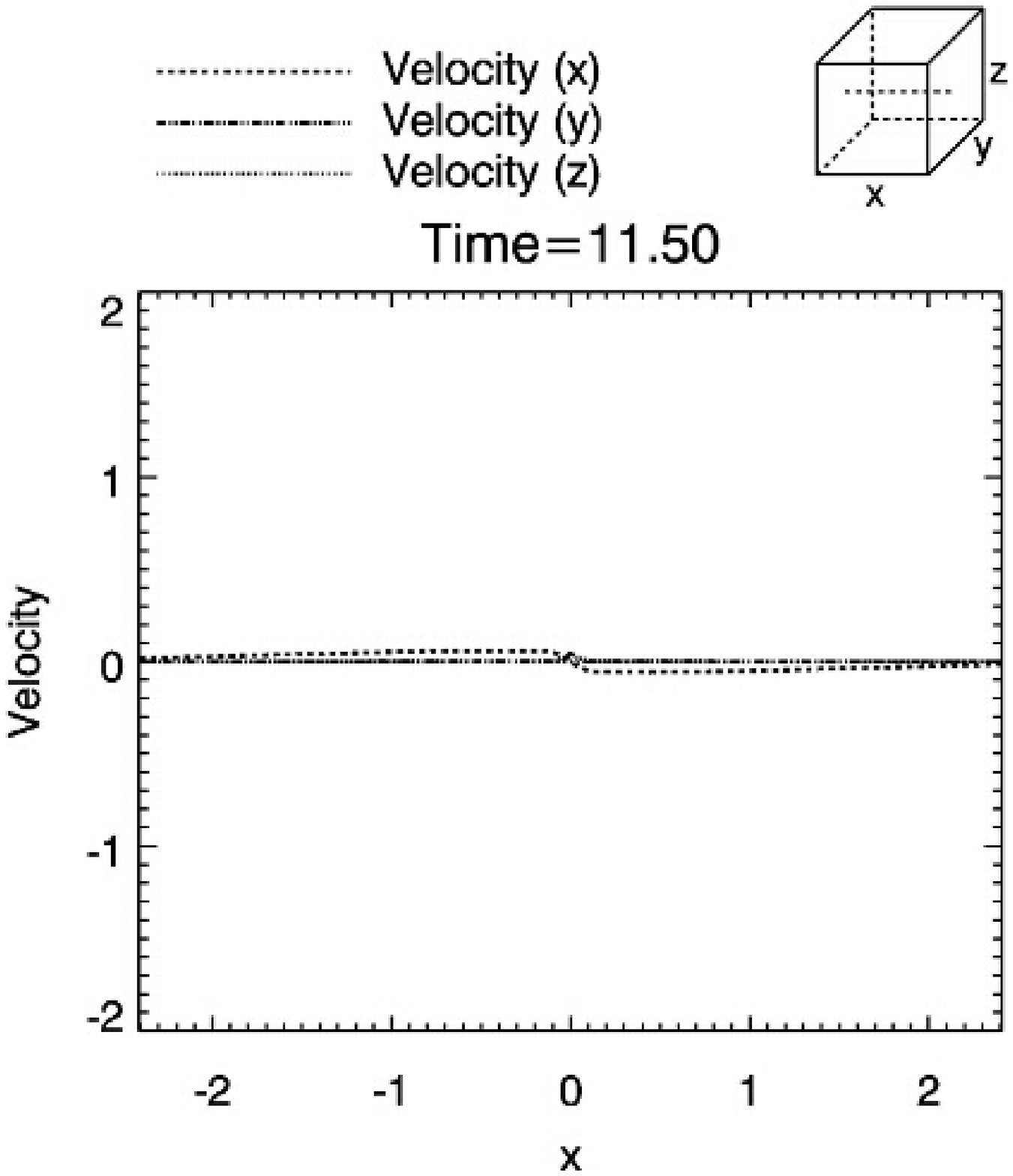}
\includegraphics[width=\ptb,trim=4mm 10mm 12mm 2mm,clip]{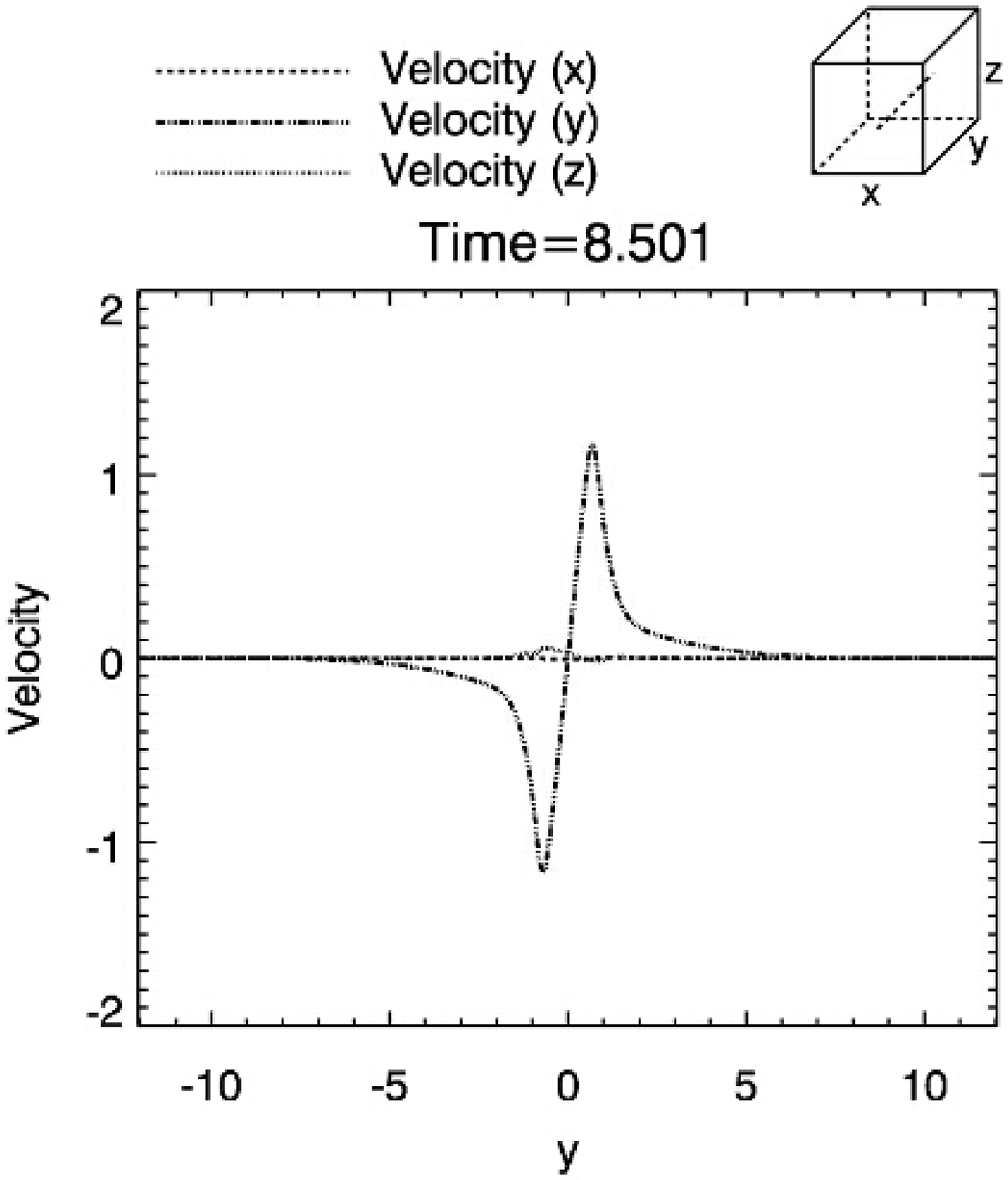}
\includegraphics[width=\ptb,trim=4mm 10mm 12mm 2mm,clip]{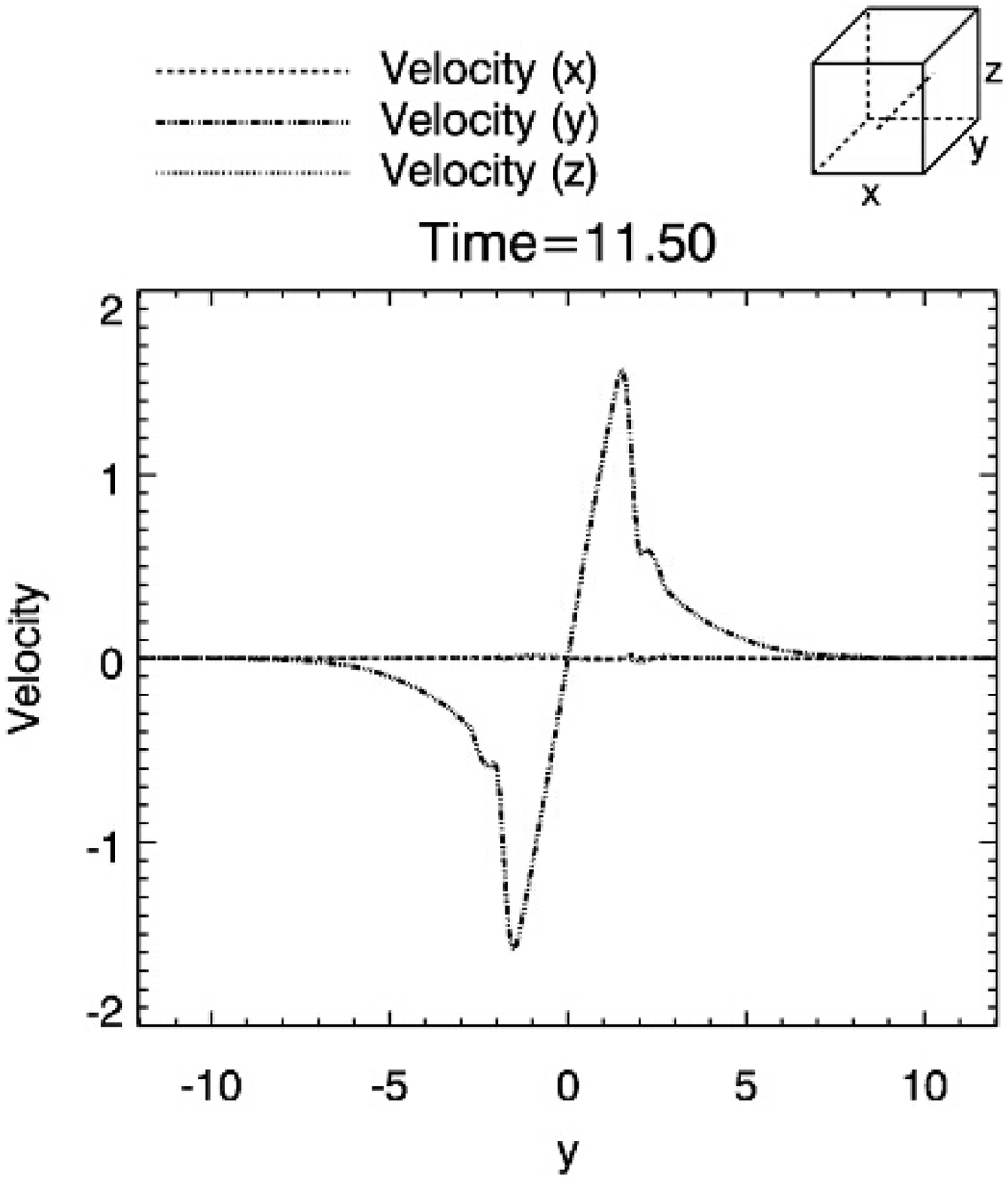}
\includegraphics[width=\ptb,trim=4mm 10mm 12mm 2mm,clip]{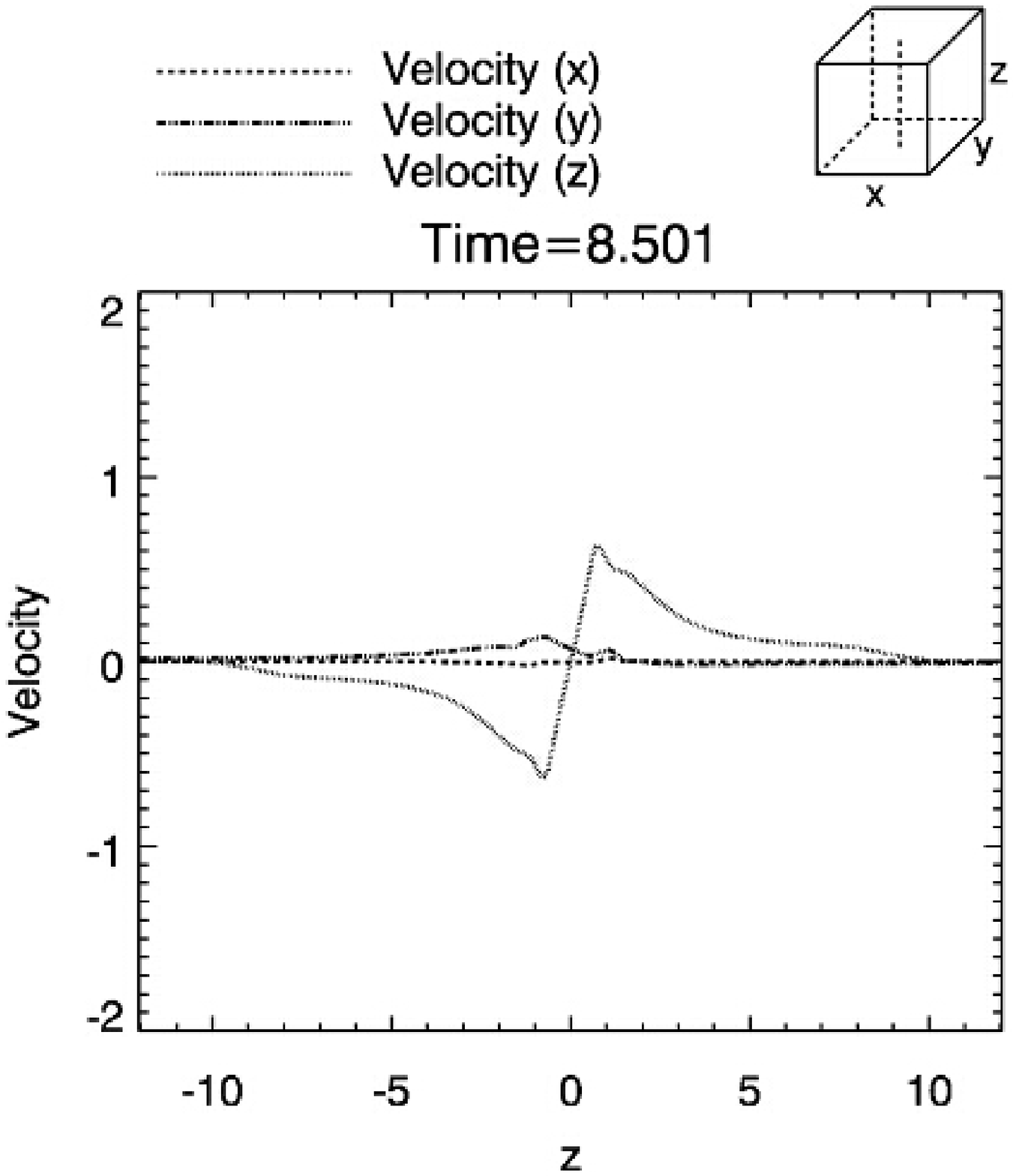}
\includegraphics[width=\ptb,trim=4mm 10mm 12mm 2mm,clip]{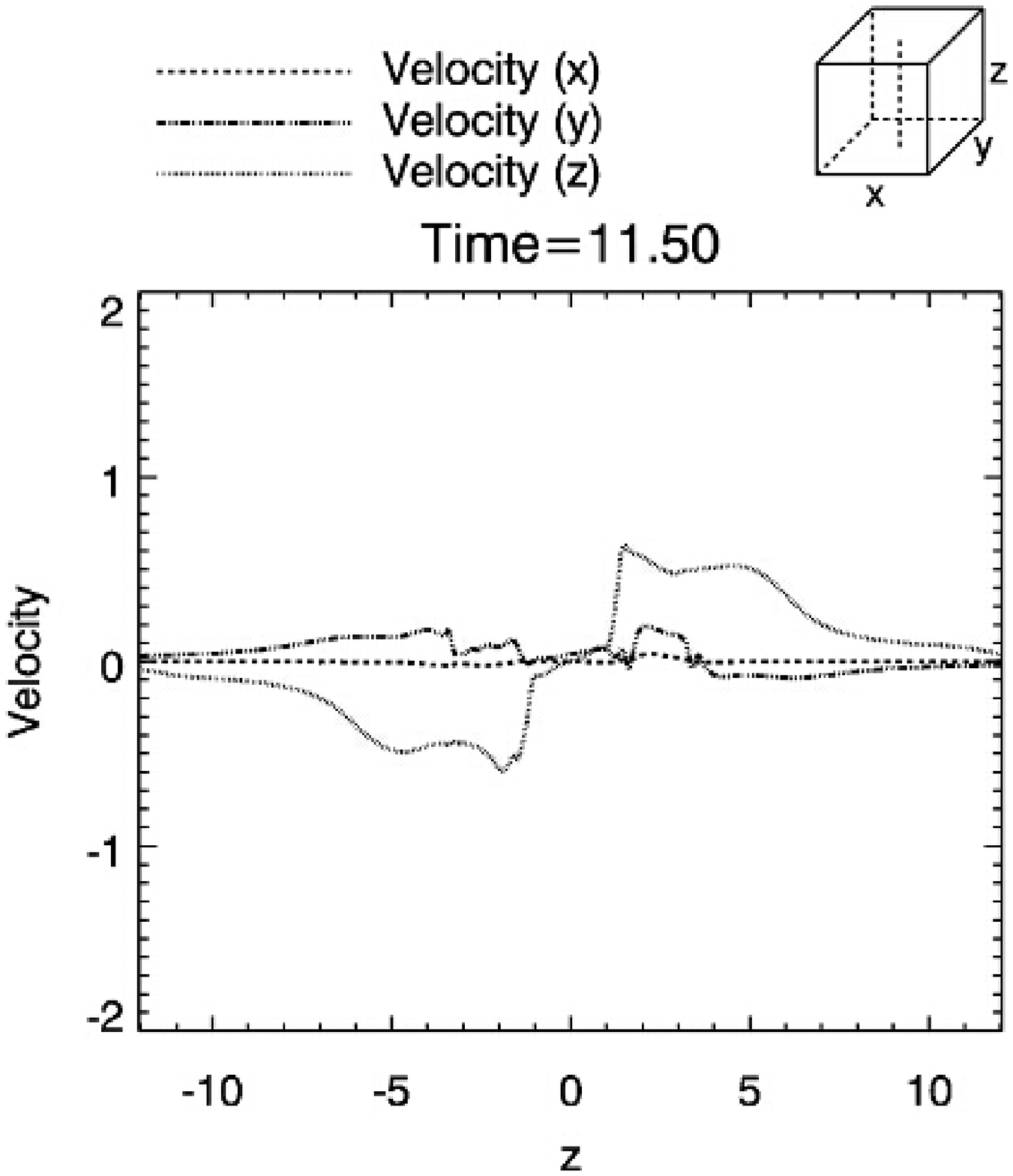}
\end{center}
\caption{Velocity distributions (typical case) along $x$, $y$, and $z$ directions at different times. The Alfv\'en speed is about 1.58.}
\label{fig10}
\end{figure}

The interesting 1D velocity distributions along $x$ ($y=0.0$, $z=0.0$), $y$ ($x=0.0$, $z=0.0$) and $z$ ($x=0.0$, $y=0.0$) directions at time $=8.5$ and $11.5$ are shown in Figure~\ref{fig10}. Note that the velocities in $x$, $y$ and $z$ directions stand for inflow speed, ordinary reconnection outflow speed and fan-shaped jet speed, respectively. The inflow speed ($x$ direction) becomes a little smaller at the later stage (time $=11.5$), which indicates that the magnetic reconnection rate (as discussed later) reaches the maximum around the time $=8.5$. At the latter stage the ordinary reconnection outflow speed is approximately the ambient Alfv\'en speed (1.58). Moreover, the maximum velocity of the fan-shaped jets along $z$ direction is about 0.6. Thus, the driving forces of the fan-shaped jets and the ordinary reconnection outflow are different.

\begin{figure}[htbp]
\begin{center}
\includegraphics[height=\ptb]{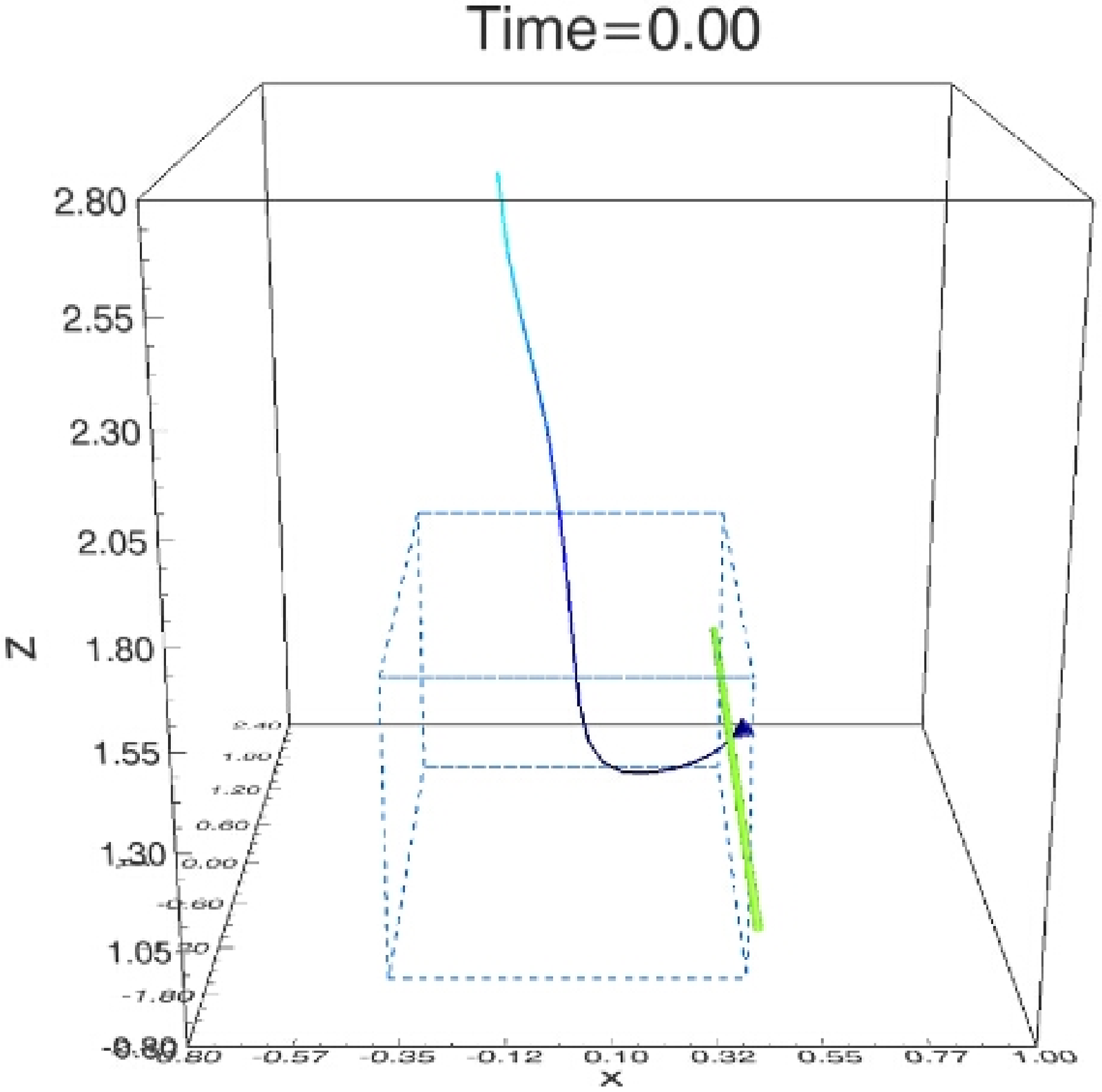}
\includegraphics[height=\ptb]{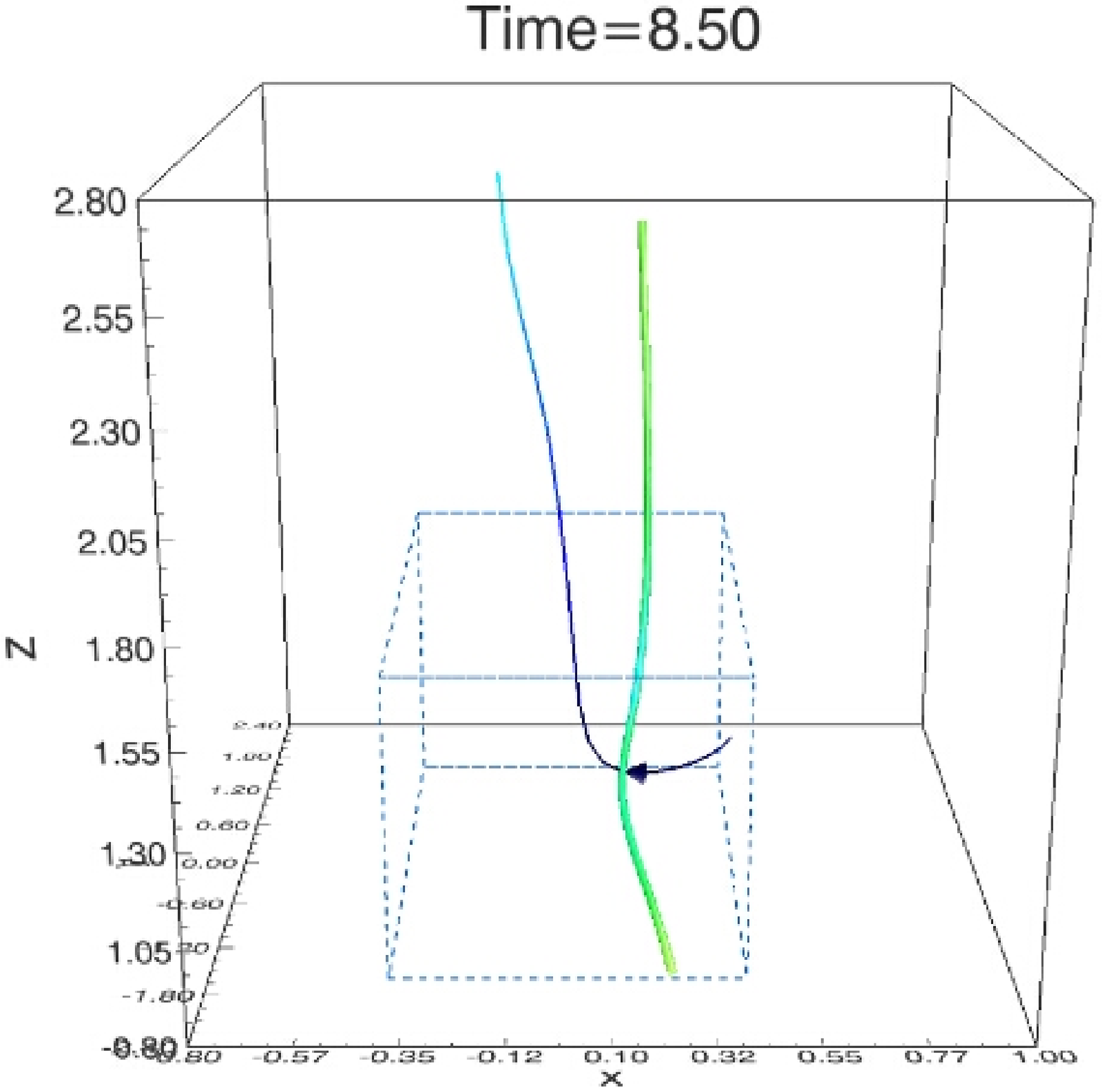}
\includegraphics[height=\ptb]{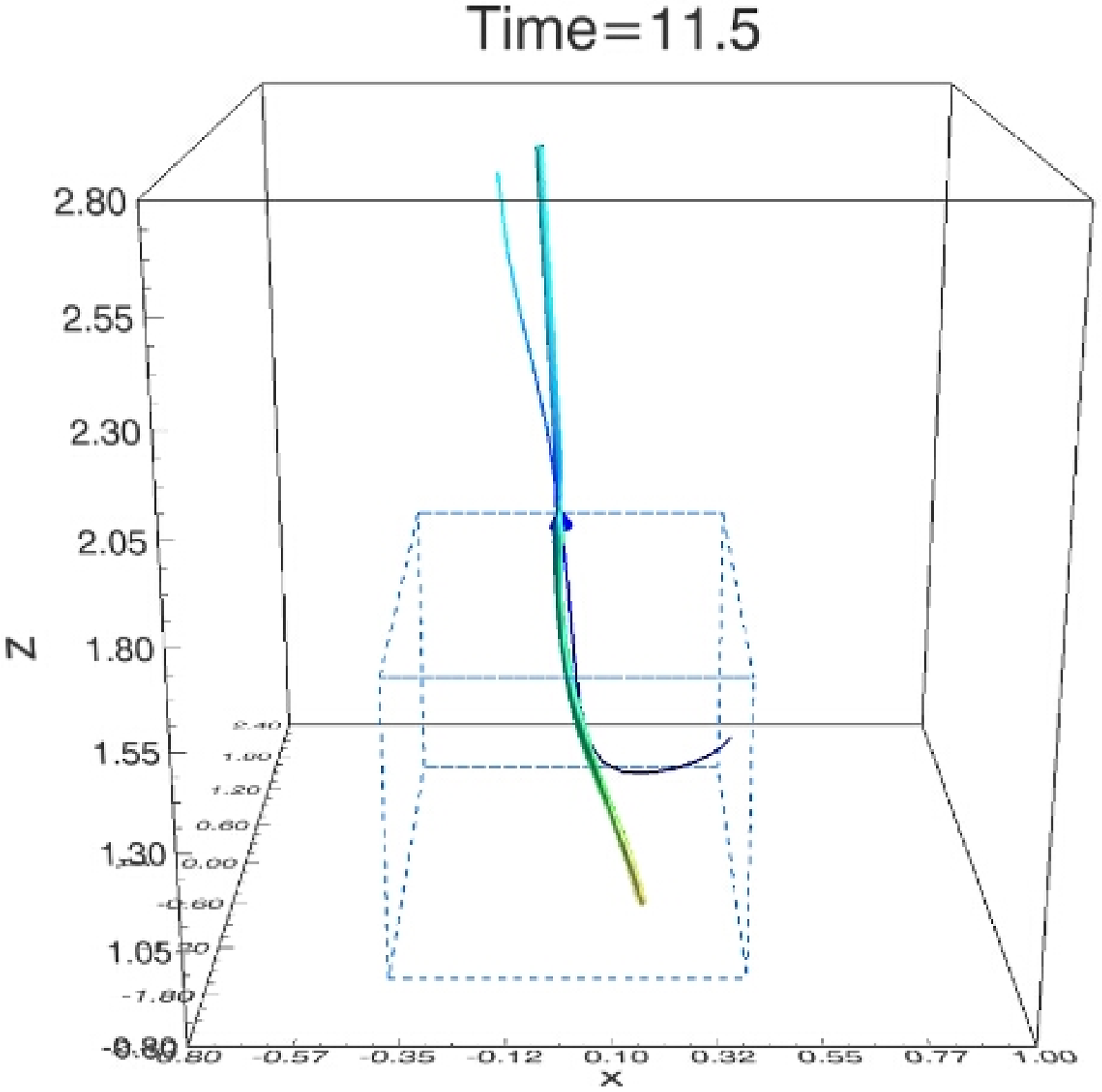}
\includegraphics[height=\ptb]{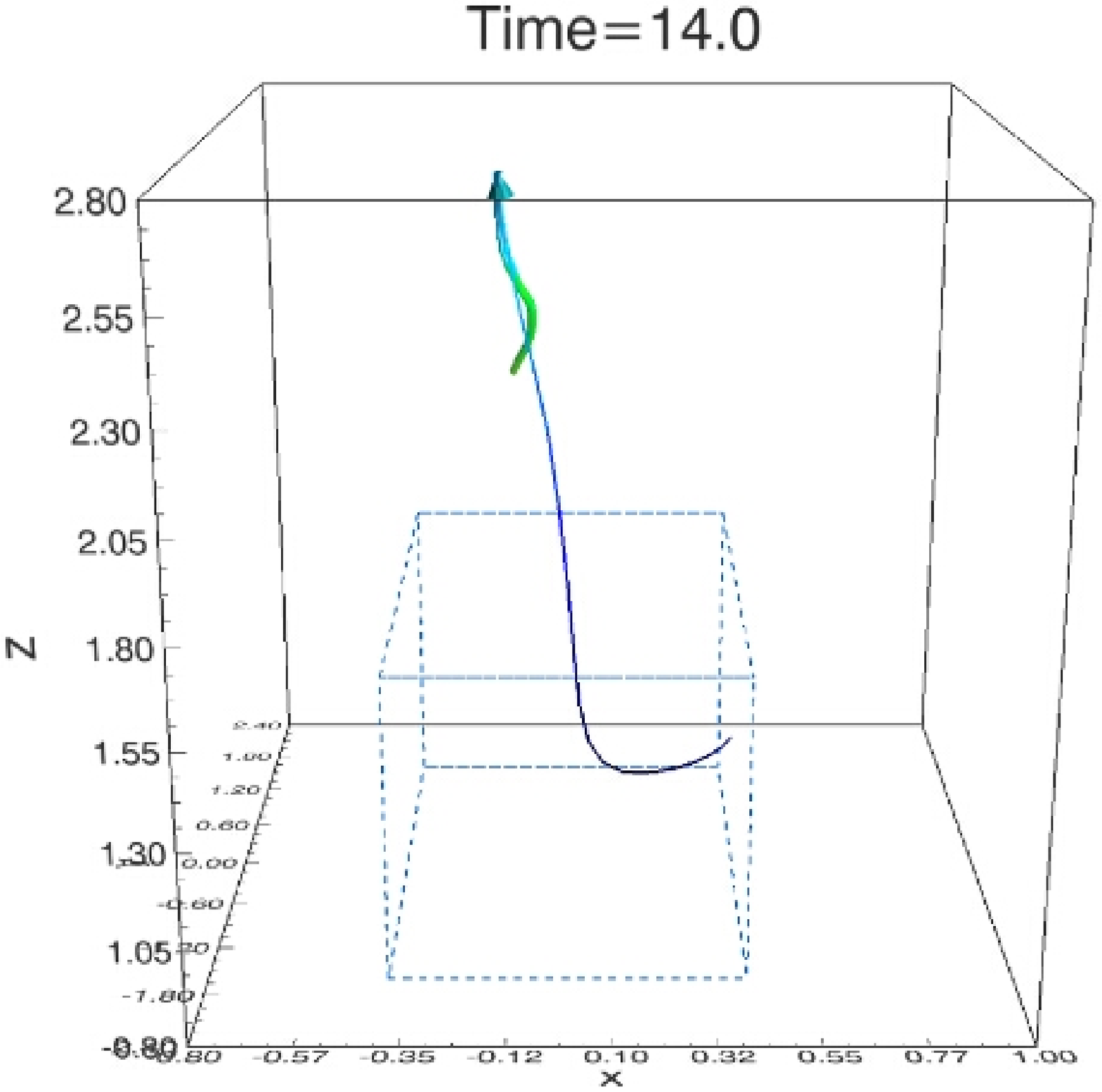}
\end{center}
\caption{The panels show the trajectories (dark lines) of a lagrangian fluid element with the magnetic field lines (typical case). The blue dashed box is the diffusion region.}
\label{fig11}
\end{figure}

\begin{figure}[htbp]
\begin{center}
\includegraphics[height=\ptb]{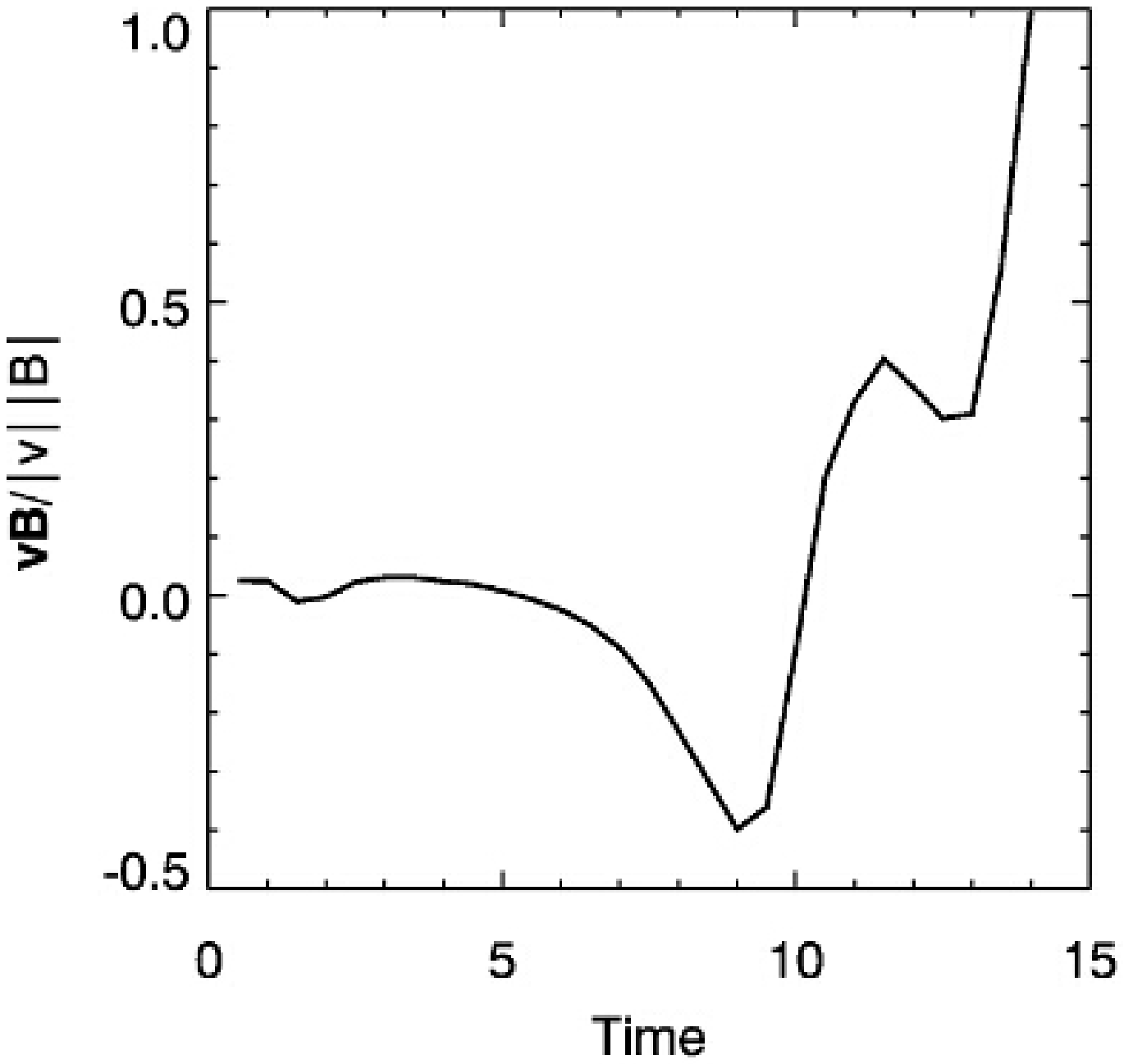}
\includegraphics[height=\ptb]{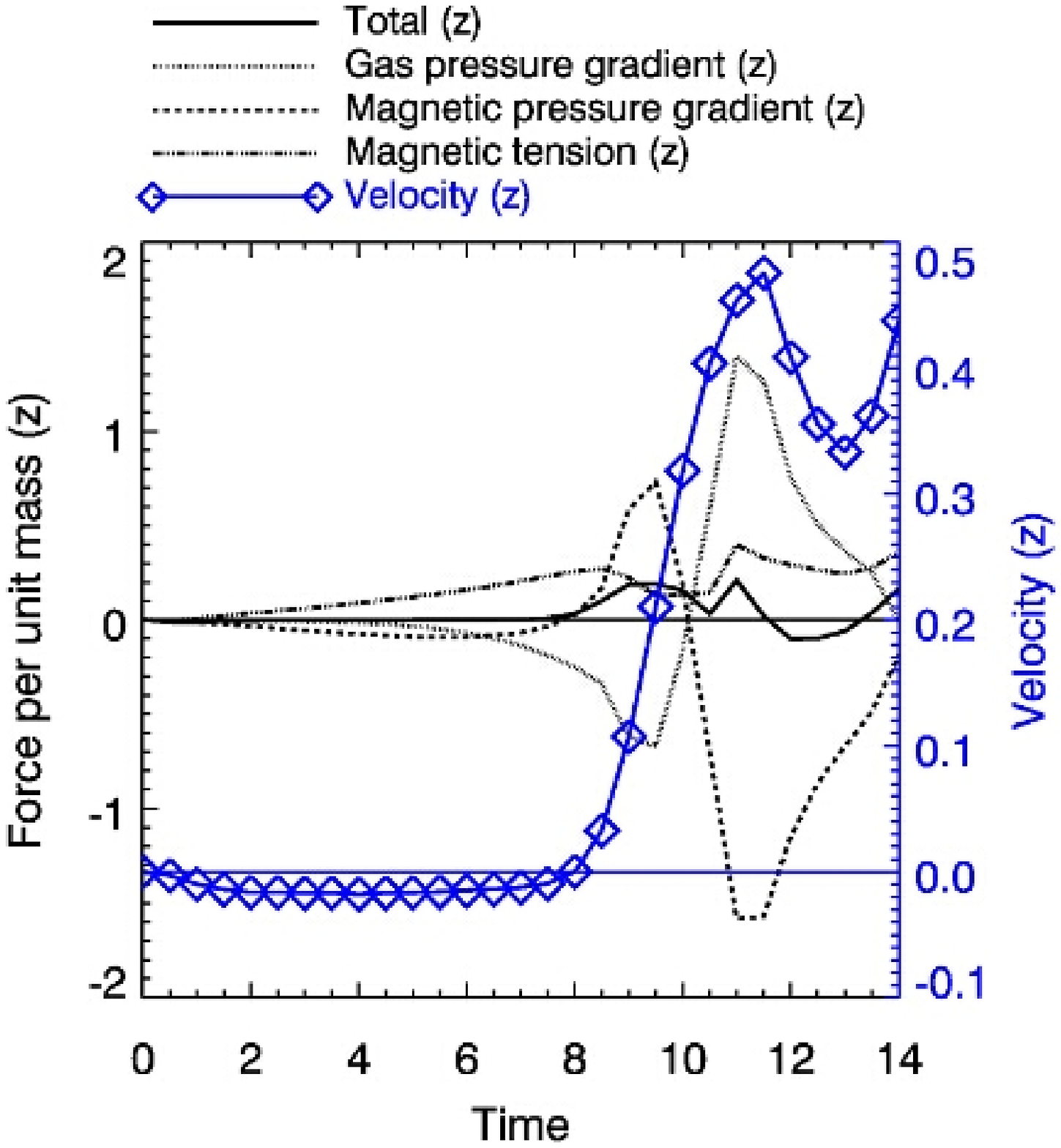}
\end{center}
\caption{Variation of the cosine values of the angle between the velocity and the magnetic field direction (left panel). Right panel shows the forces per unit mass and velocity in $z$ direction.}\label{fig12}
\end{figure}

\begin{figure}[htbp]
\begin{center}
\includegraphics[height=\ptb]{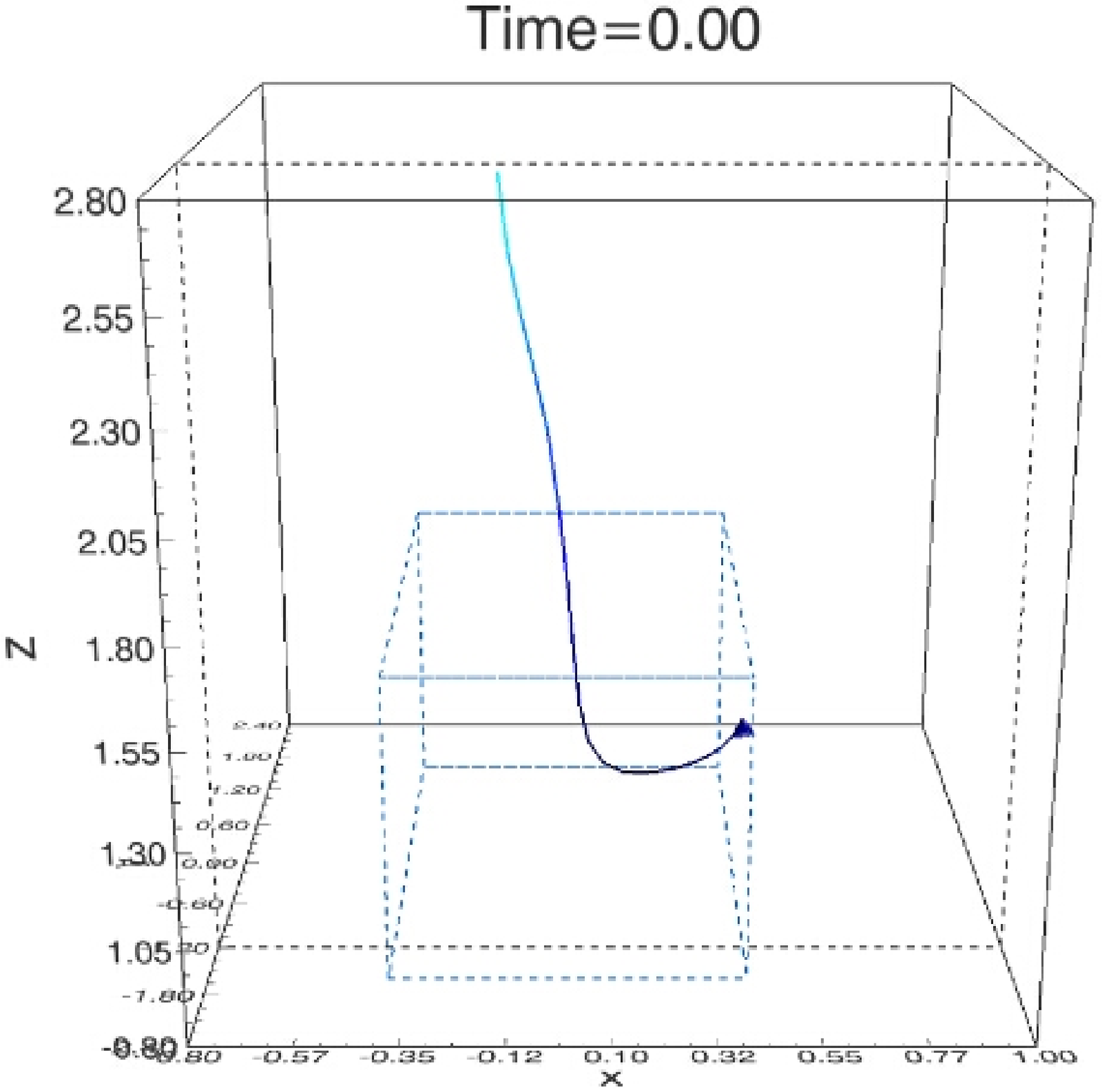}
\includegraphics[height=\ptb]{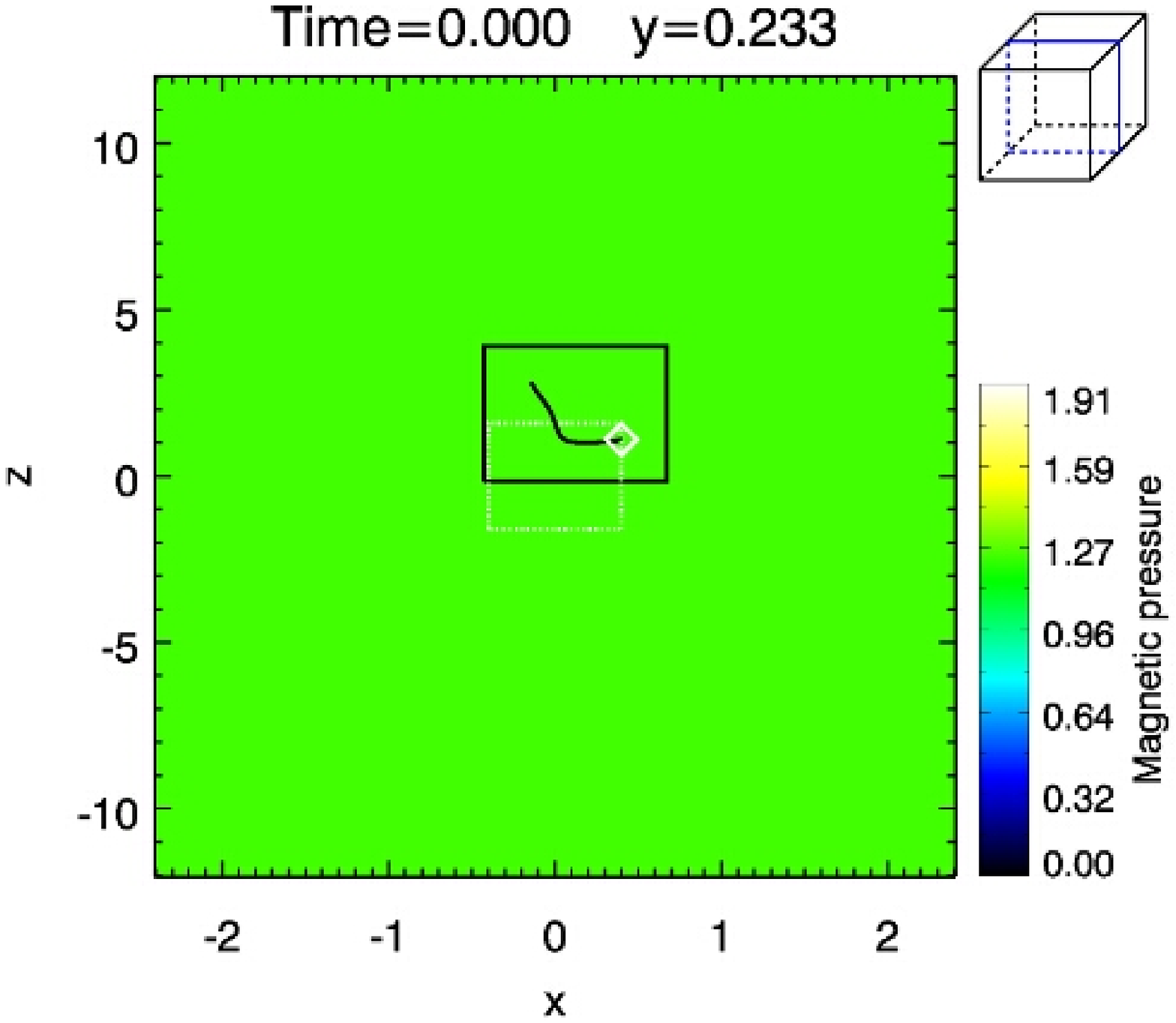}
\includegraphics[height=\ptb]{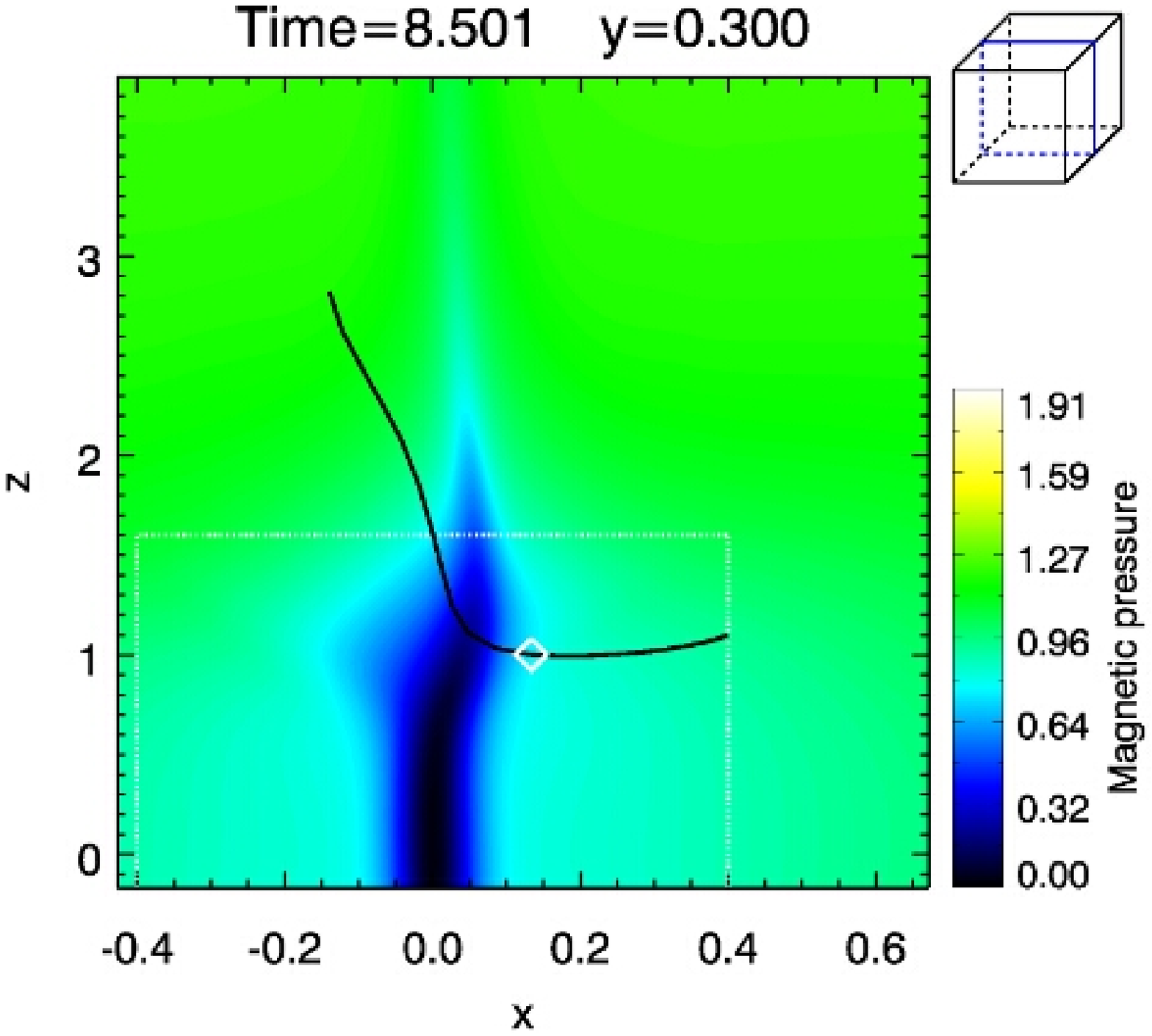}
\includegraphics[height=\ptb]{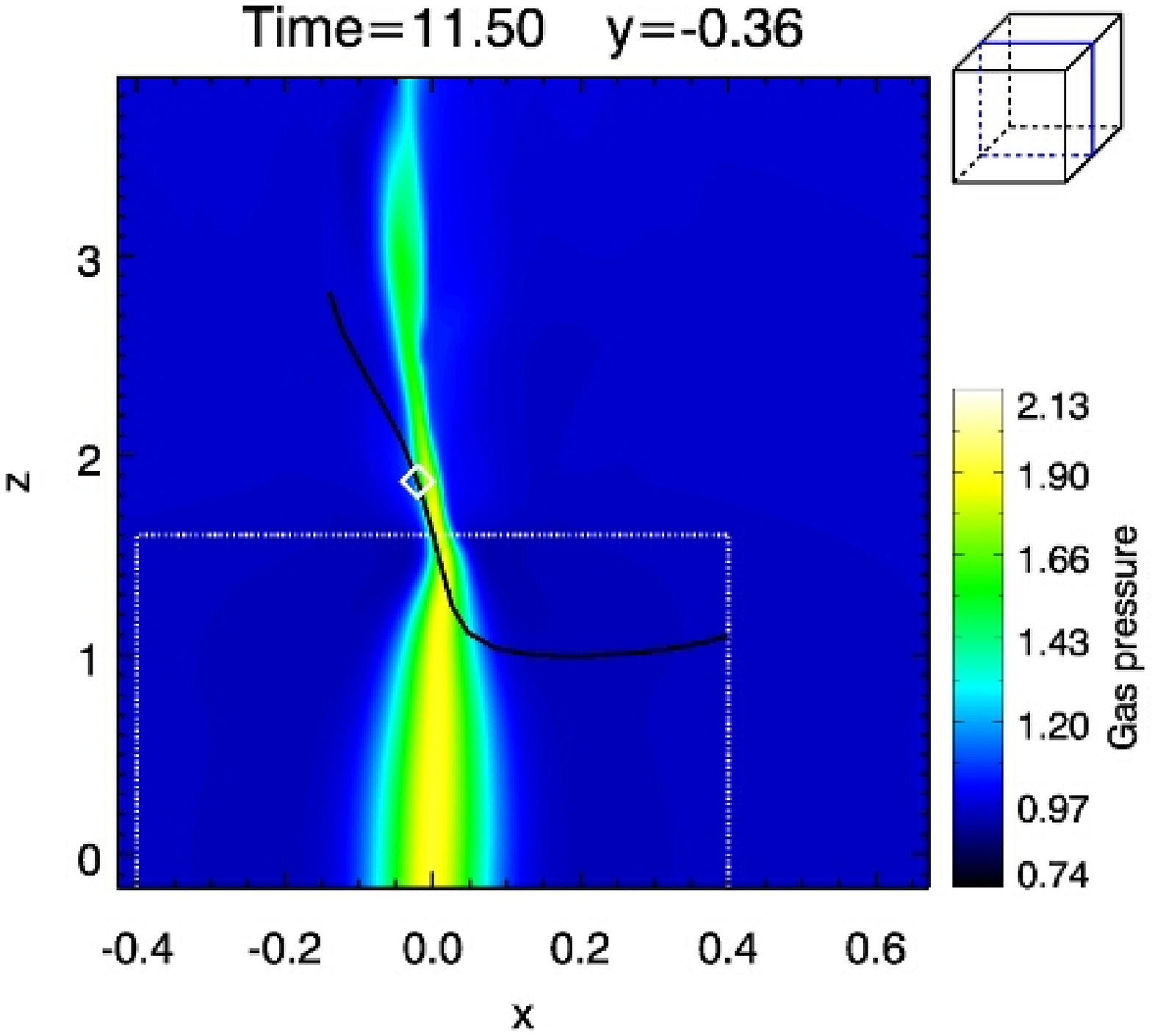}
\end{center}
\caption{The upper left panel shows the trajectory (dark line) of a lagrangian fluid element (typical case), the blue dashed box is the diffusion region and a 2D distribution in the bigger dark dashed plane in the upper left panel is shown in the upper right panel. The lower panels depict the gas and magnetic pressure distributions of the small dark box shown in the upper right panel at the time $=11.50$. The fluid element is marked by the white diamond symbol.} \label{fig13}
\end{figure}

In order to investigate the driving force of the fan-shaped jets, we examine the trajectories and the forces of Lagrangian fluid elements (test particles). This is also a method to show how plasma moves and what kind of force drives plasma to move (\citealt{Shibata1986, Kato2002}). Figure~\ref{fig11} shows the trajectory of a typical fluid element in the fan-shaped jet. This fluid element is located at the edge of the diffusion region at the initial time and the magnetic field is perpendicular to the trajectory (i.e., perpendicular to the velocity direction). After that, this element is ejected as a jet at the time $=8.5$ and eventually moves along the guide field line. The cosine value of the angle between velocity (\textbf{v}) and magnetic field (\textbf{B}) vectors is shown in the left panel of Figure~\ref{fig12}. At the initial time, the velocity of the particle is perpendicular to the magnetic field and almost parallel to that at the time $=14$. The right panel of Figure~\ref{fig12} shows the forces in $z$ direction on this element. It shows that there are two stages for acceleration. In the first stage (time $=8-10$), the magnetic tension force drives this element since the gas pressure gradient force balances the magnetic pressure gradient force in the inflow region as we mentioned before. After that it is dominated by the gas pressure gradient in the later stage (time $=10-11.5$). The trajectory in the $x-z$ plane is shown in Figure~\ref{fig13}. Note that the $y$ direction is different at different times to follow the trajectory. At the time $=8.5$, the element is driven as a inflow and at the time $=11.5$ as a fan-shaped outflow.

\begin{figure}[htbp]
\centering
\includegraphics[width=\ptb]{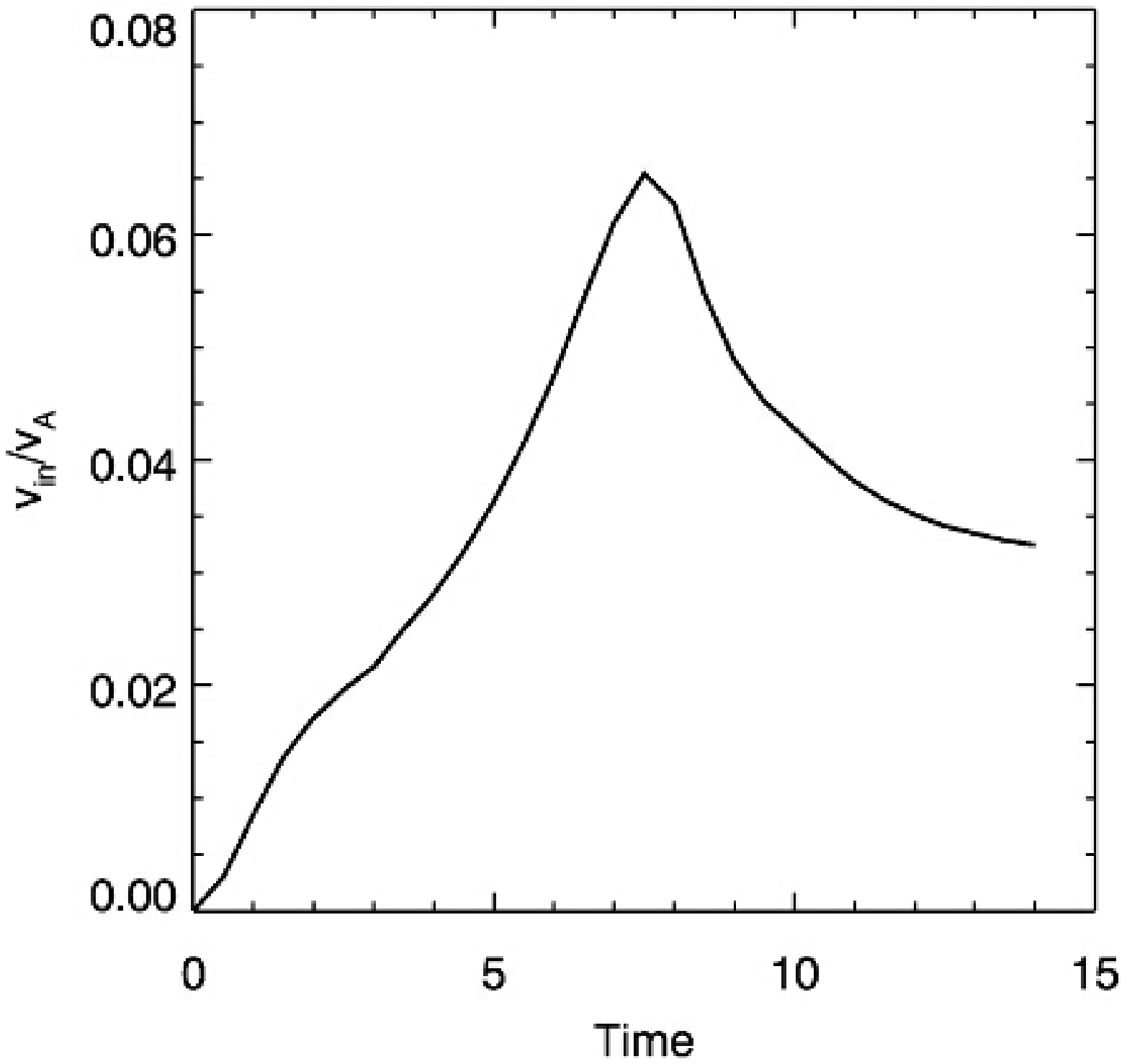}
\includegraphics[width=\ptb]{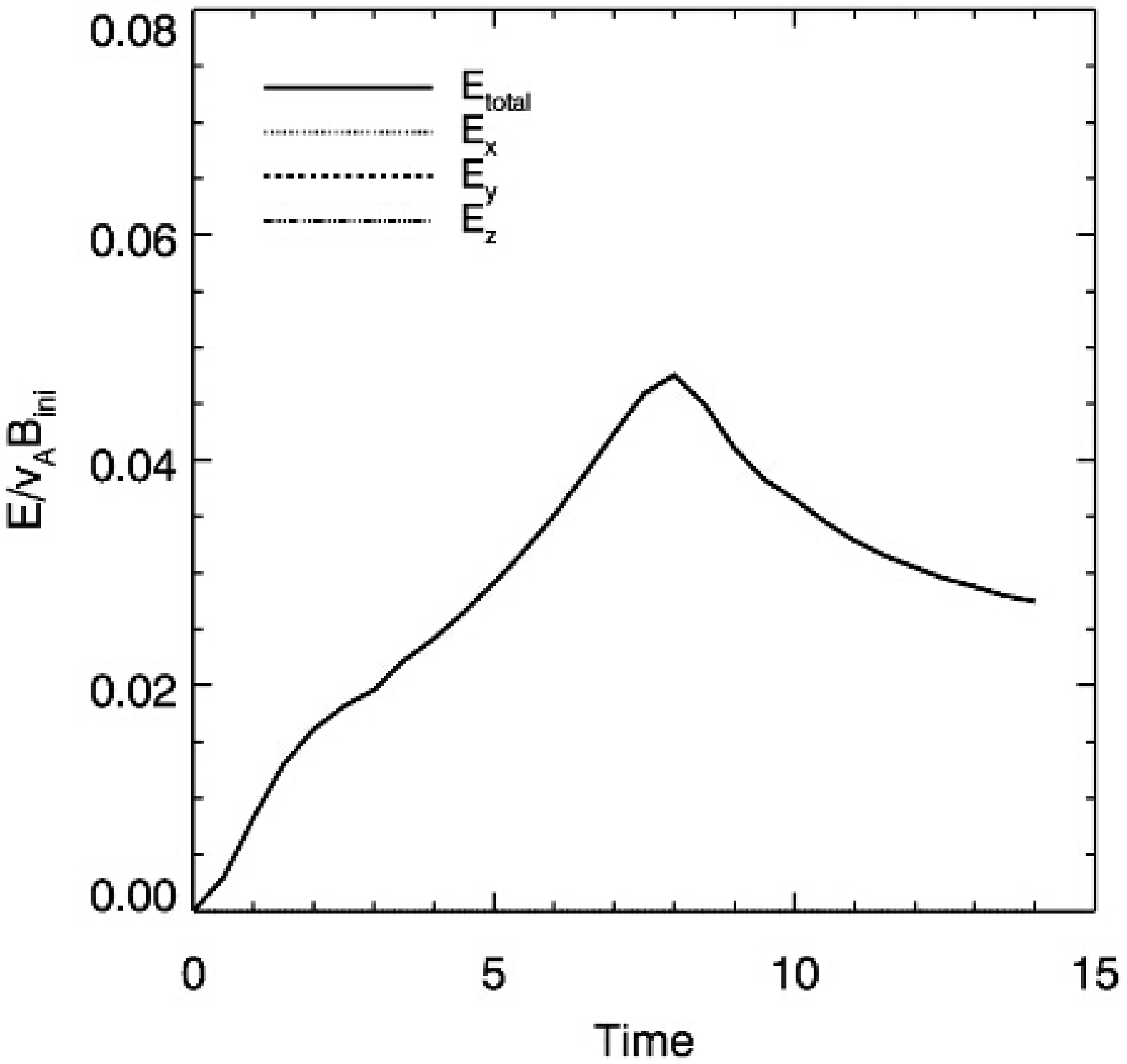}
\includegraphics[width=\ptb]{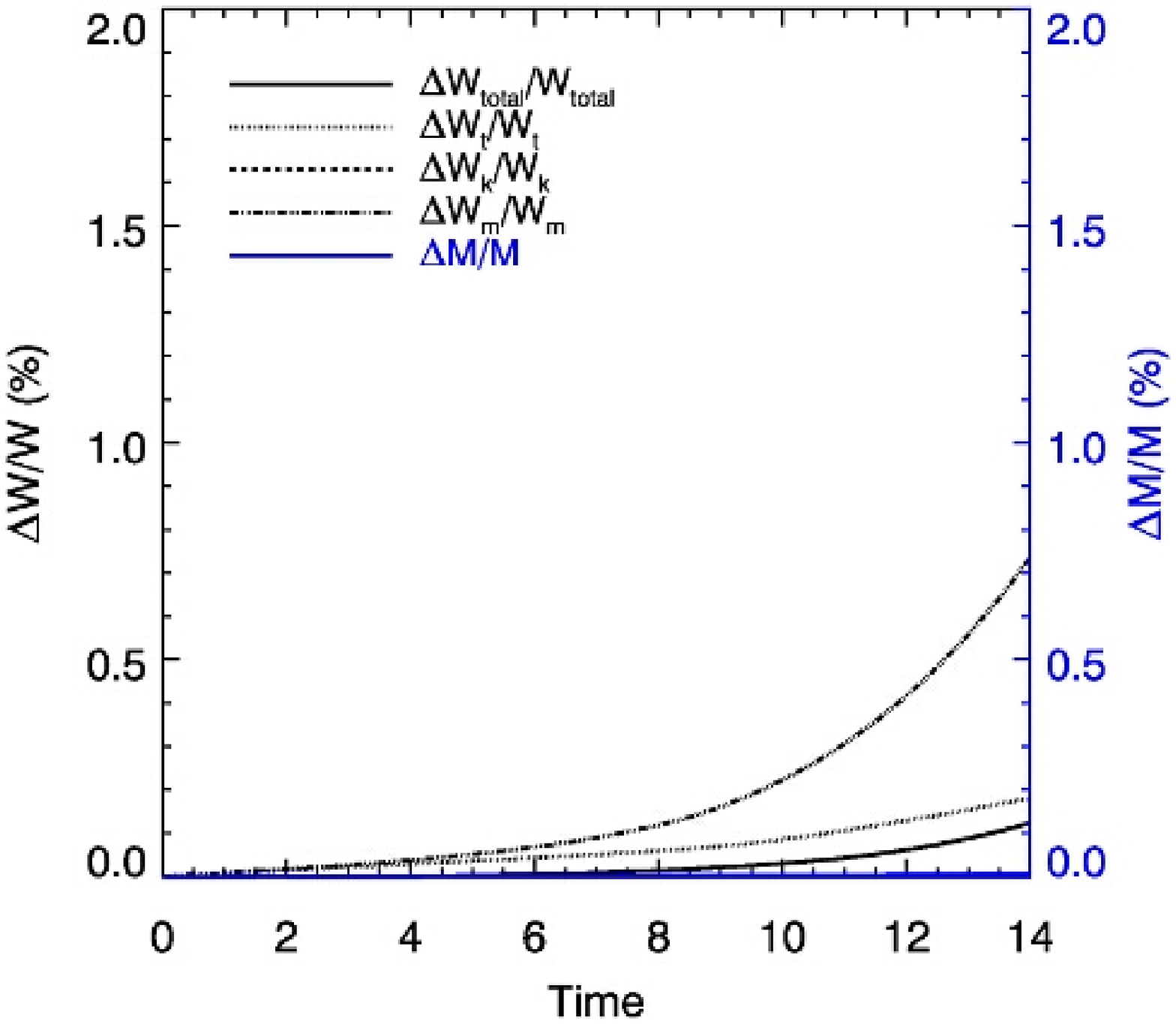}
\includegraphics[width=\ptb]{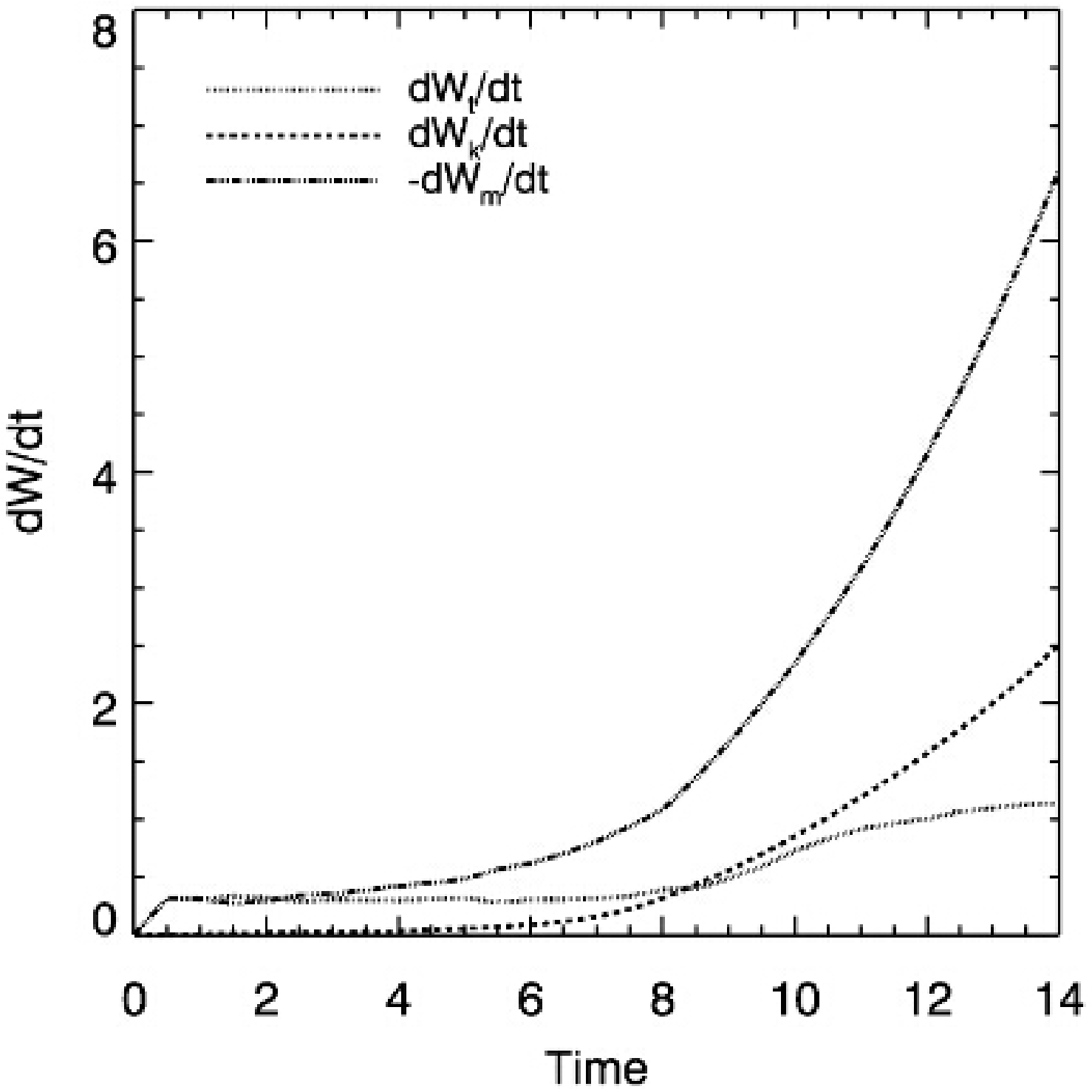}
\caption{Variations of the reconnection rate (upper left panel), electronic field at the x point (upper right panel), energy and mass conservations (lower left panel) and energy release rate (lower right panel) in the typical case. $E_x$ and $E_y$ are almost zero, so we can not see them in the upper right panel. $W_t$, $W_k$ and $W_m$ in the lower right panel are the total thermal energy, total kinetic energy and total magnetic energy, respectively.} \label{fig14}
\end{figure}

Because of the complexity in measuring the reconnection rate in 3D simulations, we give two kinds of measurement for the reconnection rate using the inflow speed and electric field as shown in Figure~\ref{fig14}. The inflow speed is the maximum one in $x$ direction (at the positions $y = 0$ and $z = 0$) and the electric field is taken at the point near the edge of the diffusion region ($E_x$ and $E_y$ are almost zero, so we can not see them in this figure). By comparing these two, one can have a overall view about how fast the magnetic field dissipates in our simulations. The upper panels in Figure~\ref{fig14} show that these two kinds of reconnection rate have a similar profile. Both of them reach the maximum almost at the same time (around 8.5) as we analyzed in 1D distributions and that is also the reason why we choose the time $=8.5$ and a later time $=11.5$ to show the 3D, 2D and 1D distributions. The lower left panel in Figure~\ref{fig14} is the mass and energy conservations with considering the flux coming in or out of the computational box. Even if CIP-MOCCT scheme is non-conservative for variables, our numerical error is still small enough to get a realistic result. $W_t$, $W_k$ and $W_m$ in the lower right panel are the total thermal energy, total kinetic energy and total magnetic energy, respectively. The magnetic energy release rate becomes faster and faster after the reconnection rate reaching the maximum because of the formation and extension of the slow mode shock (\citealt{Nitta2001}). Actually, most of magnetic energy is released by the slow mode shock (\citealt{Petschek1964}).

\begin{figure}[htbp]
\centering
\includegraphics[width=\ptb,trim=4mm  4mm  4mm  6mm, clip]{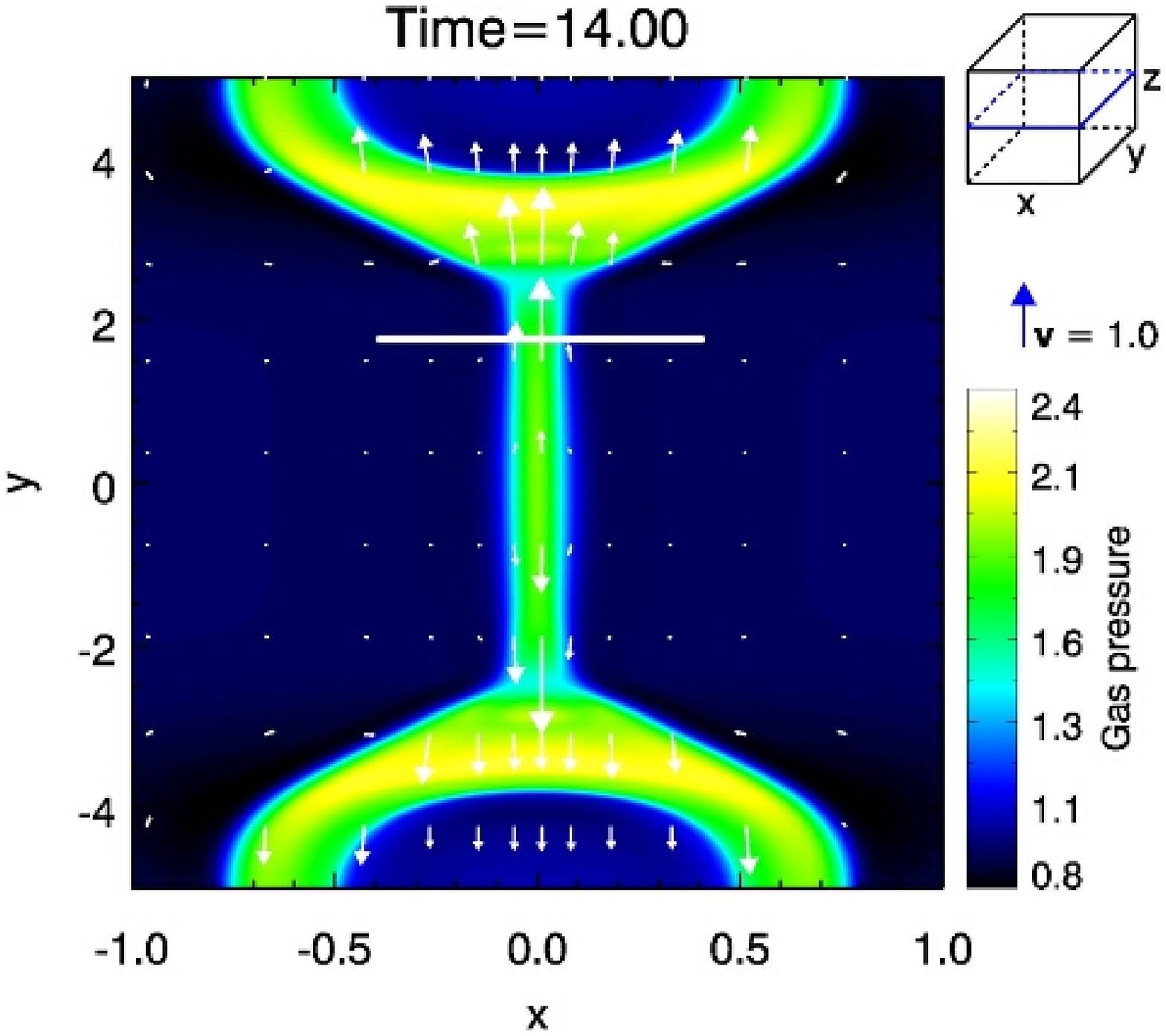}
\includegraphics[width=\ptb,trim=4mm  4mm  4mm  6mm, clip]{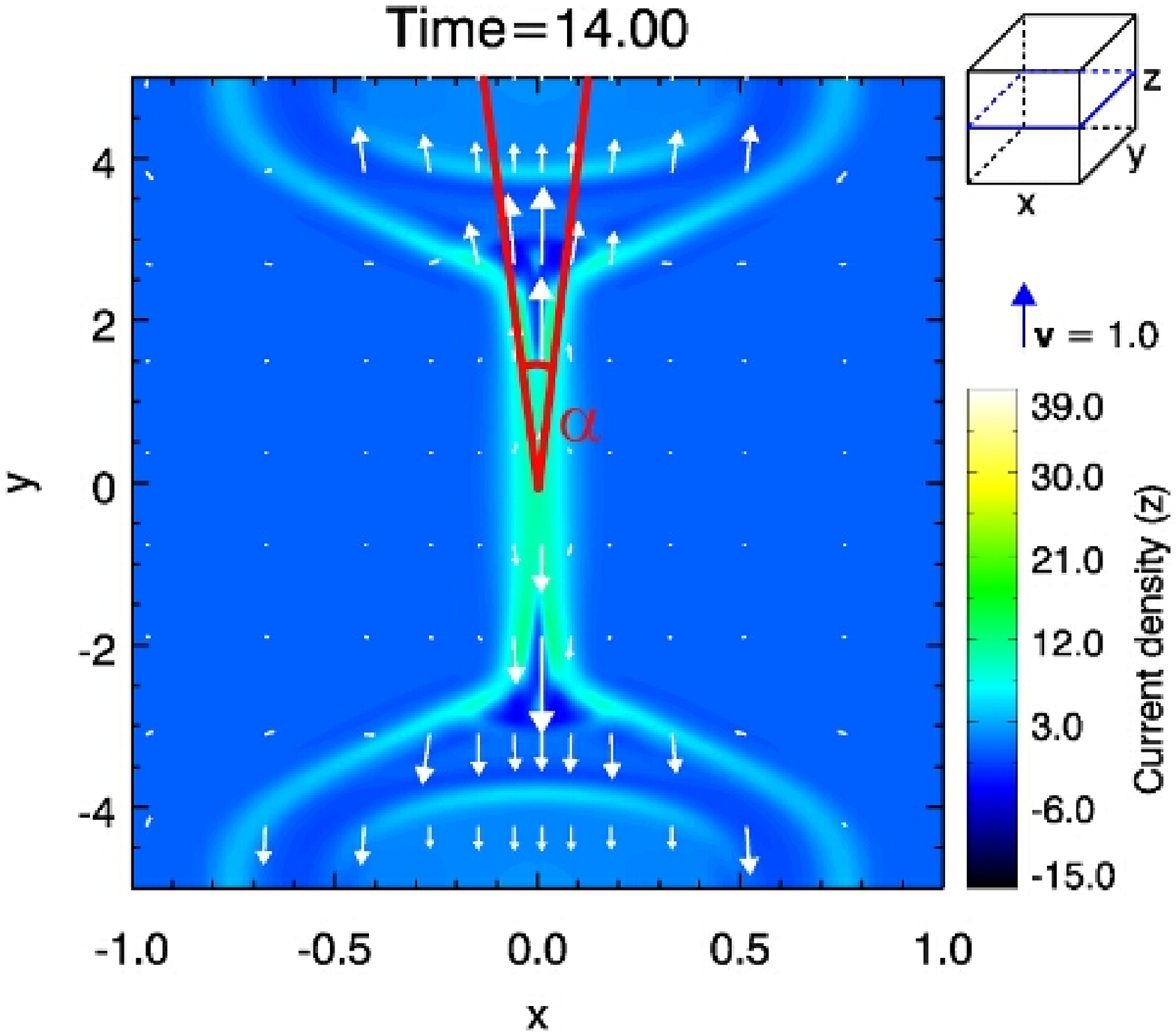}
\includegraphics[width=\ptb,trim=4mm  4mm  8mm  0mm, clip]{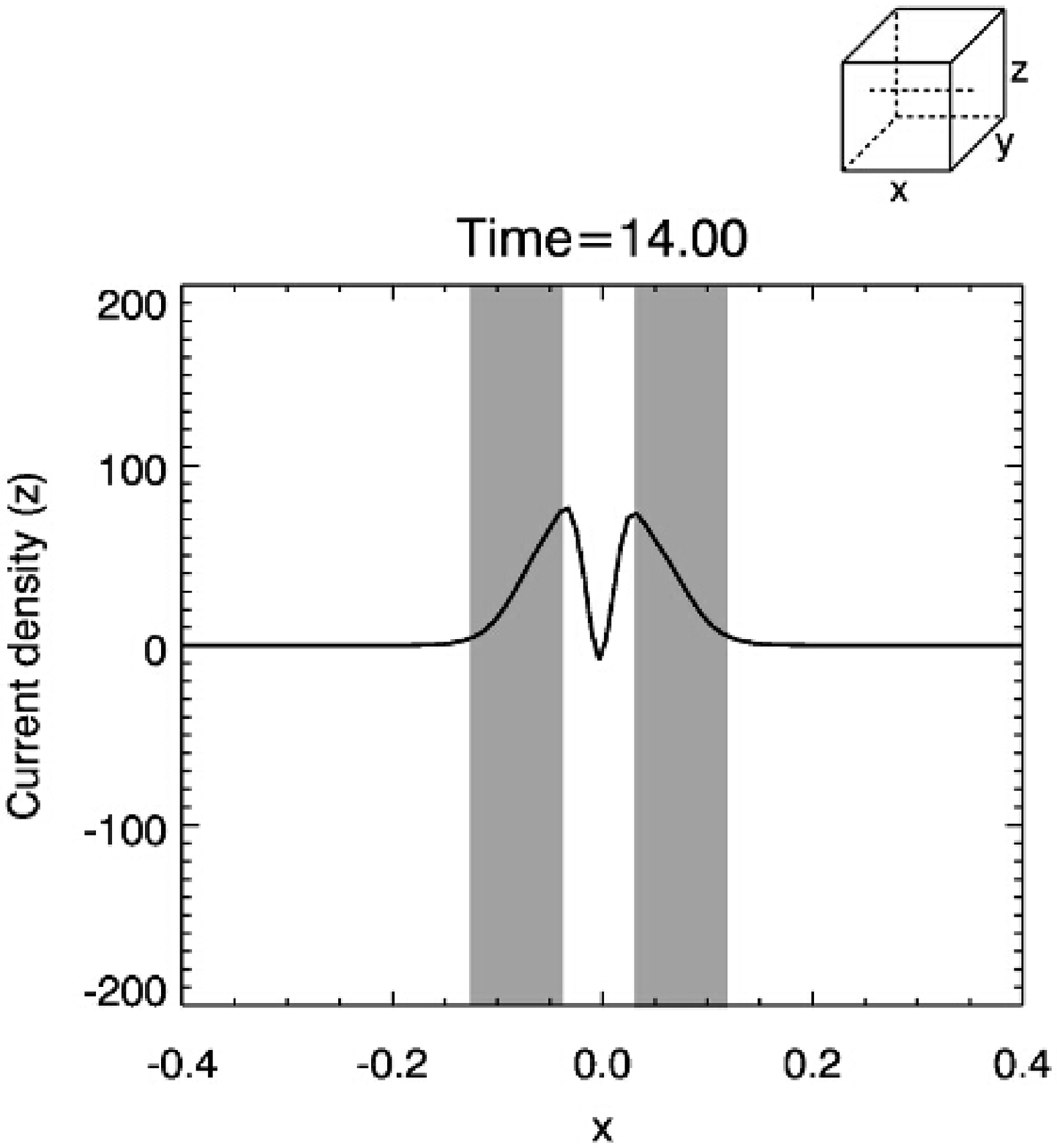}
\includegraphics[width=\ptb,trim=4mm  4mm  8mm  0mm, clip]{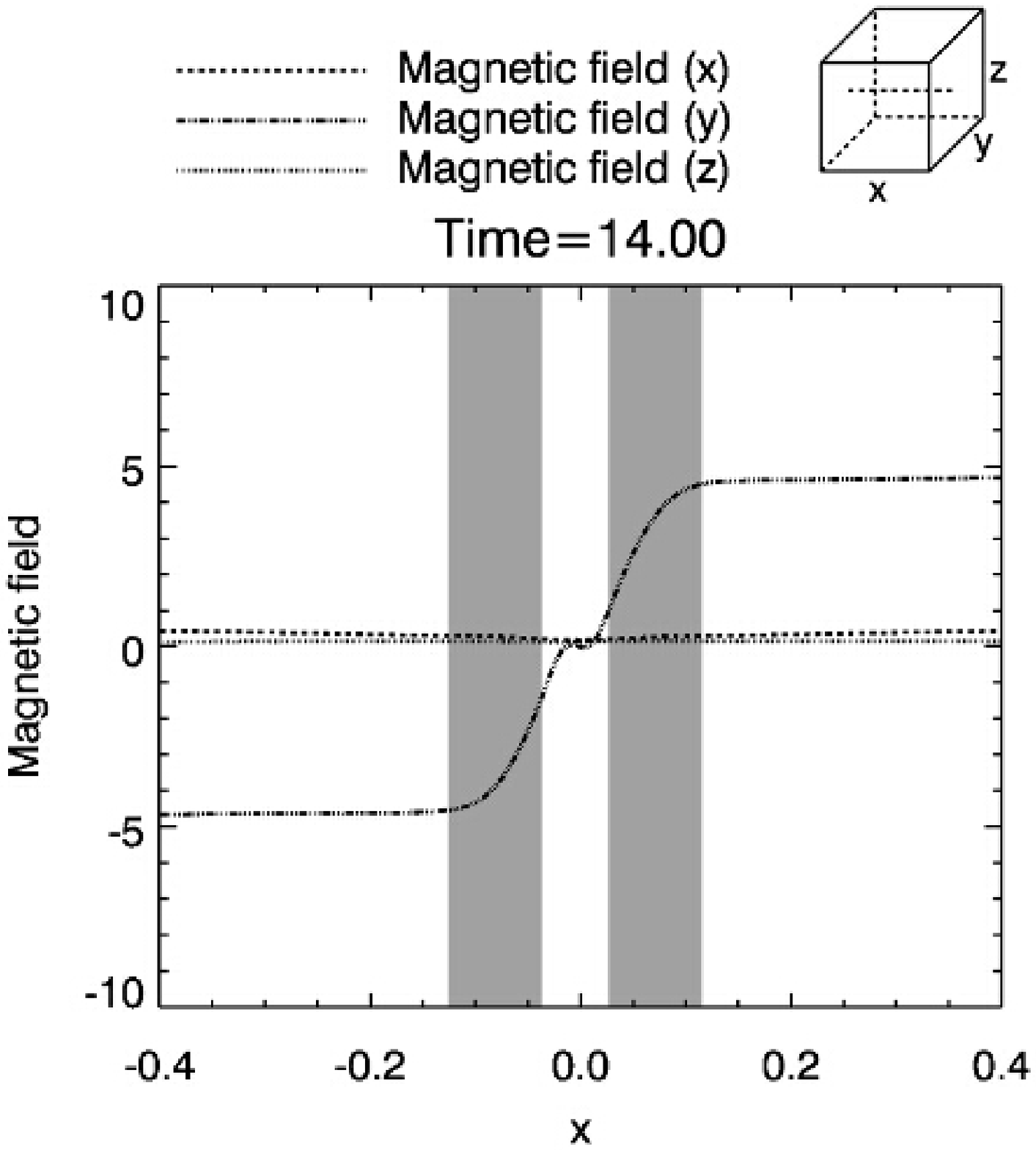}
\includegraphics[width=\ptb,trim=4mm  4mm  8mm  0mm, clip]{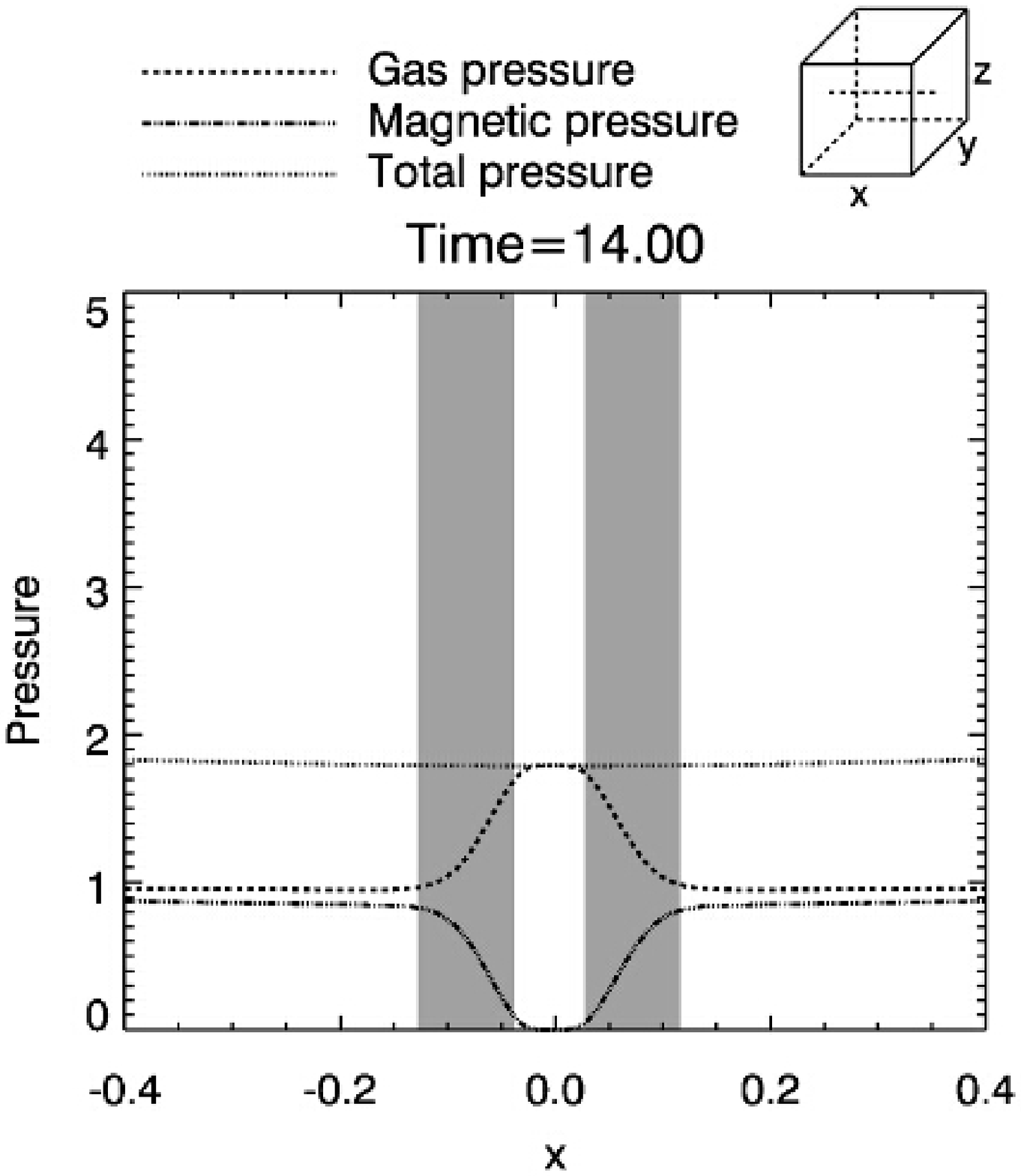}
\includegraphics[width=\ptb,trim=4mm  4mm  8mm  0mm, clip]{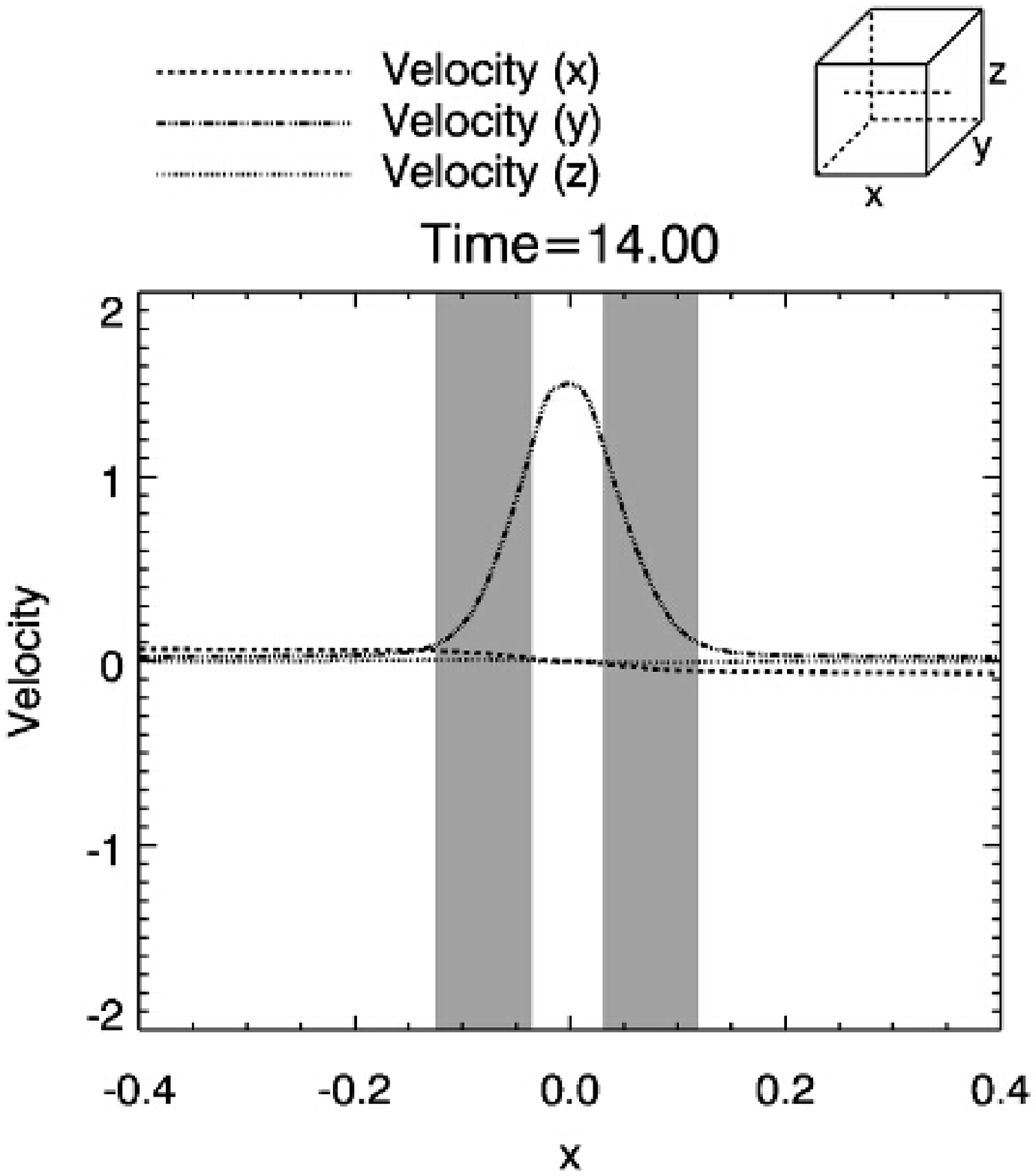}
\caption{Upper panels show the gas pressure and current density distributions in $x-y$ plane, which is also the central part of Figure~\ref{fig04}. Other panels depict one dimensional plots along the white line in the upper left panel at the time $=14.0$. A pair of slow mode shocks are shown by the current density distribution and the shadows in the 1D distributions. The red lines are used to measure the angle ($\alpha$) between the two shocks.} \label{fig15}
\end{figure}

The upper panels of Figure~\ref{fig15} depict the gas pressure and current density distributions in $z$ direction, which is only the central part of Figure~\ref{fig04}. Other panels show one dimensional plot along the white line indicated in the upper left panel at the time $=14.0$. We can see a pair of slow mode shocks in the current density distribution and the 1D panels. The value jumps between two sides of the shock are shown by the shadow parts in the 1D distributions. Maybe the shocks are not sharp enough, which is due to the lack of grid points (about 65 grid points along the white line). However, we can compare this with the distribution of current density. The clear X-shaped structure is seen around the X-point. Moreover, the angle marked by $\alpha$ is related to the reconnection rate, because we have the formula $\tan (\alpha/2) = v_{in}/v_A$ according to the Petschek reconnection mechanism (\citealt{Petschek1964}). From the upper right panel, $\tan (\alpha/2)$ is about 0.026, while the reconnection rate is approximate 0.032 given by the upper left panel of Figure~\ref{fig14} at the time $=14.0$. The reconnection rate is a bit larger than the value measured from the angle between two slow mode shocks. Considering the error of the measurement of the angle and the complexity in 3D MHD reconnection, we think that the two values are comparible.

\section{Parameter Dependence}
\label{dependence}

In this section, we have two parameters to analyze. One is the reconnection angle (i.e., $\theta$, shown in Figure~\ref{fig01}), the other is the resistivity (i.e., $\eta_{ini}$). Dependence on the reconnection angle can show us whether there is a relationship between the velocity of fan-shaped jets and the reconnecting component of magnetic field. For instance, if the reconnection angle is $\pi/2$, the maximum reconnecting component of magnetic field is only $\mathbf{B}_{ini}\sin{(\pi/4)}$ (for other reconnection angle case, $\mathbf{B}_{ini}\sin{(\theta/2)}$), and the velocity of the ordinary reconnection outflow ejected by the magnetic tension force has a similar form as $v_A\sin{(\pi/4)}$ (for other reconnection angle case, $v_A\sin{(\theta/2)}$). We know that the dependence of the velocity on the resistivity value, if it is localized, is not so remarkable in 2D magnetic reconnection simulations (\citealt{Jiang2010}). However, the driving force for these jets seems different from the magnetic tension force. Thus, we have to check such a dependence. The dependence on other parameters will be studied in the near future.

\begin{figure}[htbp]
\centering
\includegraphics[height=\pts,trim=4mm 10mm 10mm 16mm,clip]{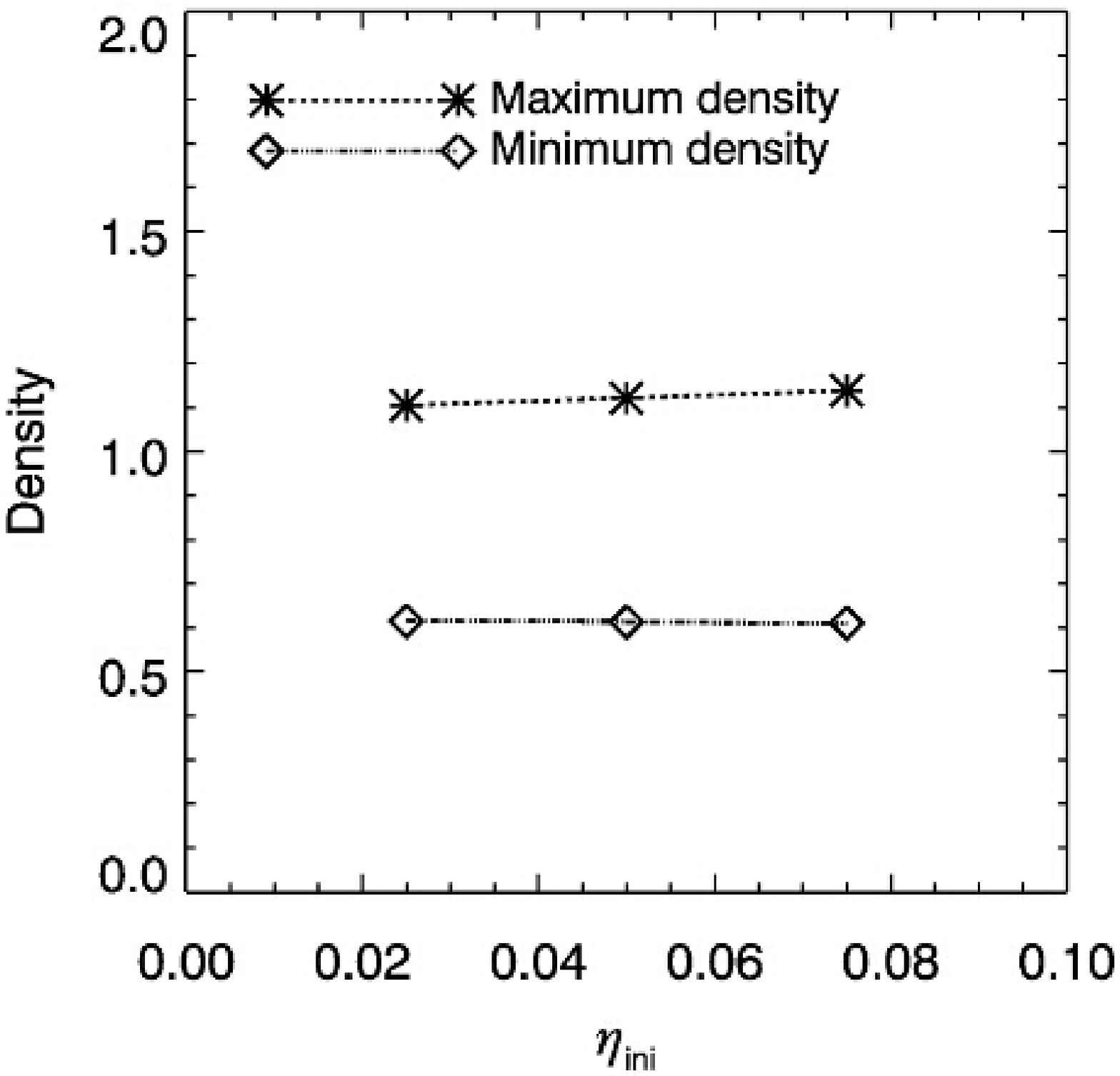}
\includegraphics[height=\pts,trim=4mm 10mm 10mm 16mm,clip]{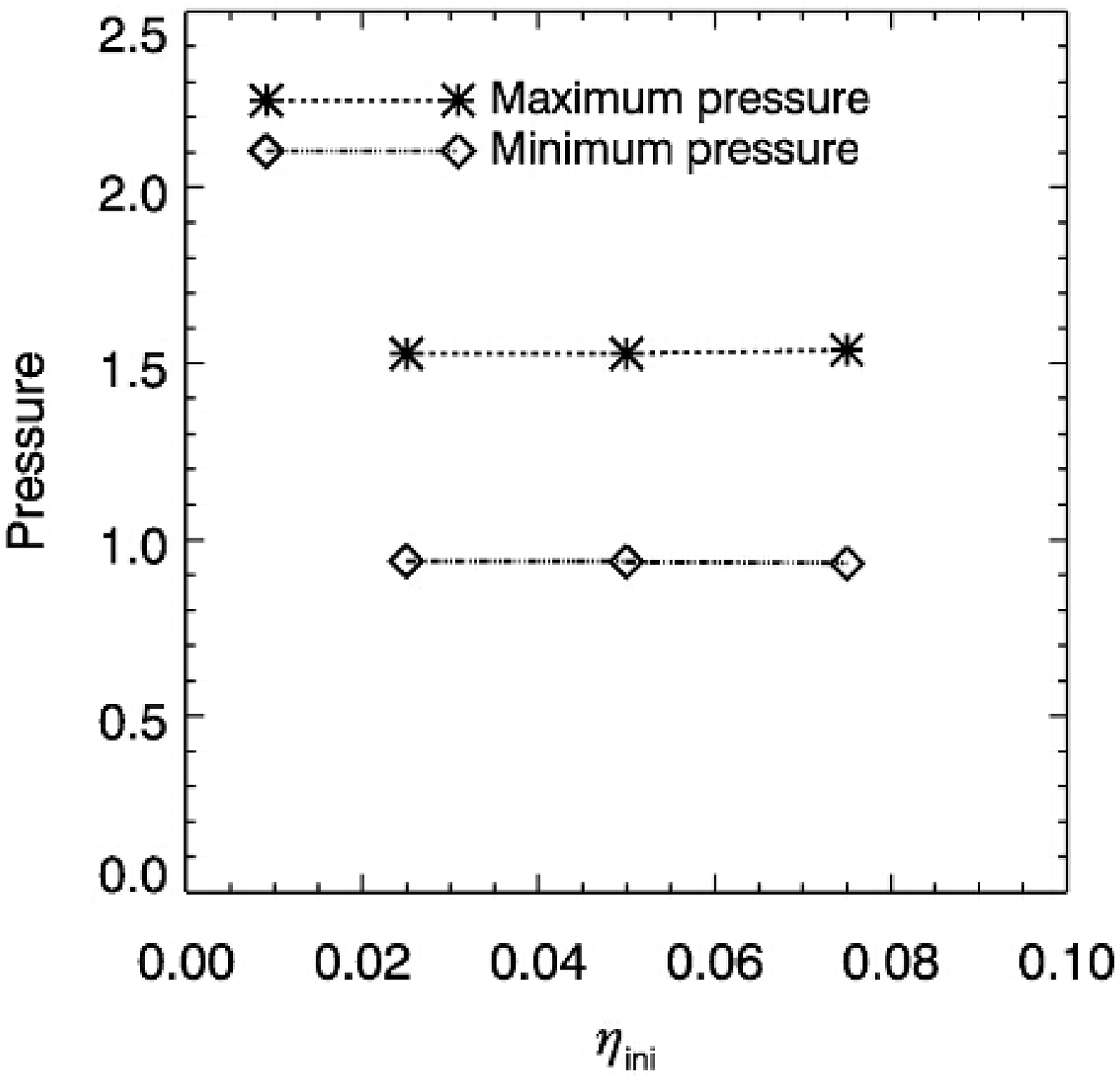}
\includegraphics[height=\pts,trim=4mm 10mm 10mm 16mm,clip]{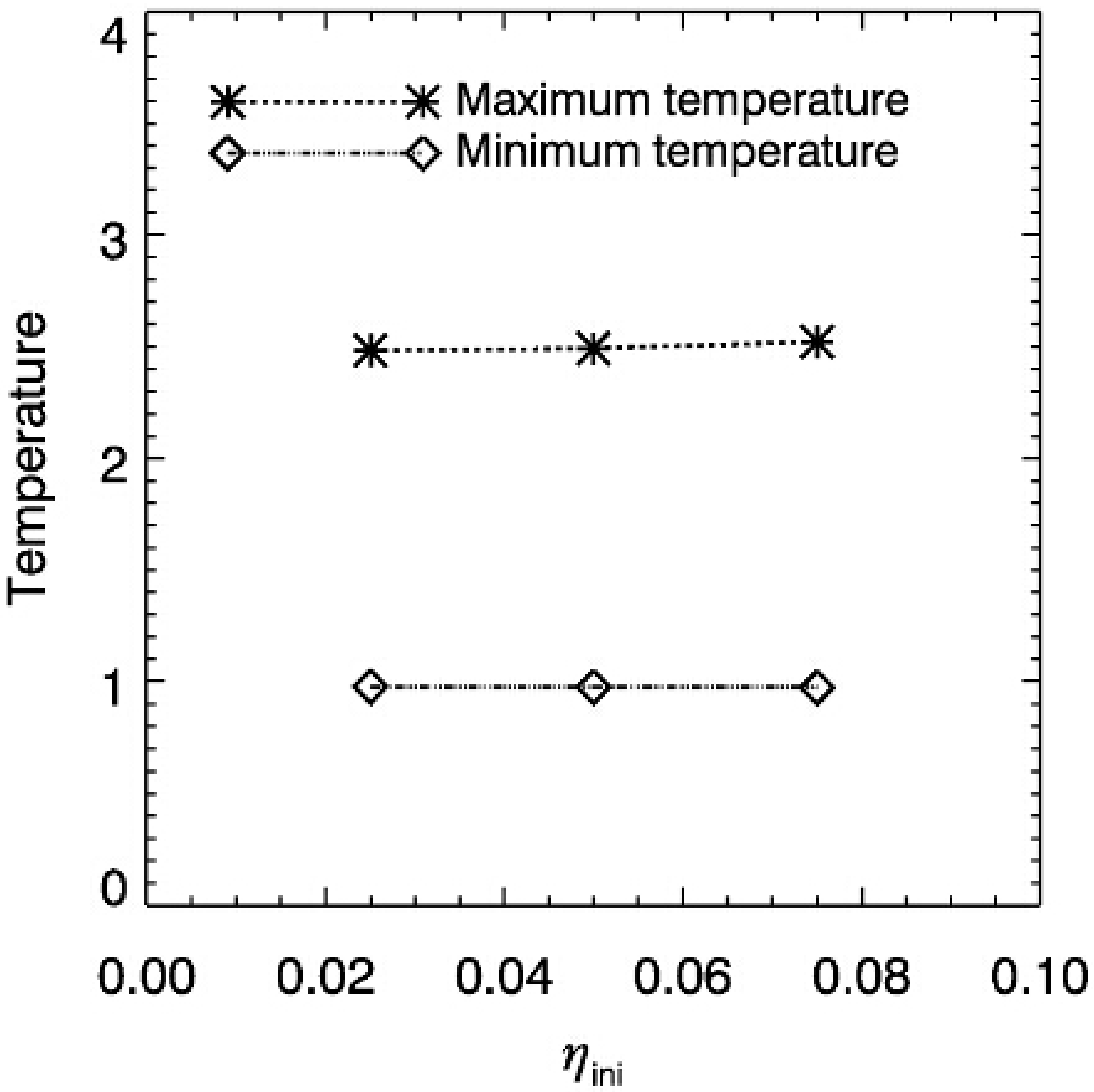}
\caption{The maximum and the minimum densities, pressures and temperatures as a function of the resistivity values.}
\label{fig16}
\end{figure}

\begin{figure}[htbp]
\centering
\includegraphics[height=\ptb,trim=4mm 10mm 10mm 16mm,clip]{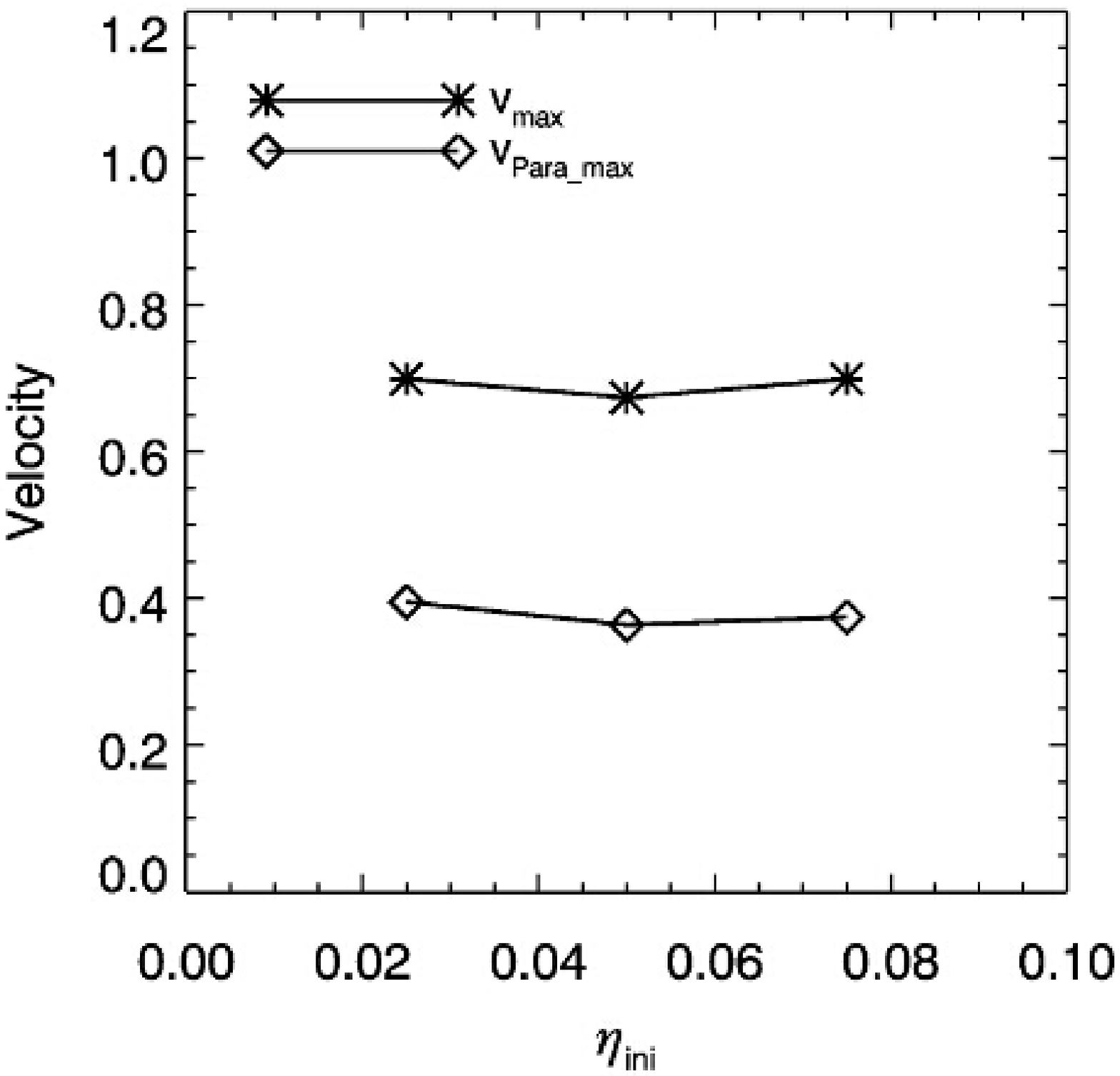}
\includegraphics[height=\ptb,trim=4mm 10mm 10mm 16mm,clip]{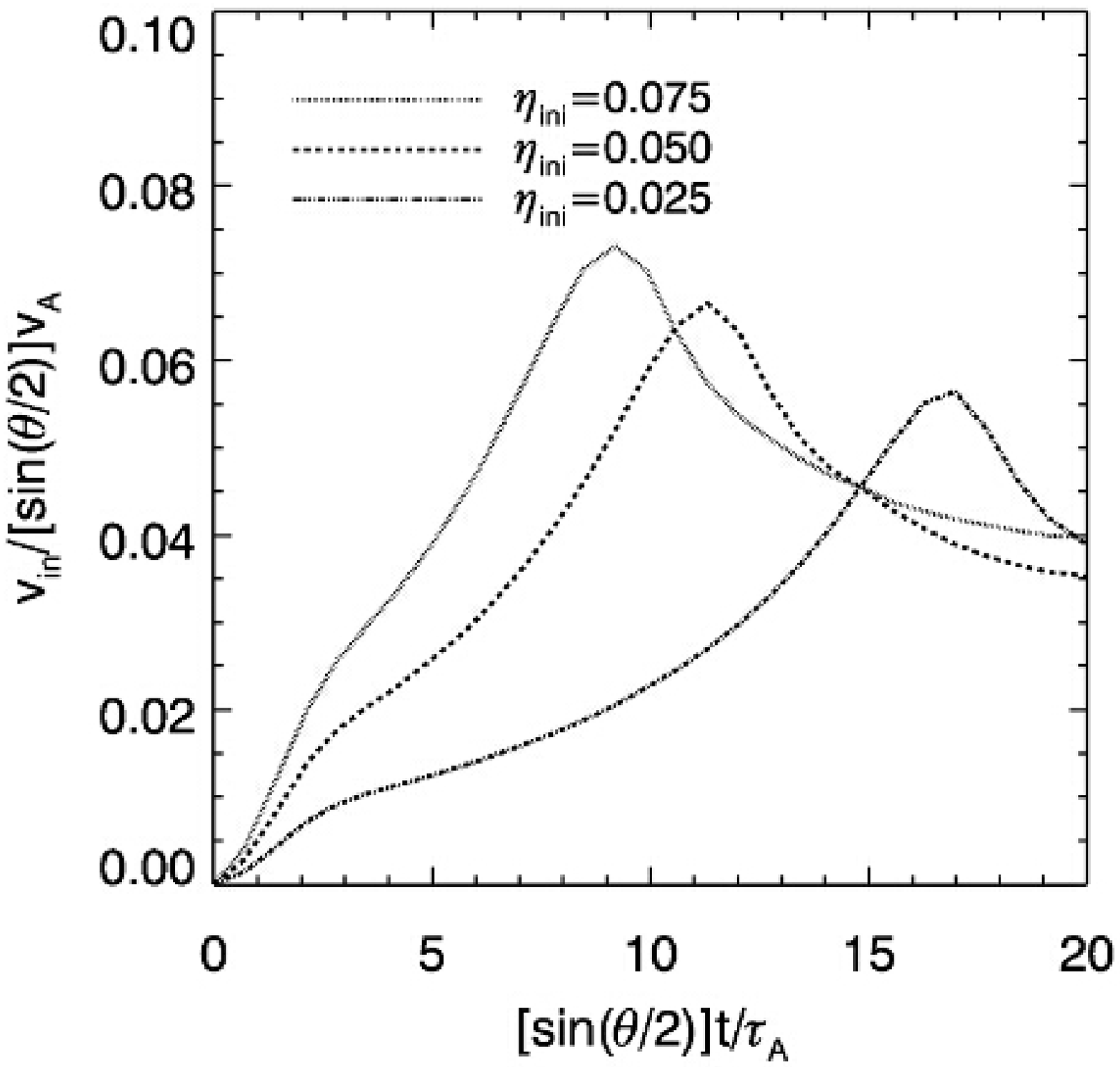}
\caption{The left panel shows the maximum velocity indicating the speed of ordinary reconnection jets) and the parallel velocity (parallel to the initial magnetic field which means the velocity of fan-shaped jets) as a function of the resistivity values. The right panel depicts the reconnection rates presented by velocity. Note that the reconnection angle ($\theta$) is $\pi$ in this case.} \label{fig17}
\end{figure}

The dependence on the resistivity is very simple. As shown in Figures~\ref{fig16} and~\ref{fig17}, the variables and velocity are constant for different resistivity values. One important thing is that there are two kinds of velocity we used in these figures. One is the maximum velocity in the computational box which stands for the velocity of the ordinary reconnection jets, the other is the maximum parallel velocity (parallel to the initial magnetic field) which indicates the speed of fan-shaped jets. Moveover, the reconnection becomes faster when the resistivity value increases. We can see that there is no remarkable dependence between simulation results (density, pressure, temperature and velocity) and the initial resistivity values.

\clearpage

\begin{figure}[htbp]
\centering
\includegraphics[width=\ptb,trim=5mm 5mm 5mm 5mm,clip]{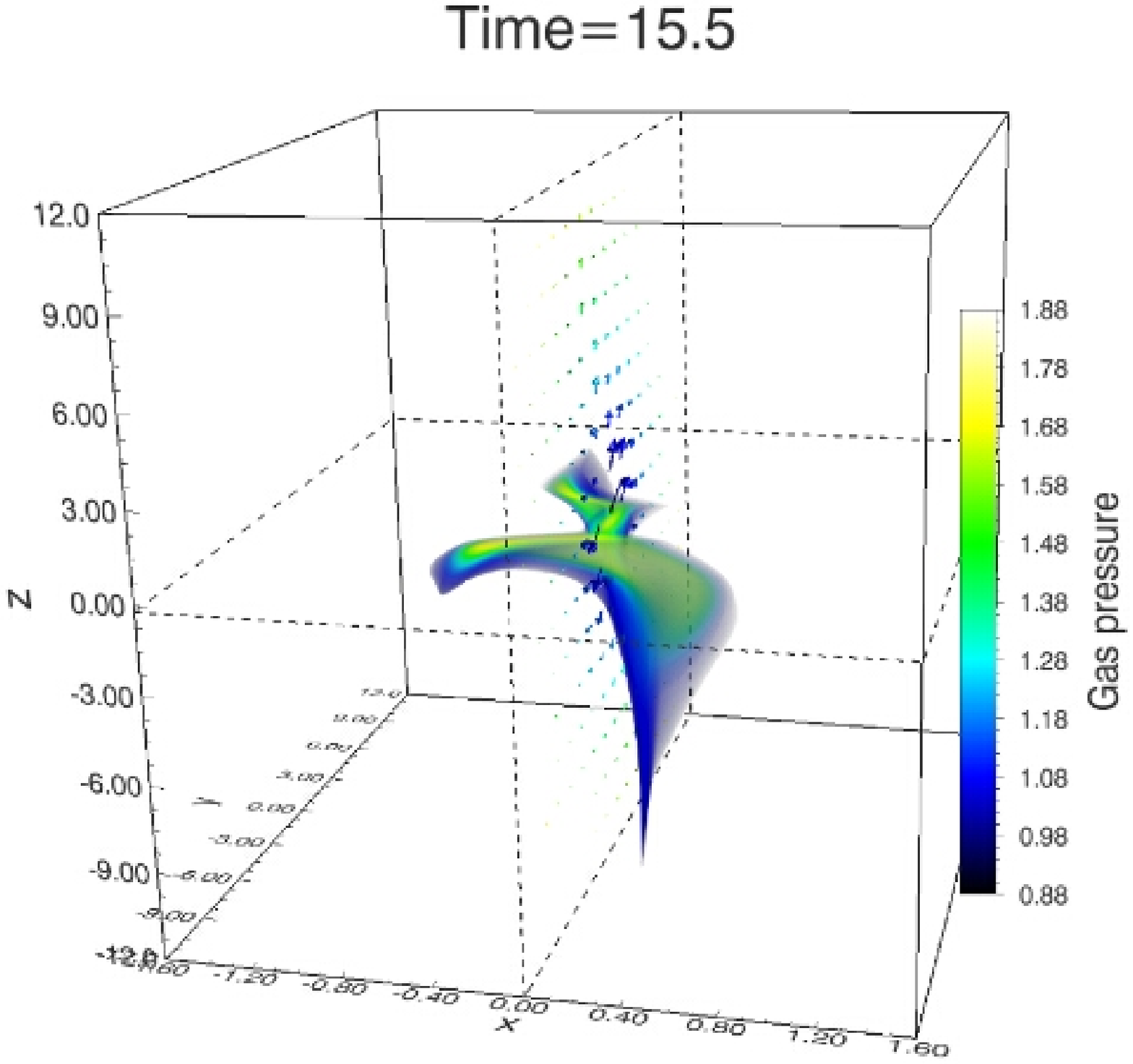}
\includegraphics[width=\ptb,trim=5mm 5mm 5mm 5mm,clip]{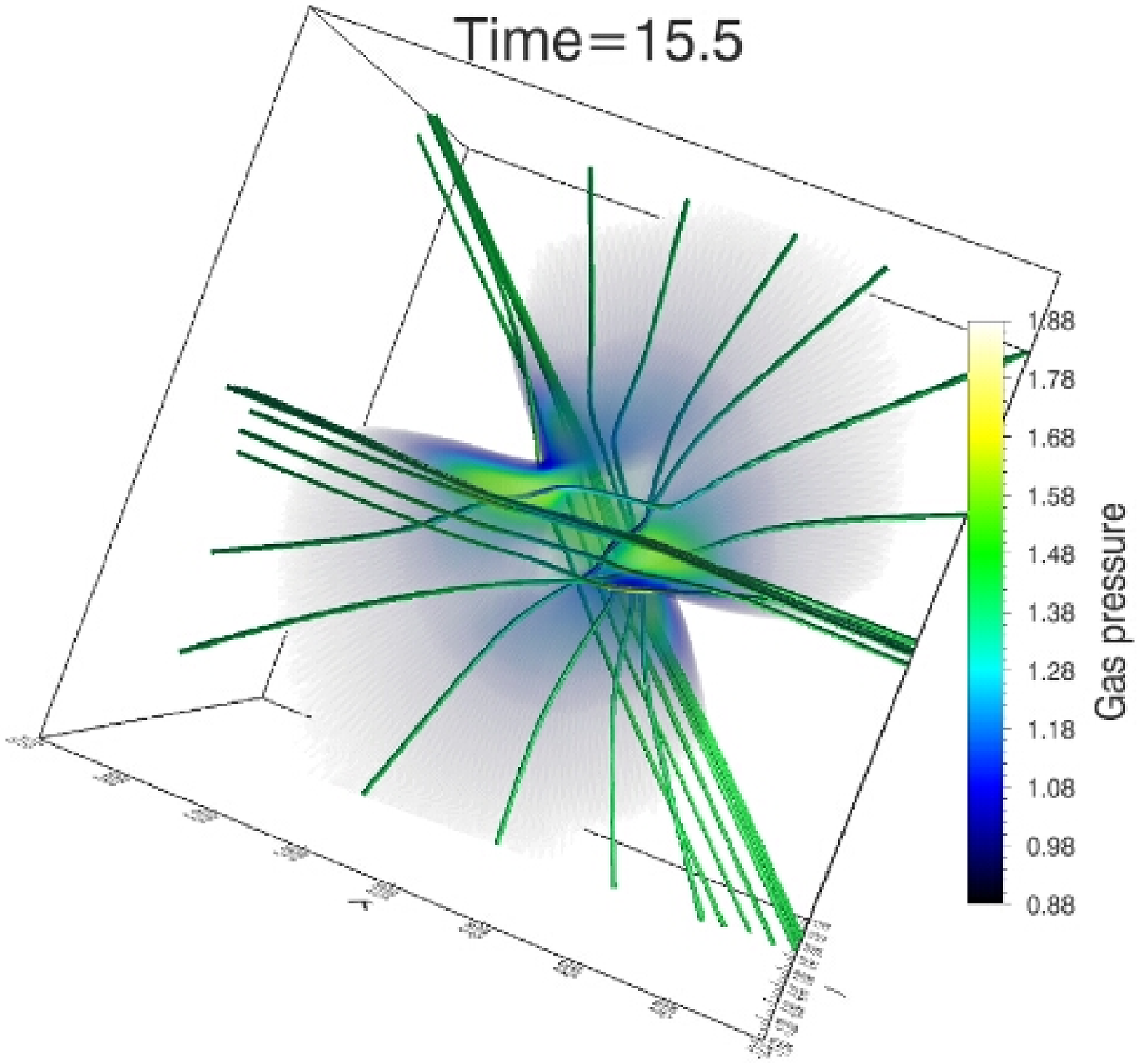}
\includegraphics[width=\ptb,trim=5mm 5mm 5mm 5mm,clip]{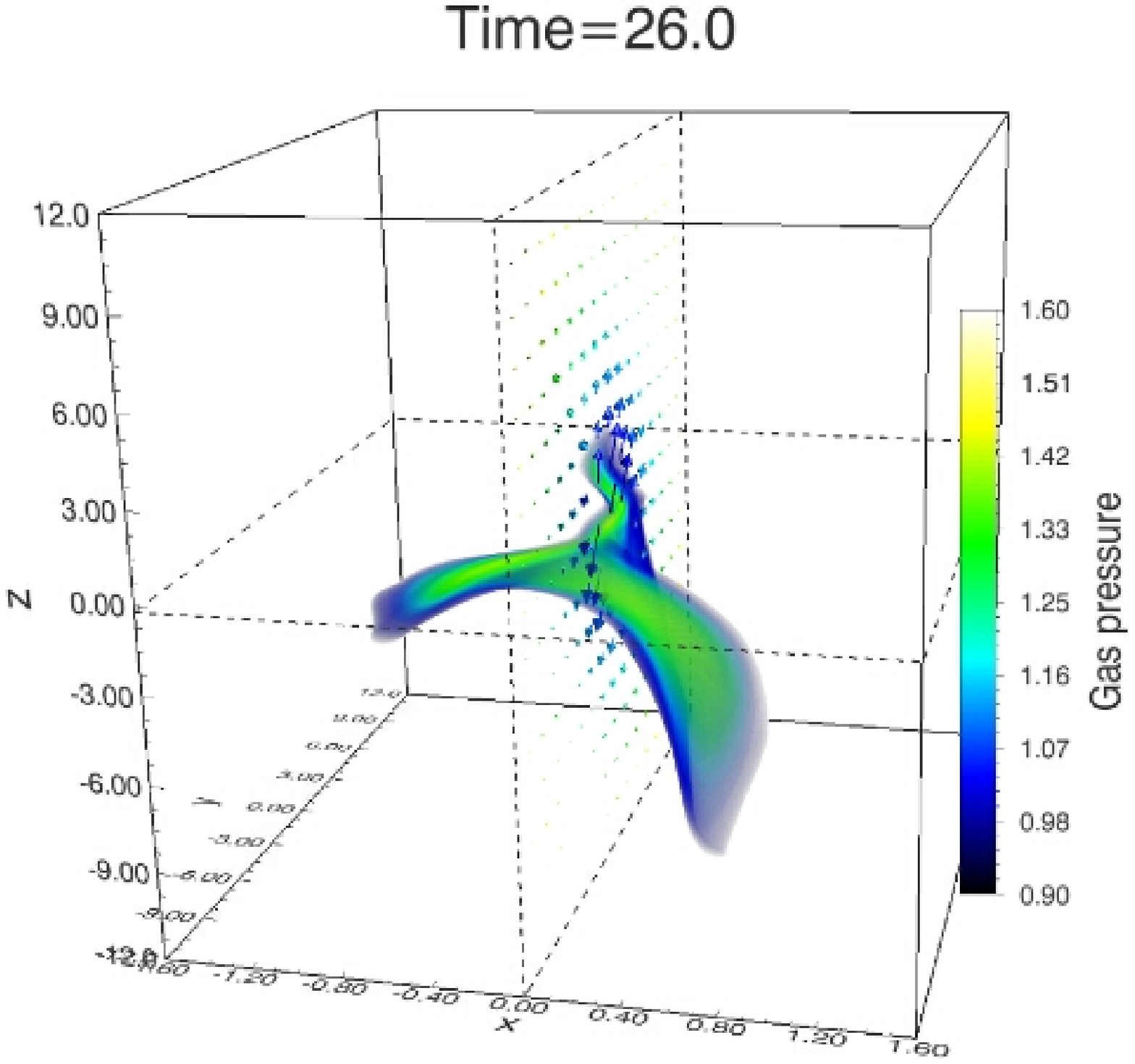}
\includegraphics[width=\ptb,trim=5mm 5mm 5mm 5mm,clip]{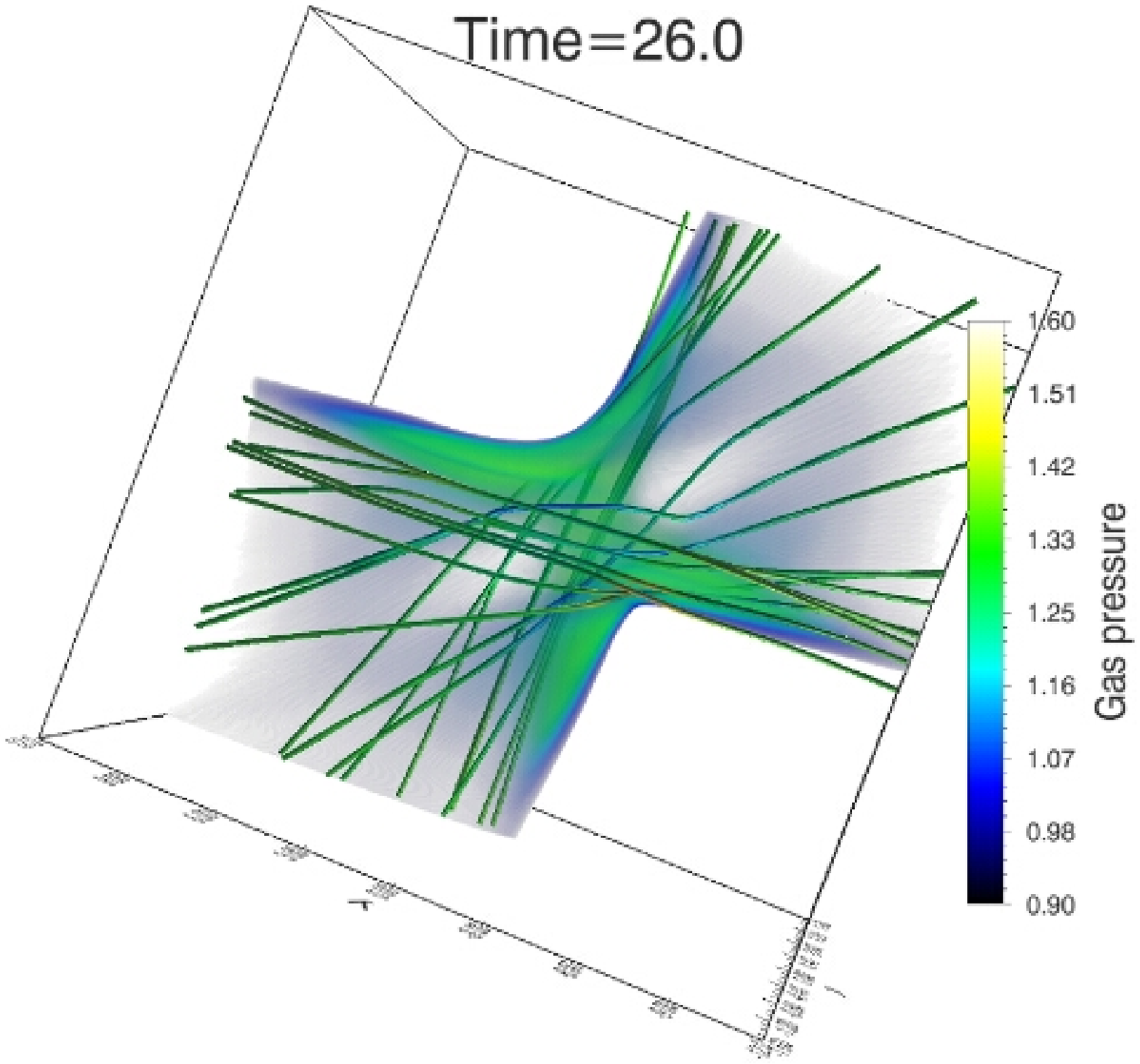}
\includegraphics[width=\ptb,trim=5mm 5mm 5mm 5mm,clip]{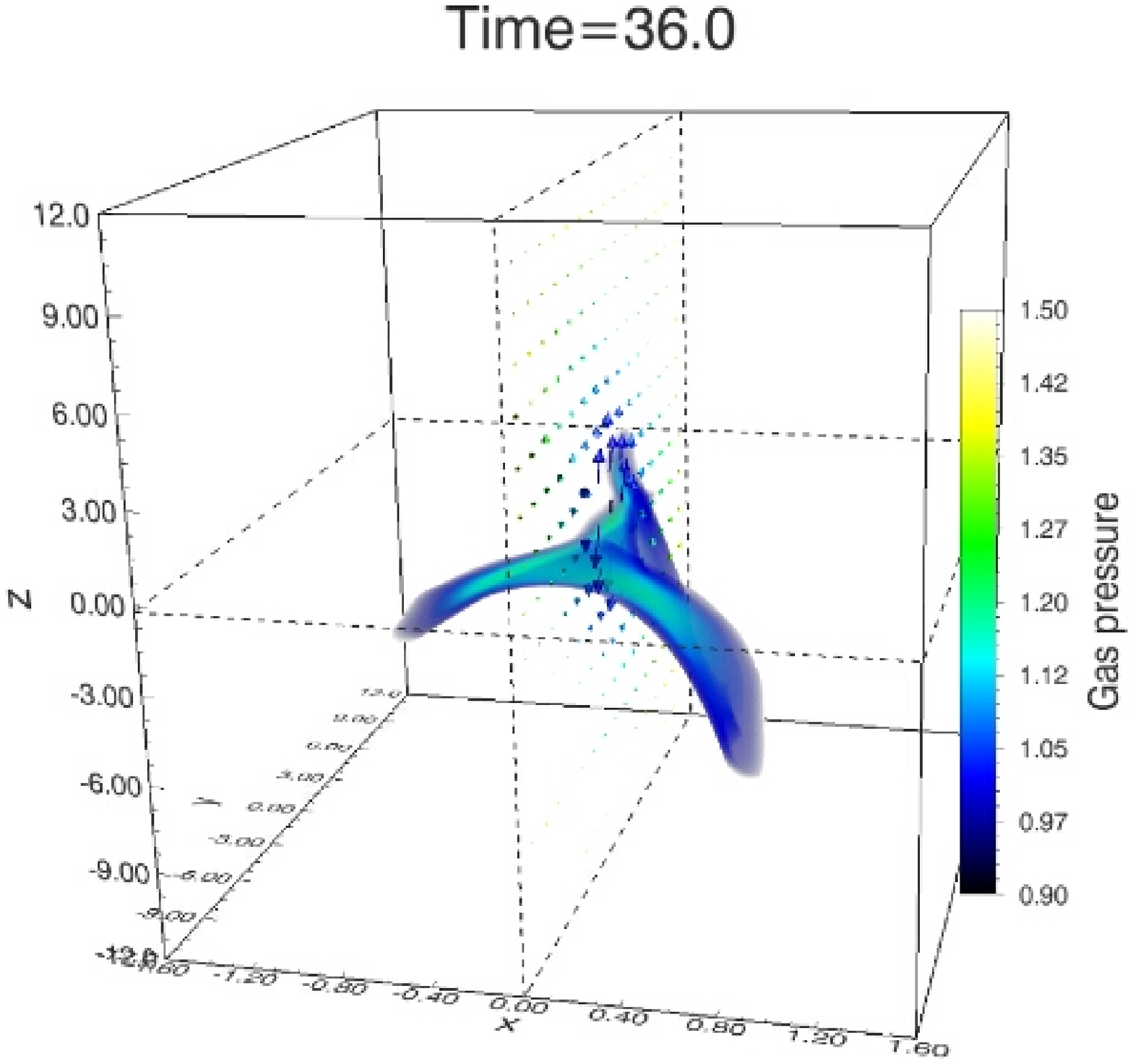}
\includegraphics[width=\ptb,trim=5mm 5mm 5mm 5mm,clip]{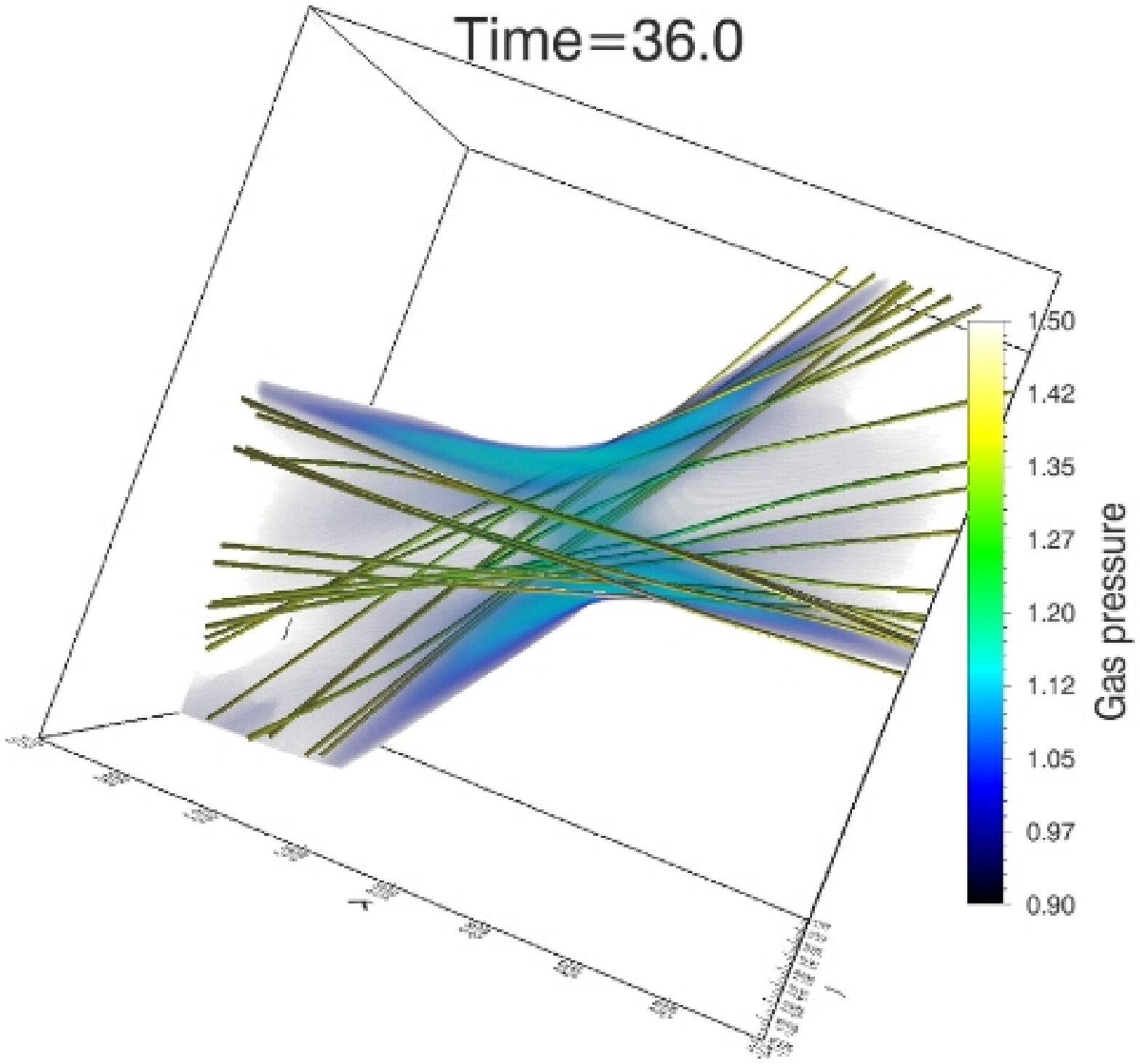}
\caption{3D visualizations of the gas pressure distributions at different times (The reconnection angles for these cases are $3\pi/4$, $\pi/2$ and $\pi/3$ from top to bottom). The bright colors and high opacity stand for the high gas pressure and the low value is transparent. A $z=0$ cross section and a velocity plane are used in the left panels. Magnetic field and a different view points are used in the right panels.} \label{fig18}
\end{figure}

\begin{figure}[htbp]
\centering
\includegraphics[height=\pts,trim=4mm 10mm 4mm 16mm,clip]{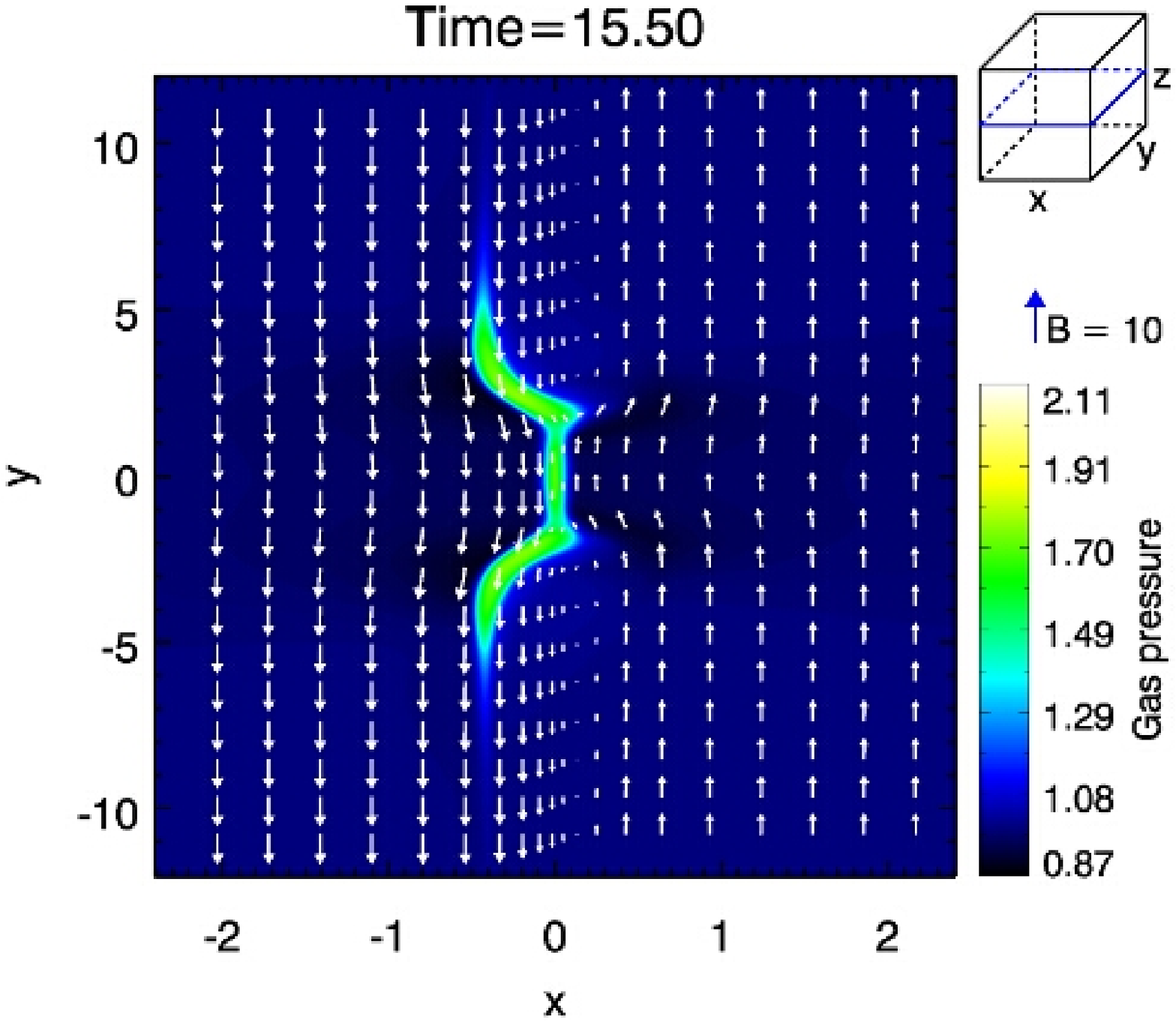}
\includegraphics[height=\pts,trim=4mm 10mm 4mm 16mm,clip]{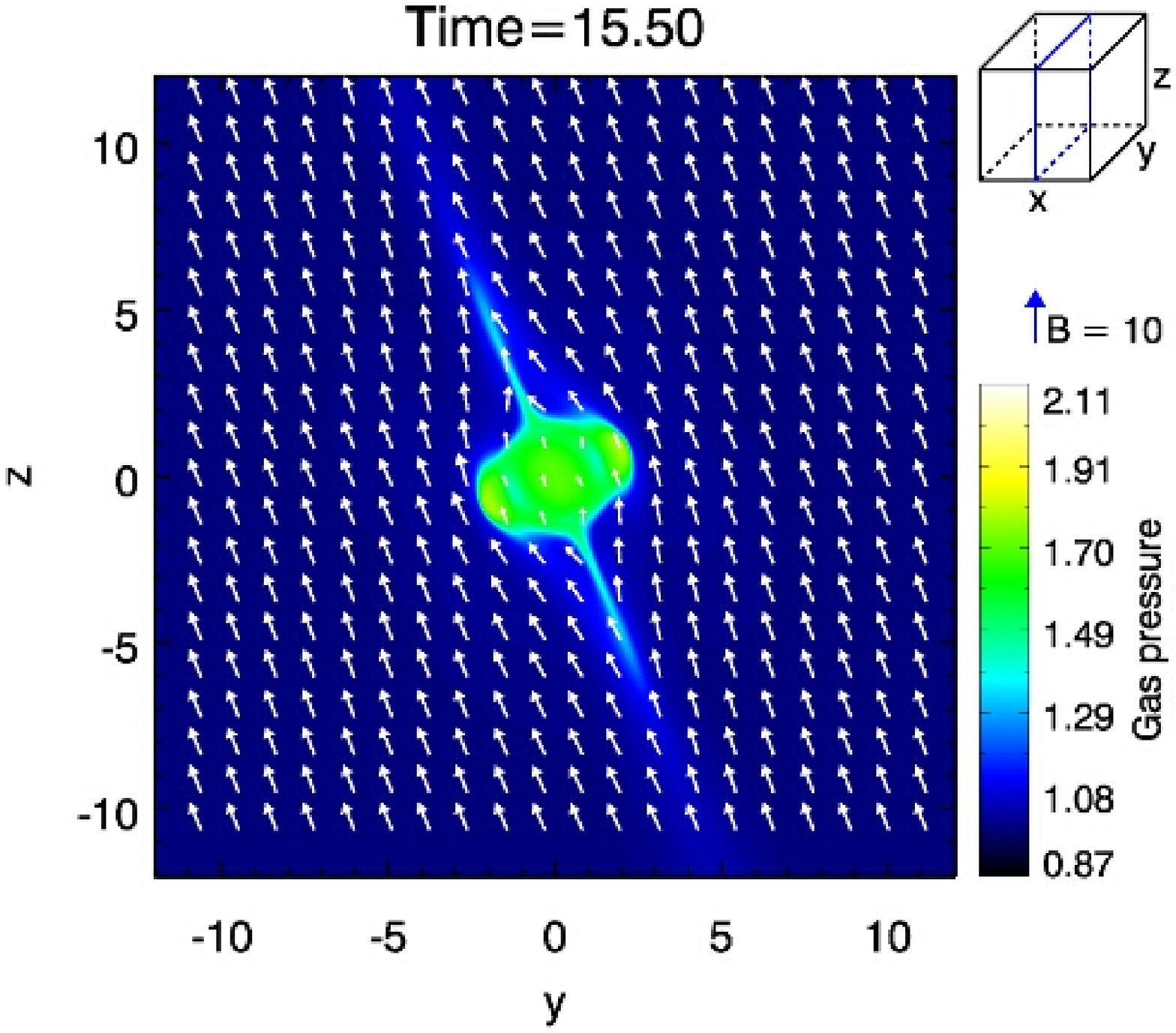}
\includegraphics[height=\pts,trim=4mm 10mm 4mm 16mm,clip]{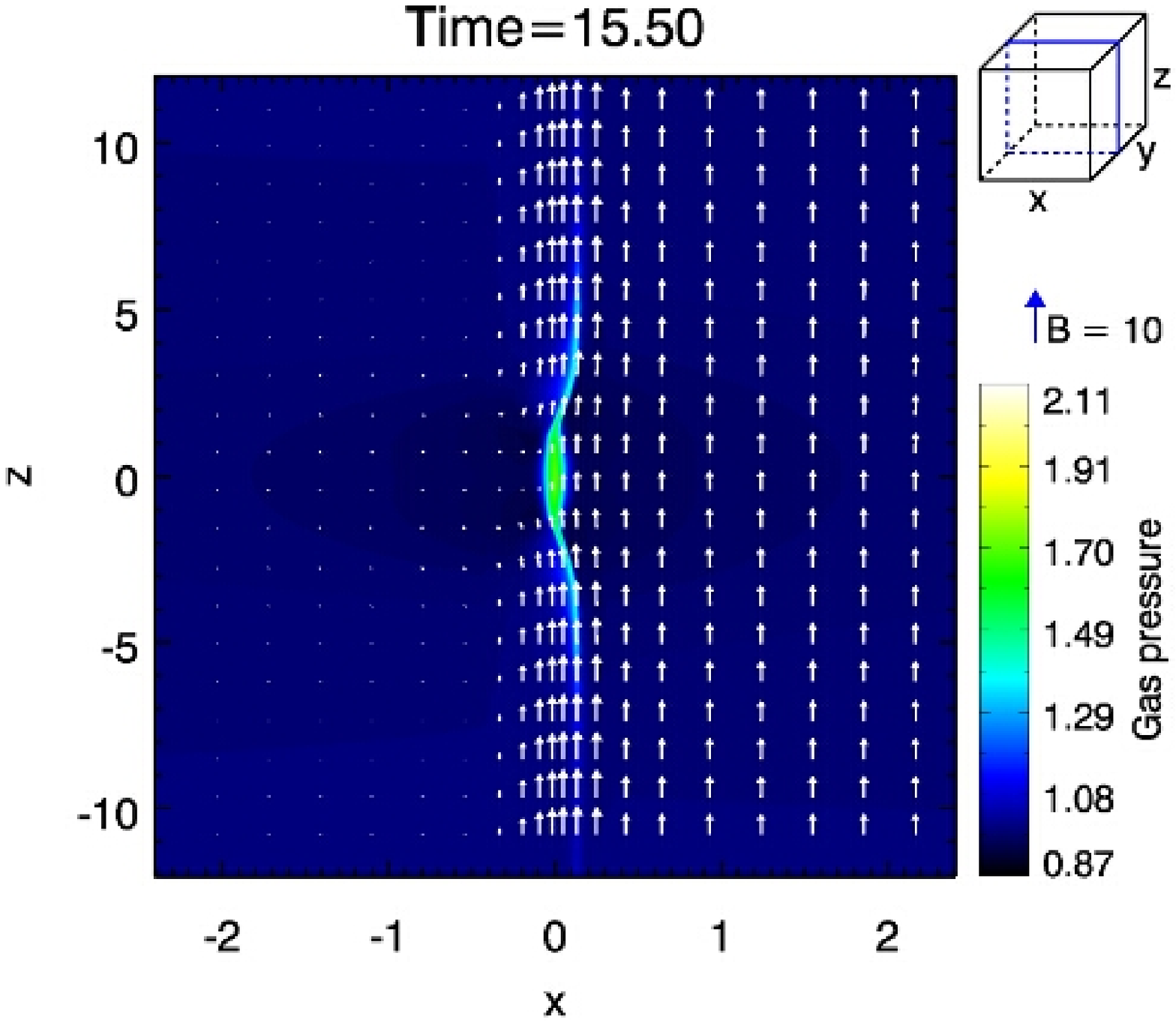}
\includegraphics[height=\pts,trim=4mm 10mm 4mm 16mm,clip]{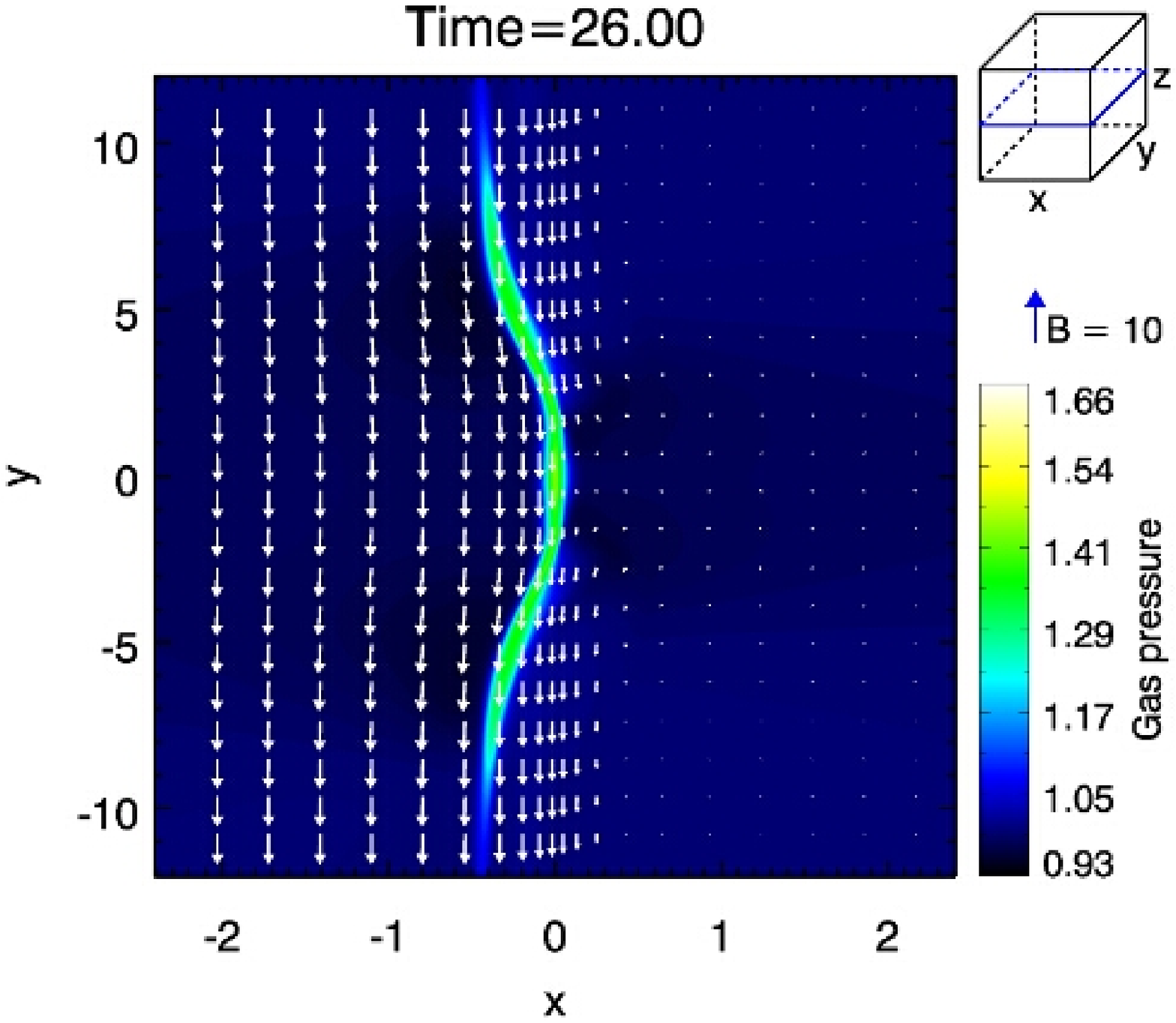}
\includegraphics[height=\pts,trim=4mm 10mm 4mm 16mm,clip]{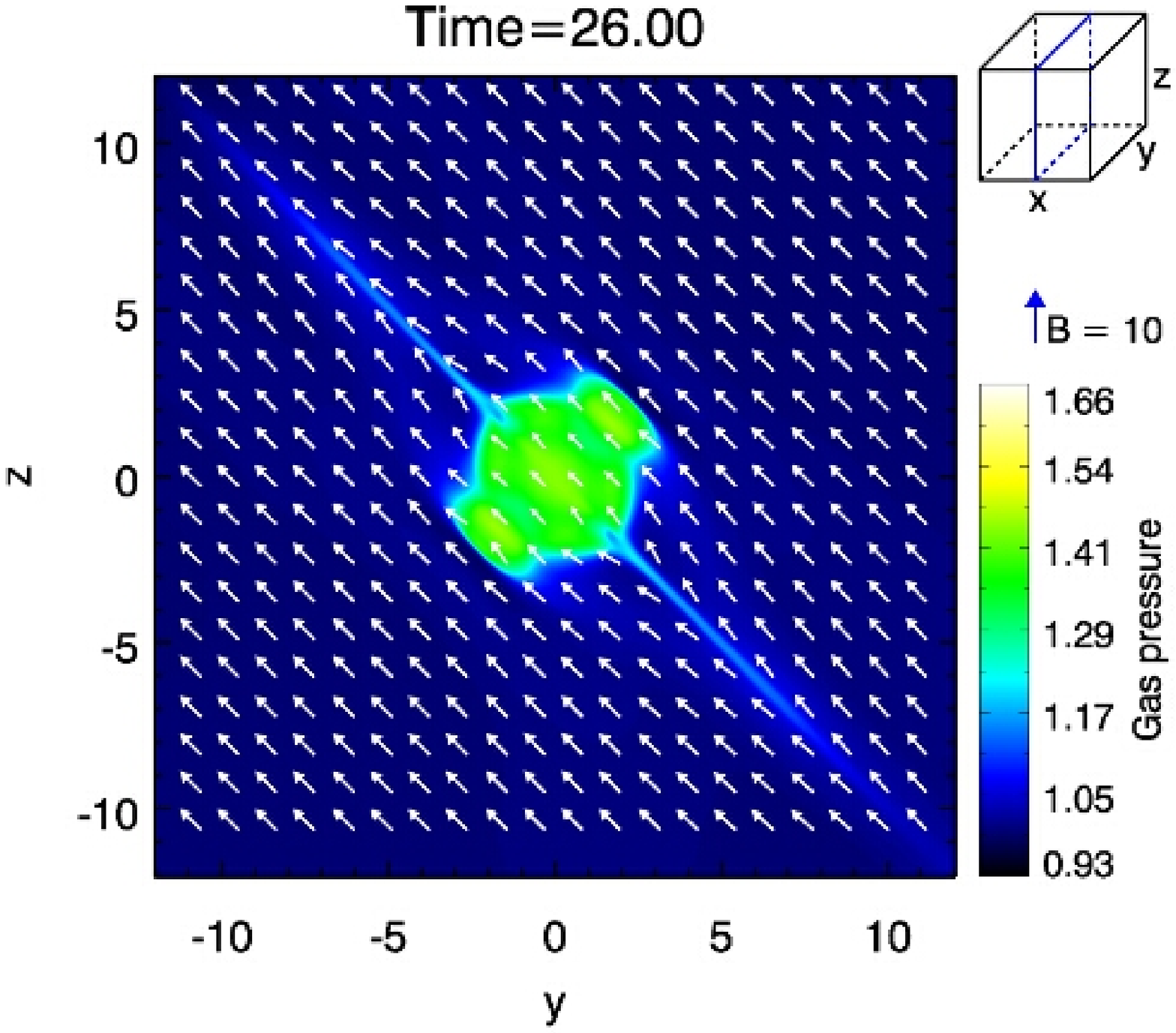}
\includegraphics[height=\pts,trim=4mm 10mm 4mm 16mm,clip]{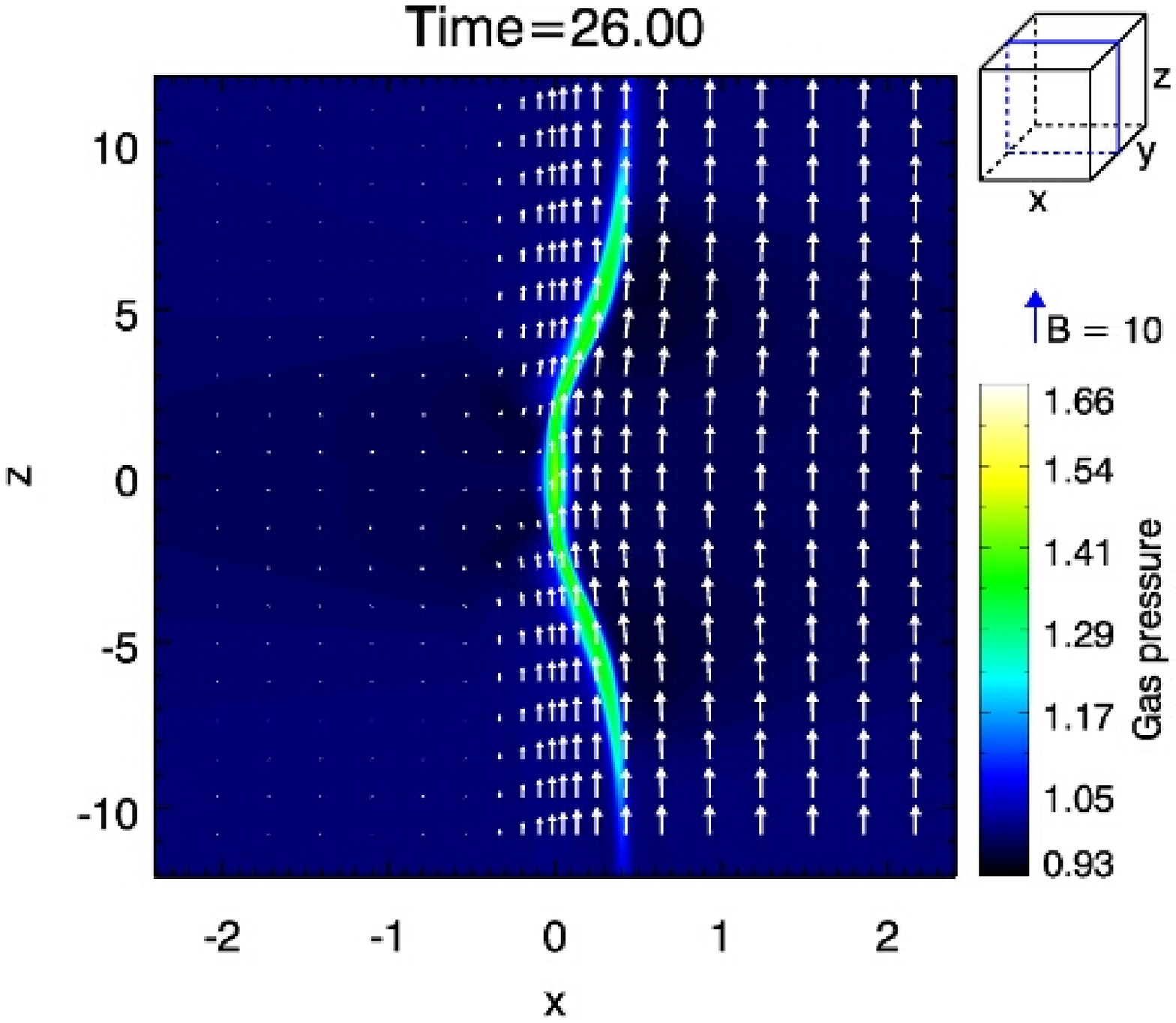}
\includegraphics[height=\pts,trim=4mm 10mm 4mm 16mm,clip]{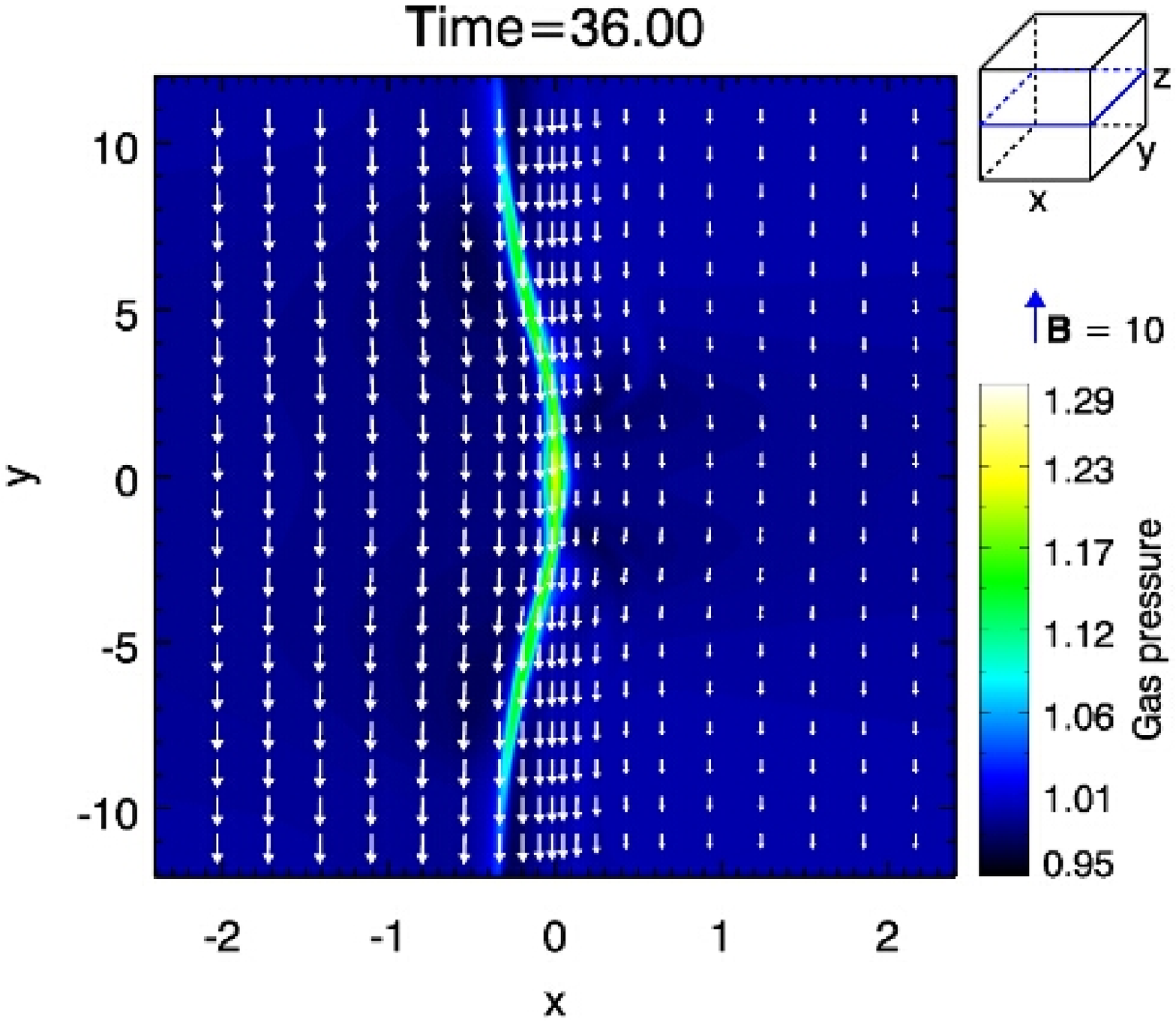}
\includegraphics[height=\pts,trim=4mm 10mm 4mm 16mm,clip]{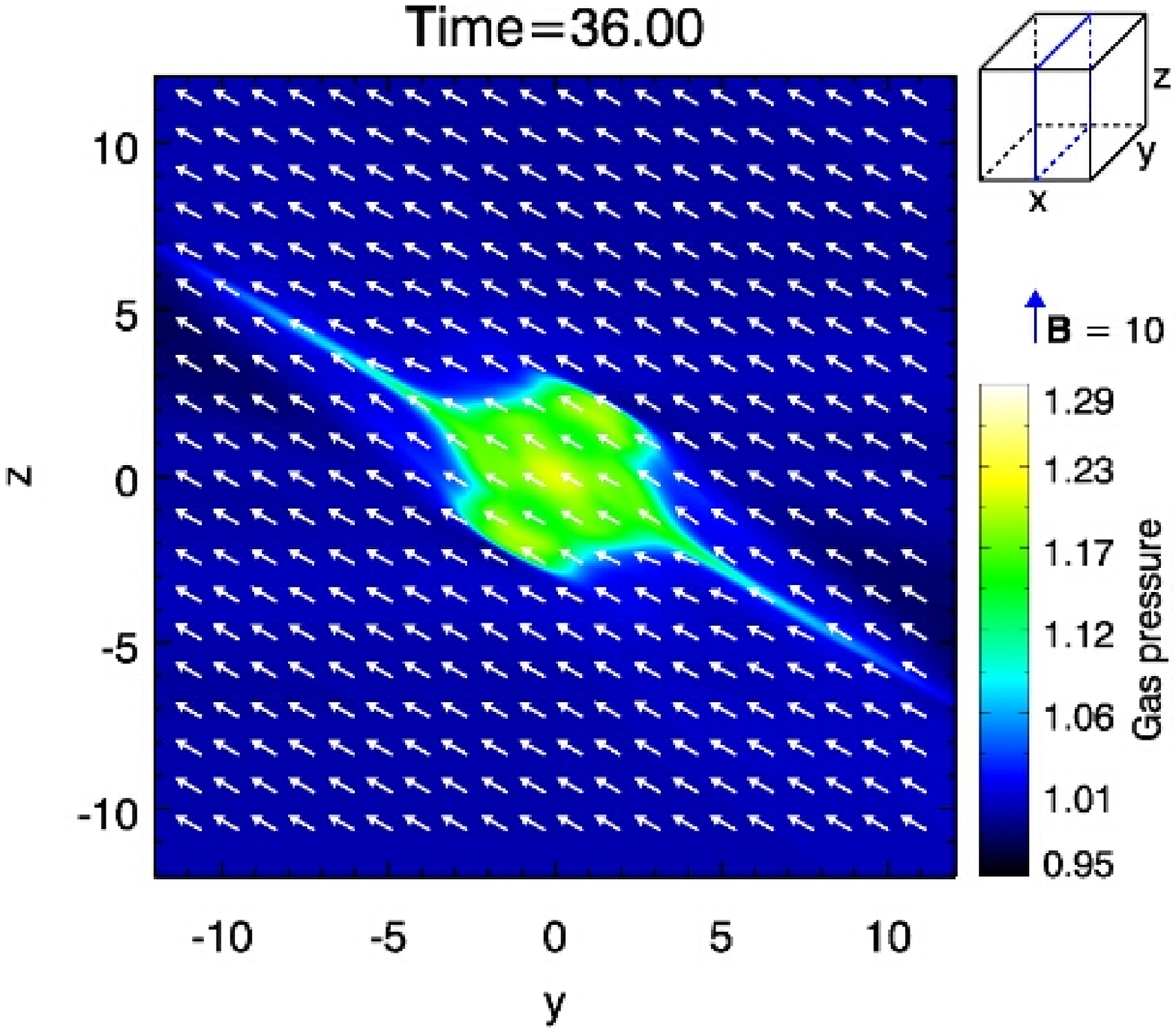}
\includegraphics[height=\pts,trim=4mm 10mm 4mm 16mm,clip]{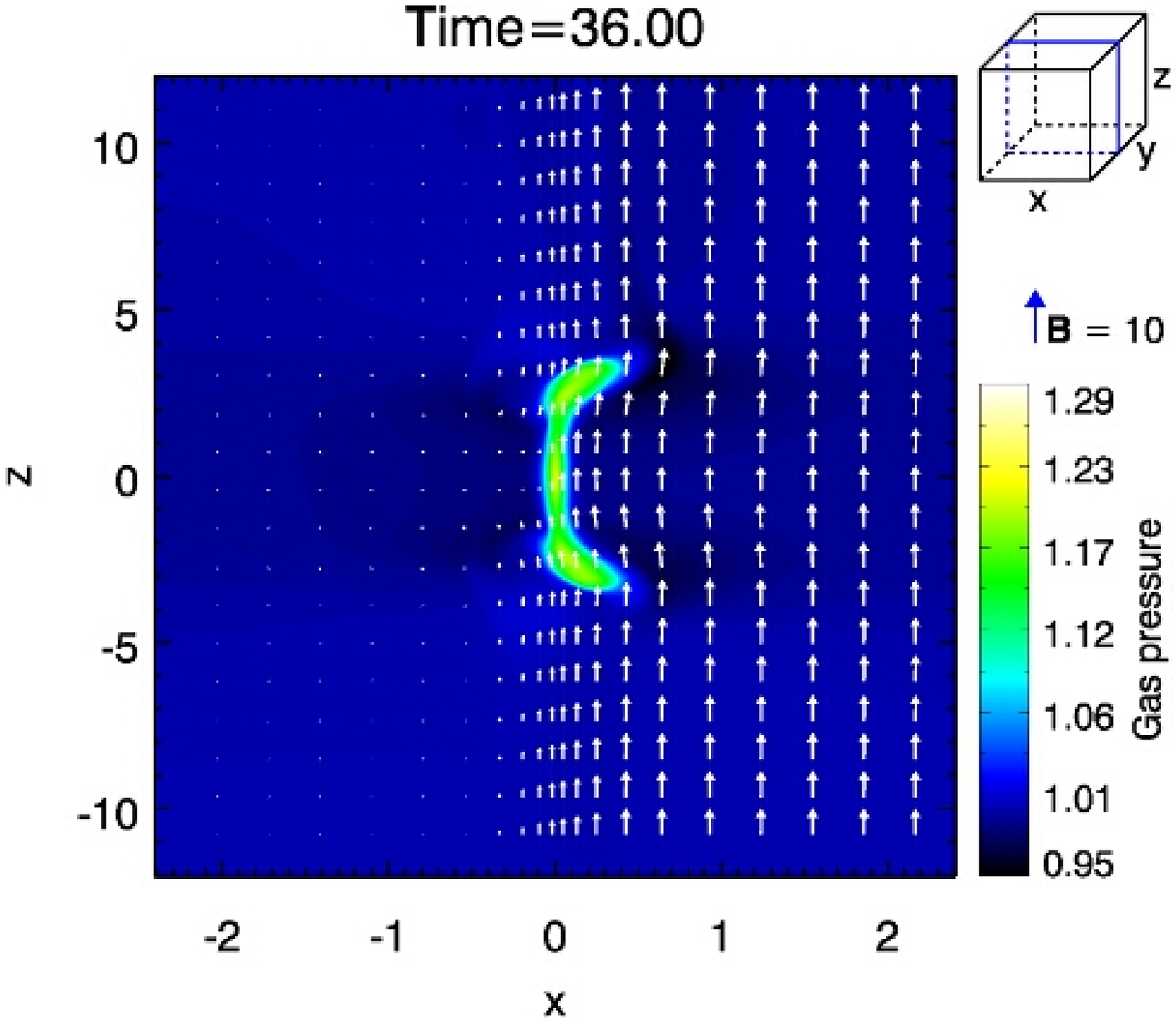}
\caption{2D distributions of gas pressure in the $x-y$, $y-z$ and $x-z$ planes for different cases (The reconnection angles for these cases are $3\pi/4$, $\pi/2$ and $\pi/3$ from top to bottom, respectively). The arrows indicate the magnetic field.} \label{fig19} \end{figure}

\begin{figure}[htbp]
\centering
\includegraphics[height=\pts,trim=4mm 10mm 4mm 16mm,clip]{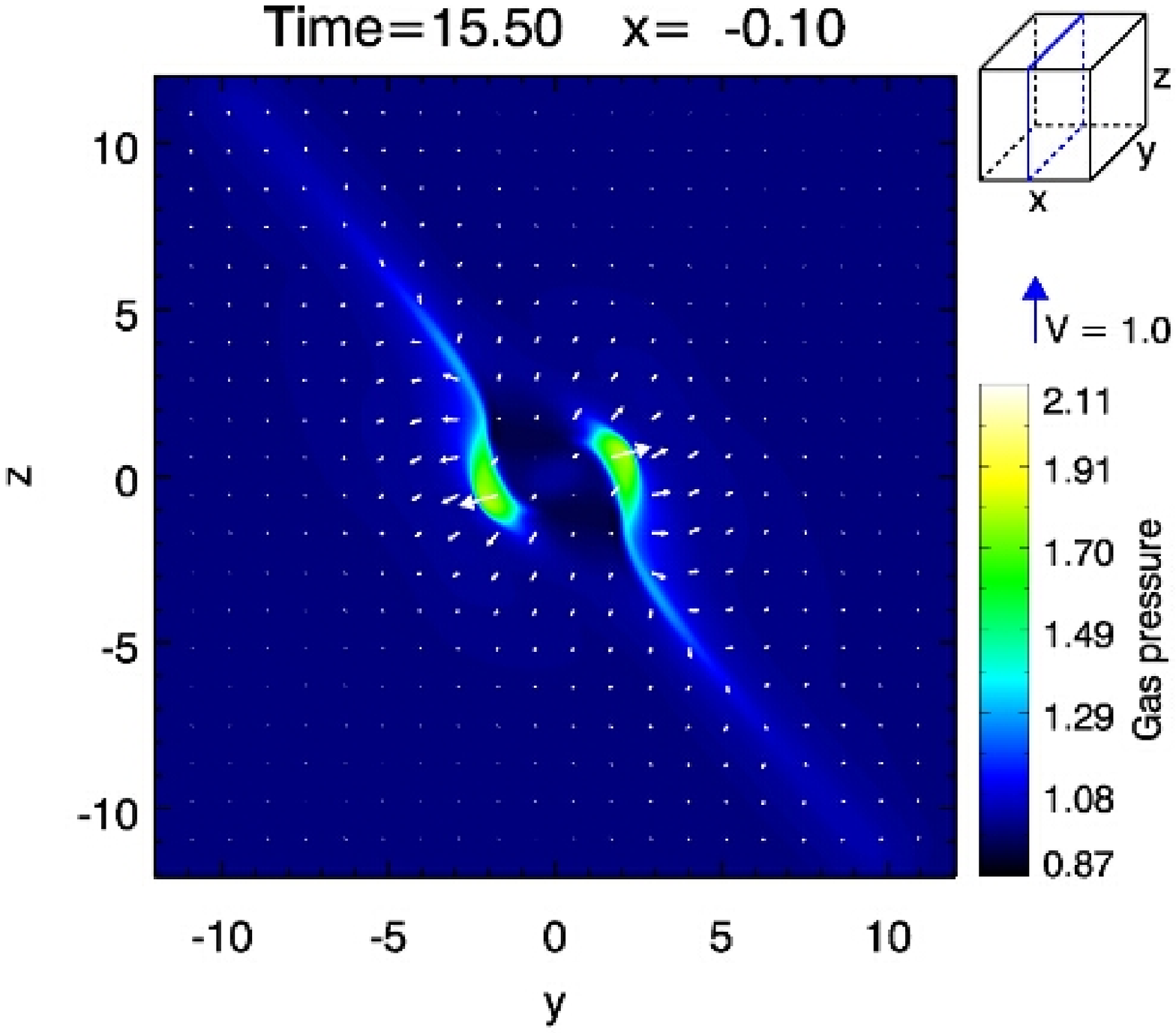}
\includegraphics[height=\pts,trim=4mm 10mm 4mm 16mm,clip]{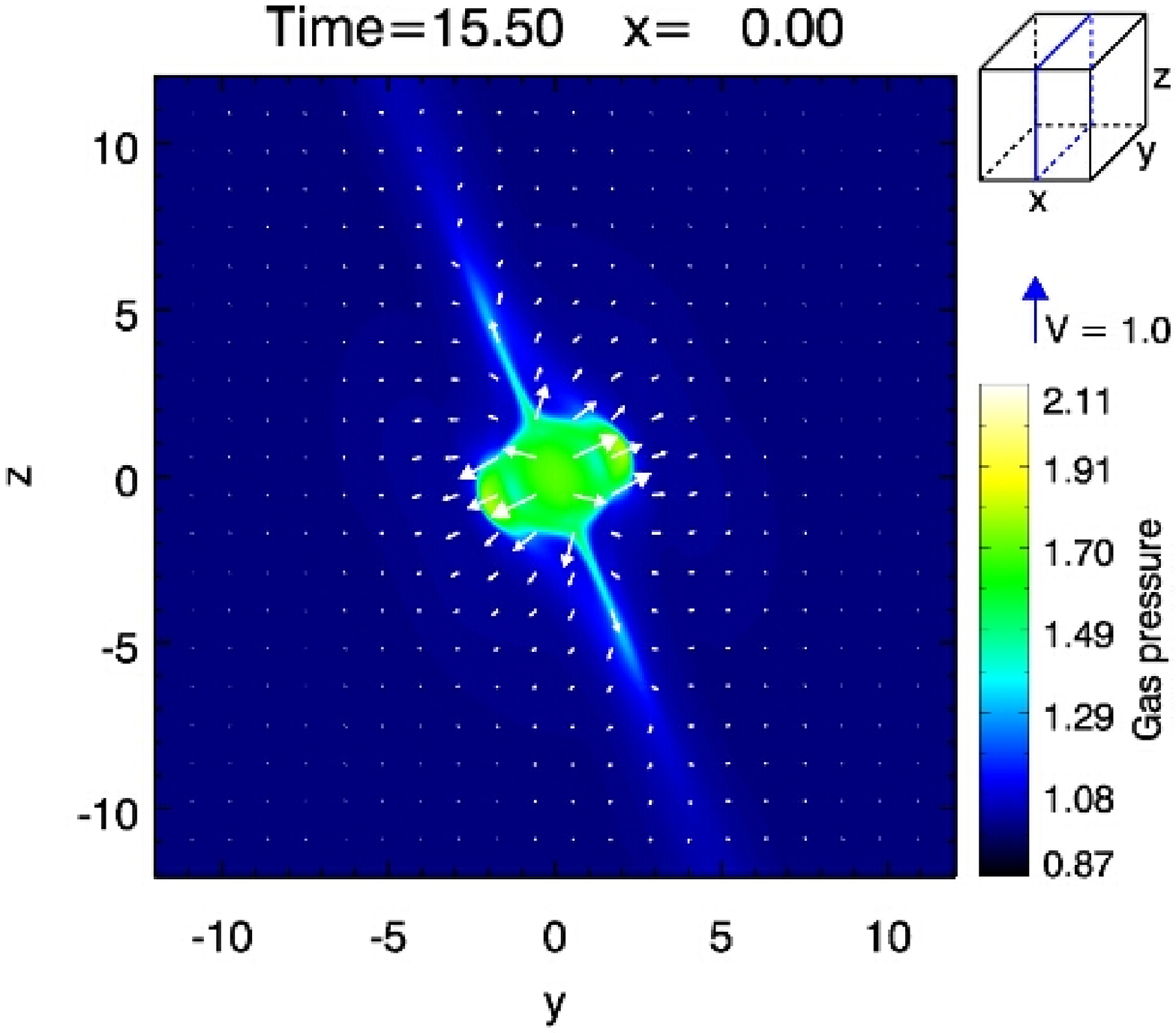}
\includegraphics[height=\pts,trim=4mm 10mm 4mm 16mm,clip]{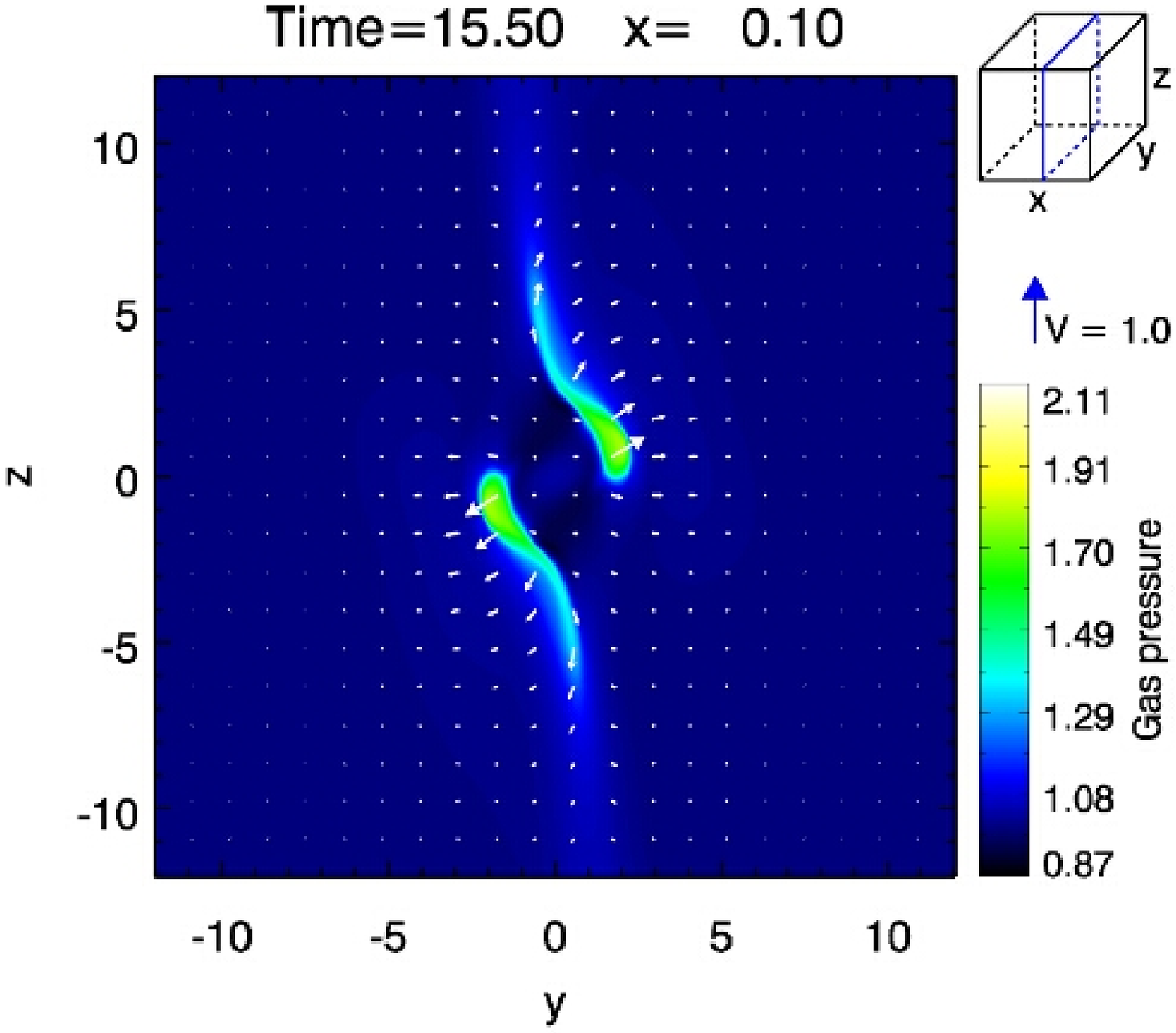}
\includegraphics[height=\pts,trim=4mm 10mm 4mm 16mm,clip]{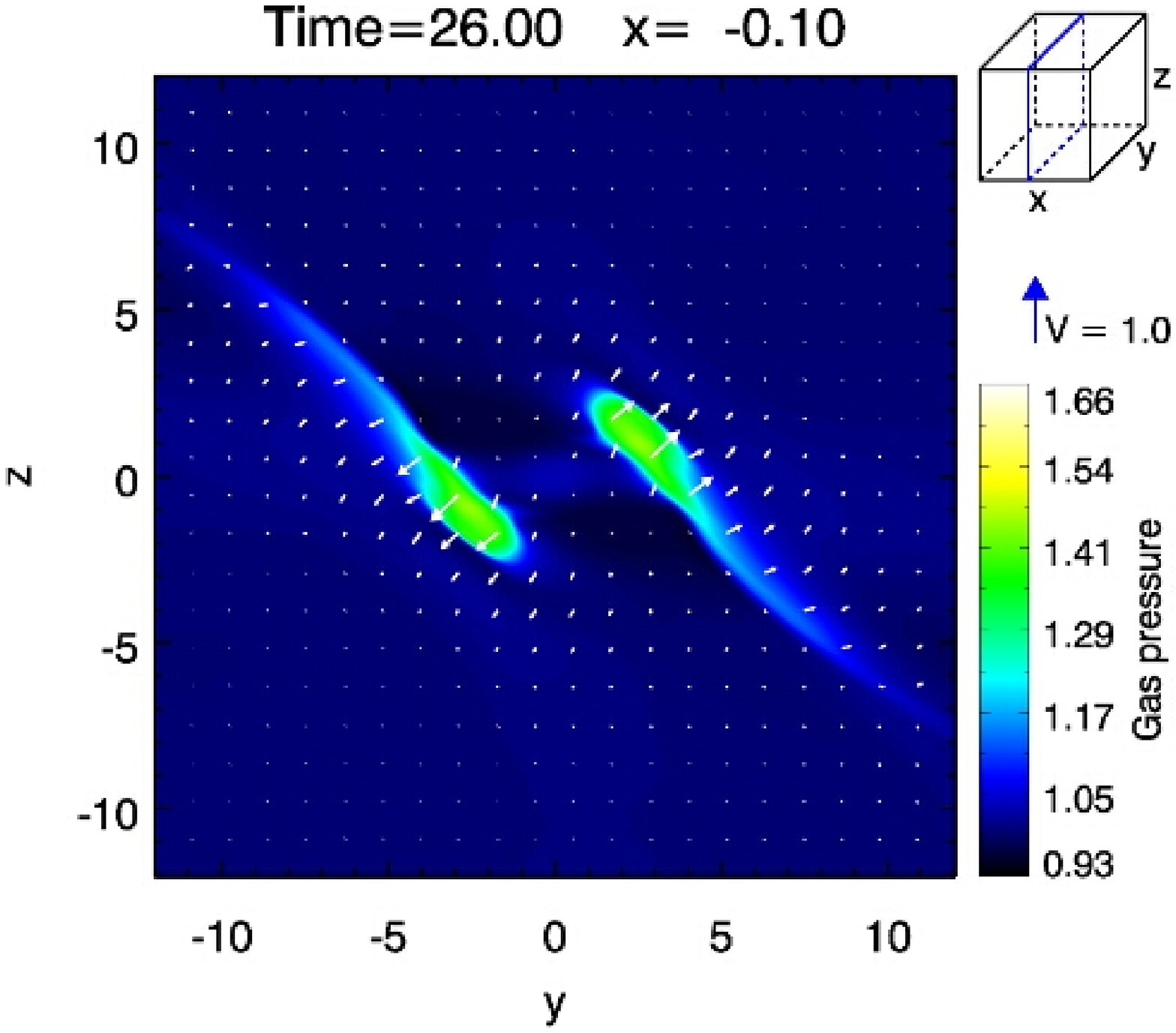}
\includegraphics[height=\pts,trim=4mm 10mm 4mm 16mm,clip]{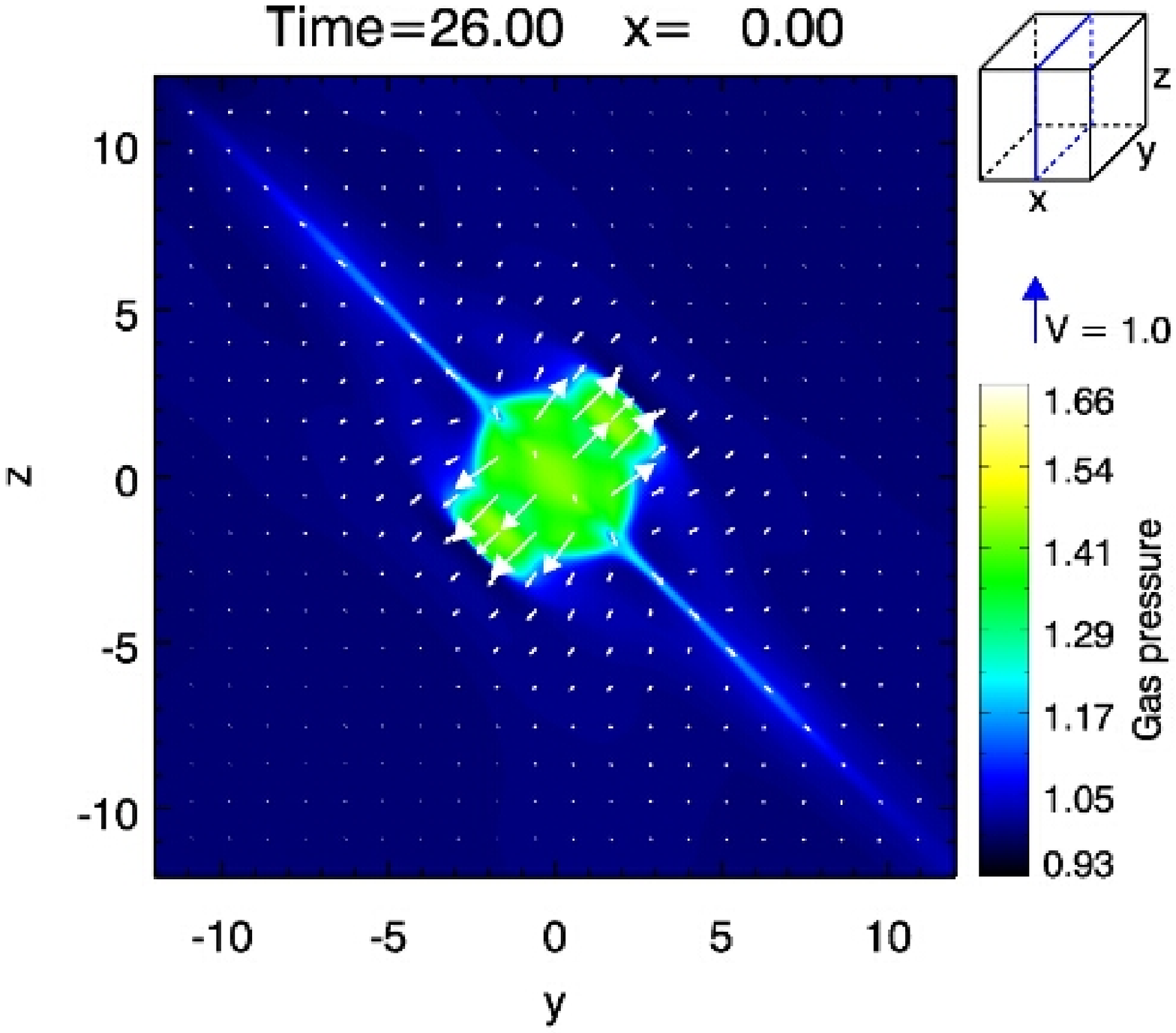}
\includegraphics[height=\pts,trim=4mm 10mm 4mm 16mm,clip]{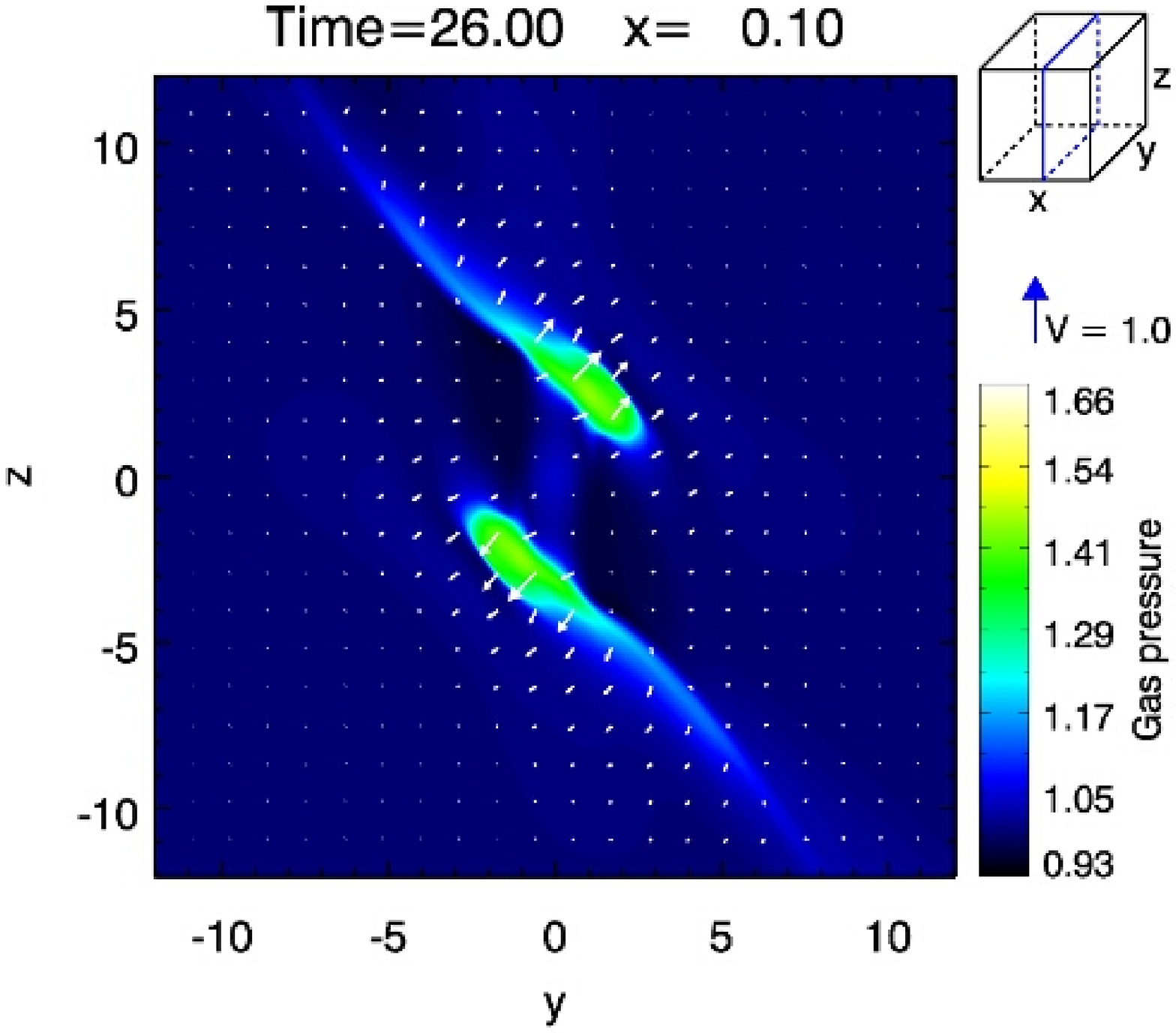}
\includegraphics[height=\pts,trim=4mm 10mm 4mm 16mm,clip]{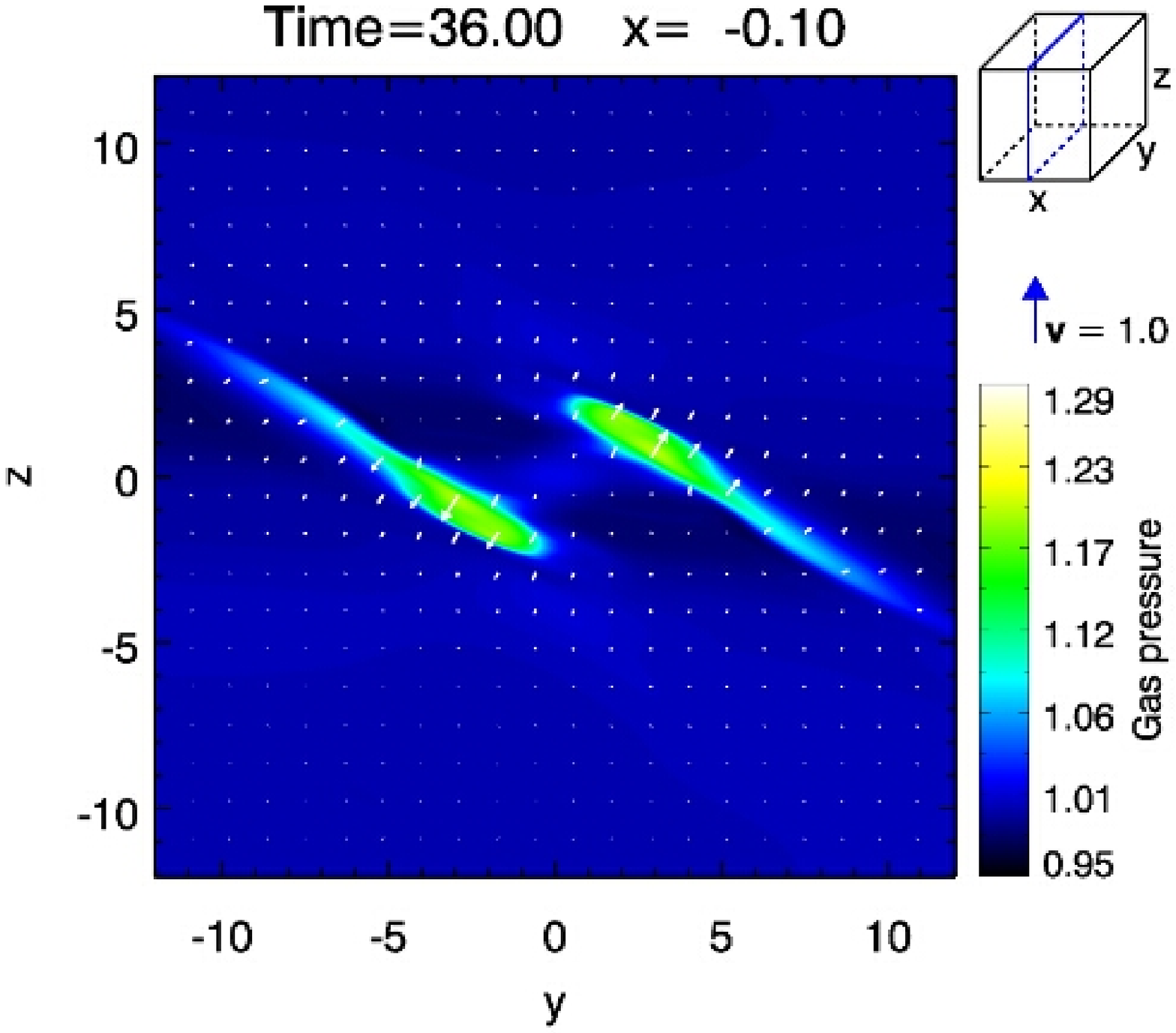}
\includegraphics[height=\pts,trim=4mm 10mm 4mm 16mm,clip]{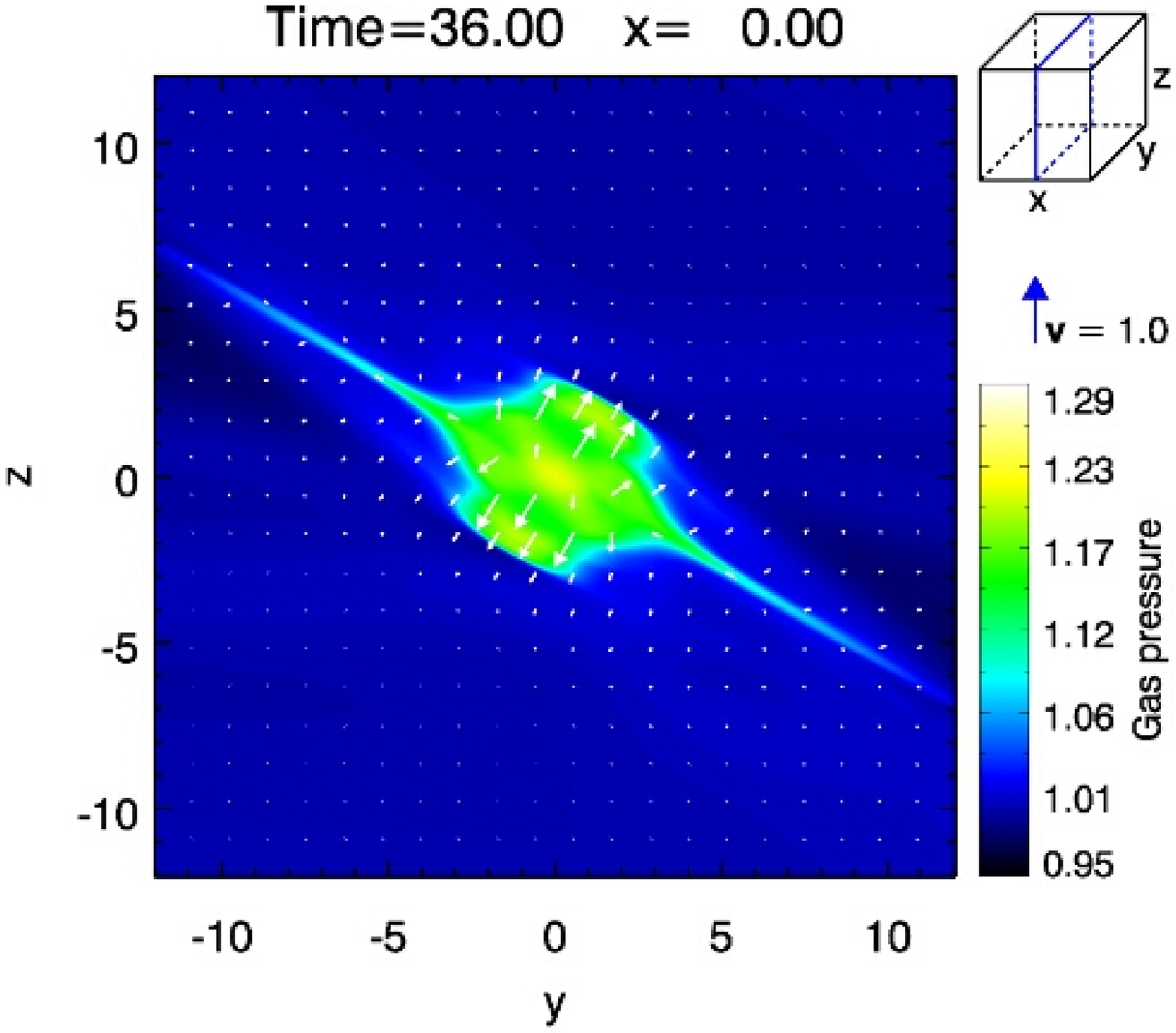}
\includegraphics[height=\pts,trim=4mm 10mm 4mm 16mm,clip]{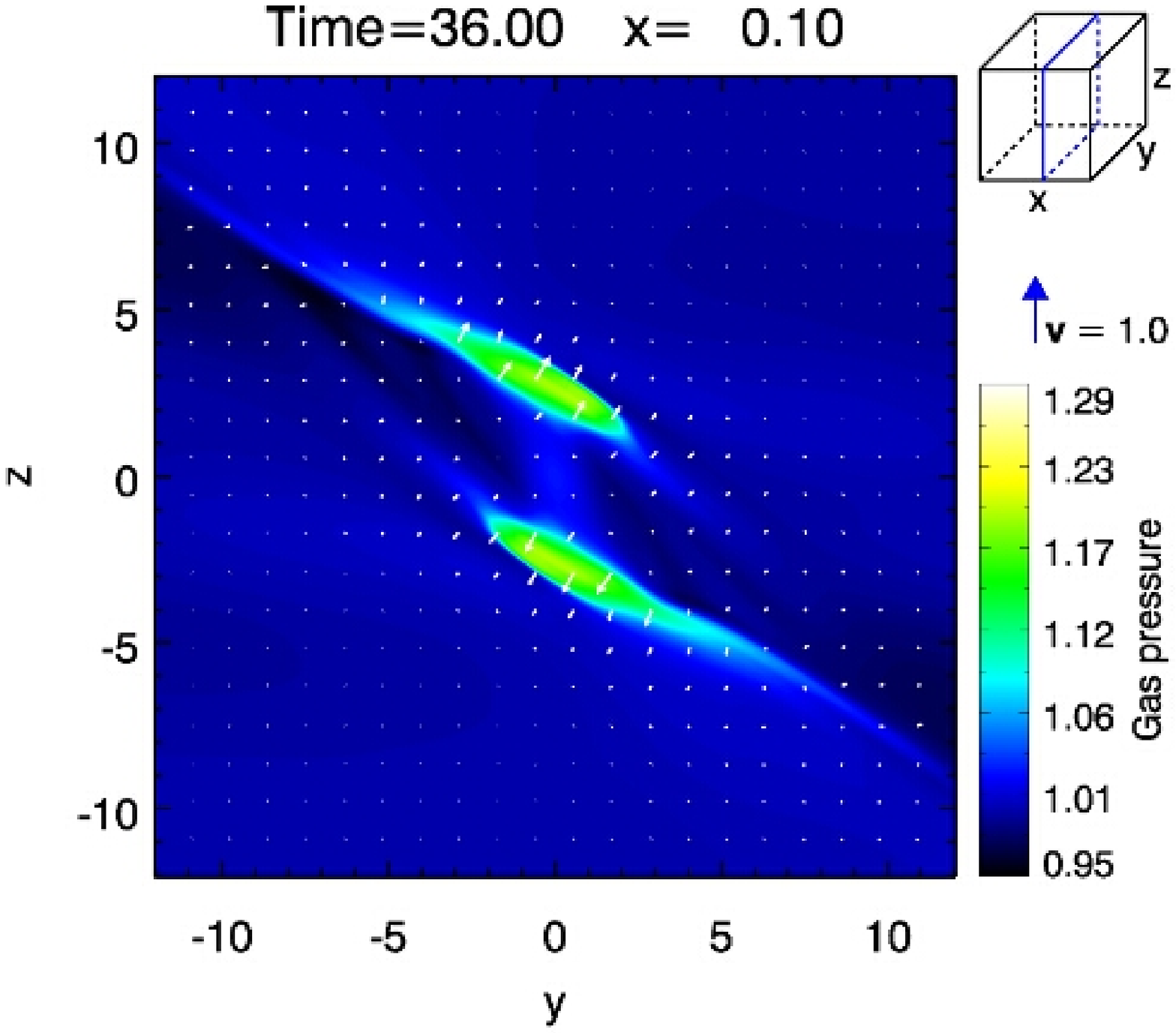}
\caption{2D distributions of gas pressure in the $y-z$ plane for different cases (The reconnection angle for these cases are $3\pi/4$, $\pi/2$ and $\pi/3$ from top to bottom, respectively). The arrows indicate the velocity field.} \label{fig20}
\end{figure}

\begin{figure}[htbp]
\centering
\includegraphics[height=\pts,trim=4mm 10mm 10mm 16mm,clip]{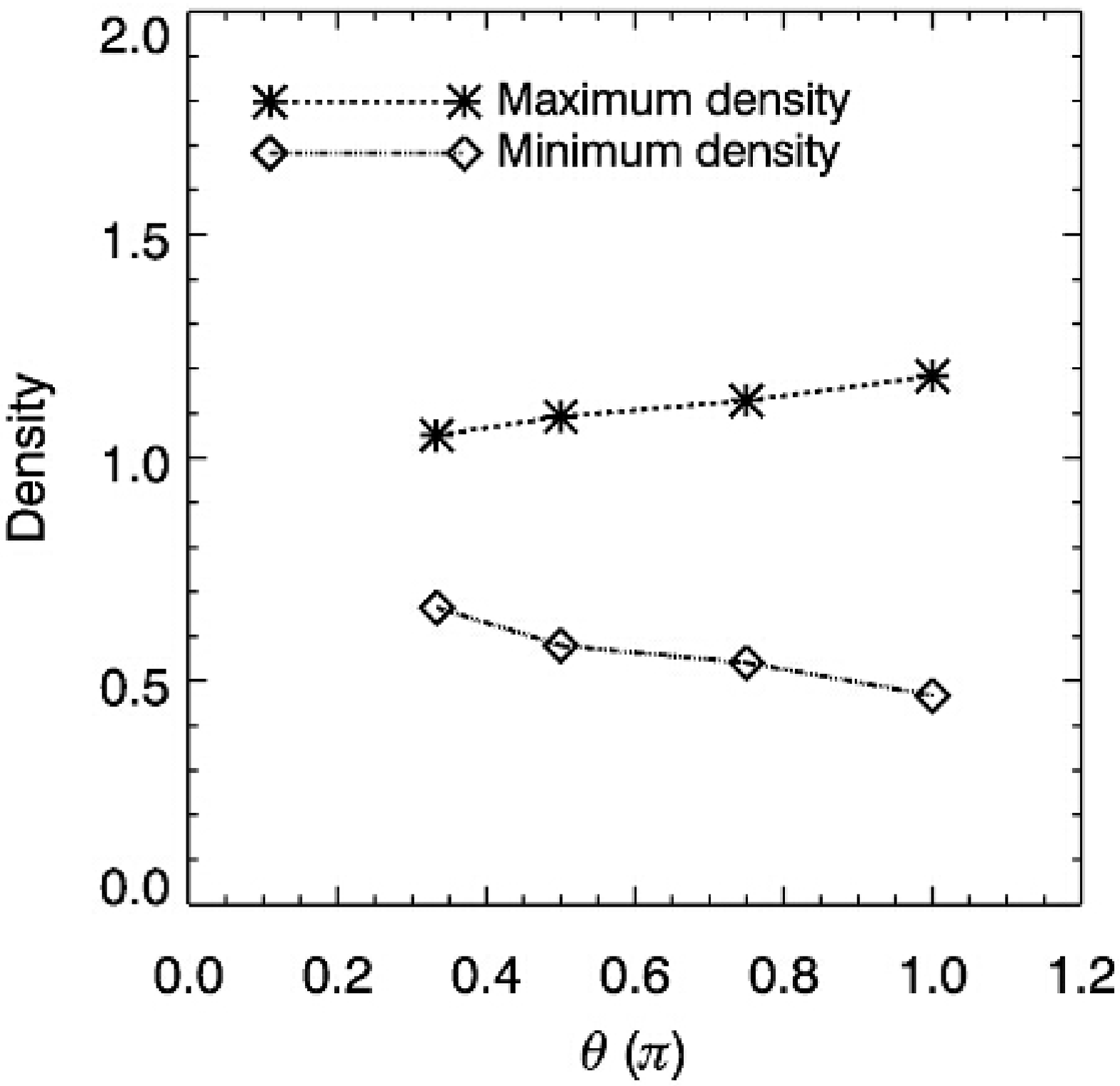}
\includegraphics[height=\pts,trim=4mm 10mm 10mm 16mm,clip]{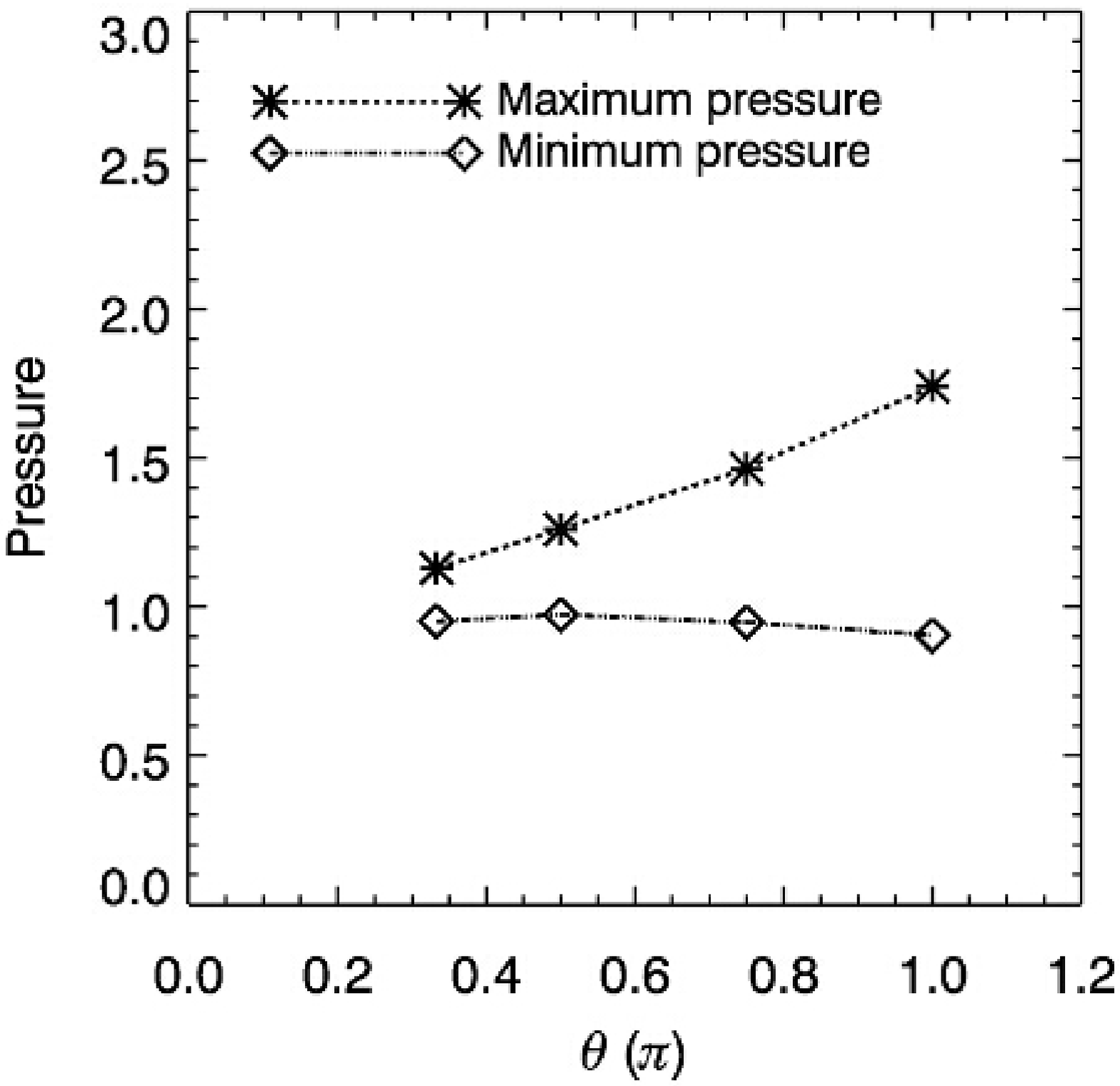}
\includegraphics[height=\pts,trim=4mm 10mm 10mm 16mm,clip]{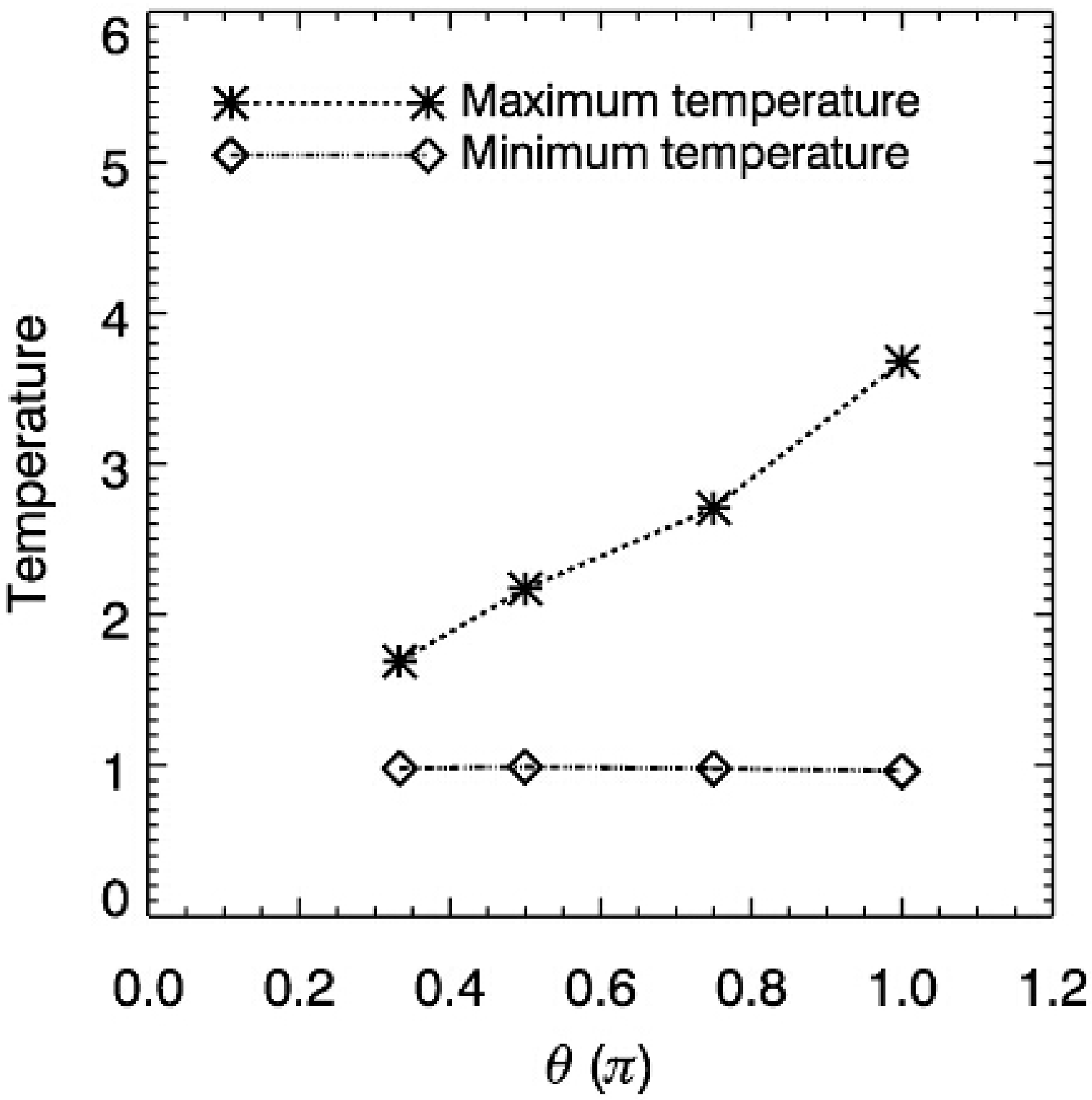}
\caption{The maximum and minimum densities, pressures and temperatures as a function of reconnection angles.}
\label{fig21}
\end{figure}

\begin{figure}[htbp]
\centering
\includegraphics[height=\pts,trim=4mm 10mm 10mm 16mm,clip]{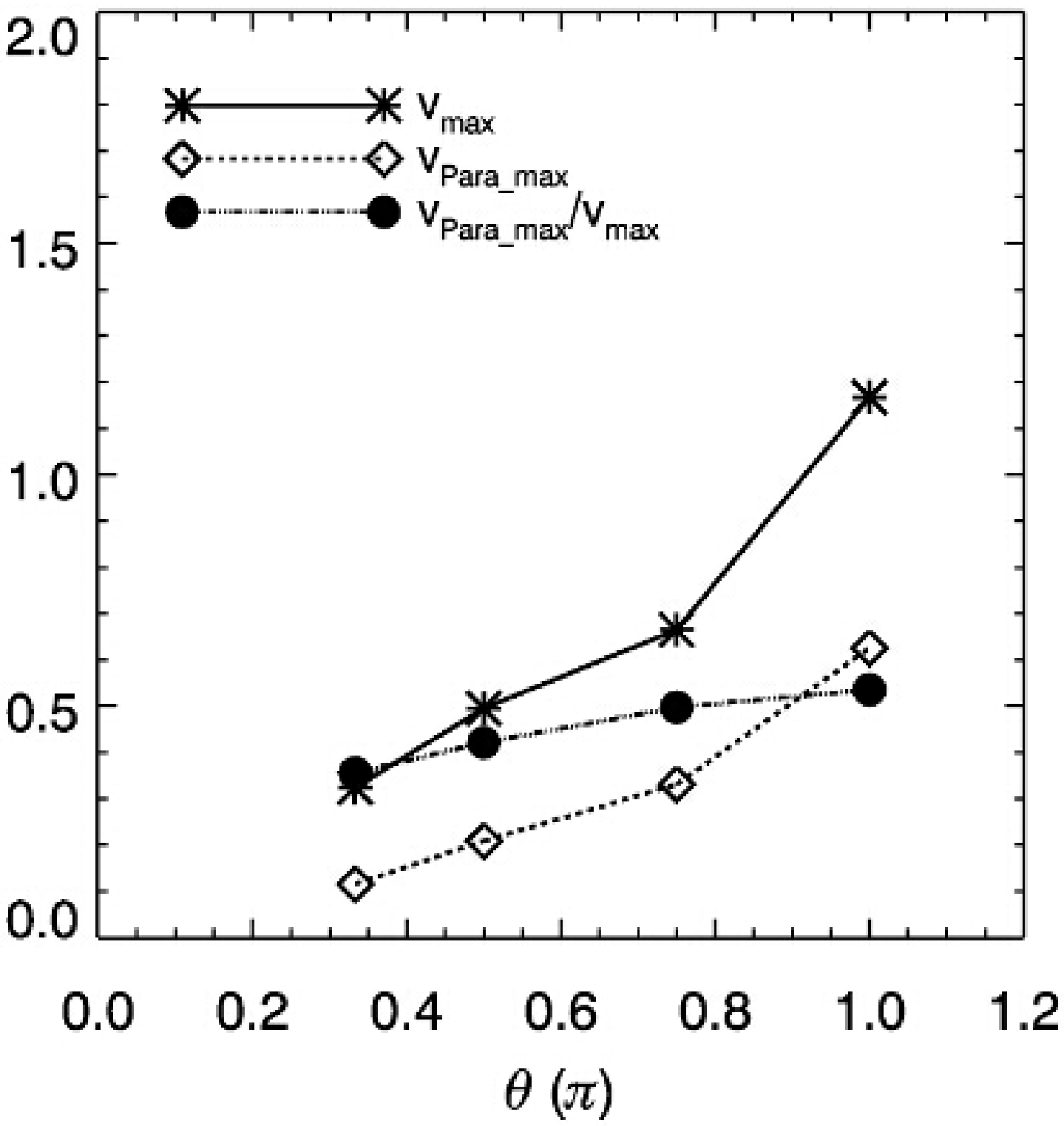}
\includegraphics[height=\pts,trim=4mm 10mm 10mm 16mm,clip]{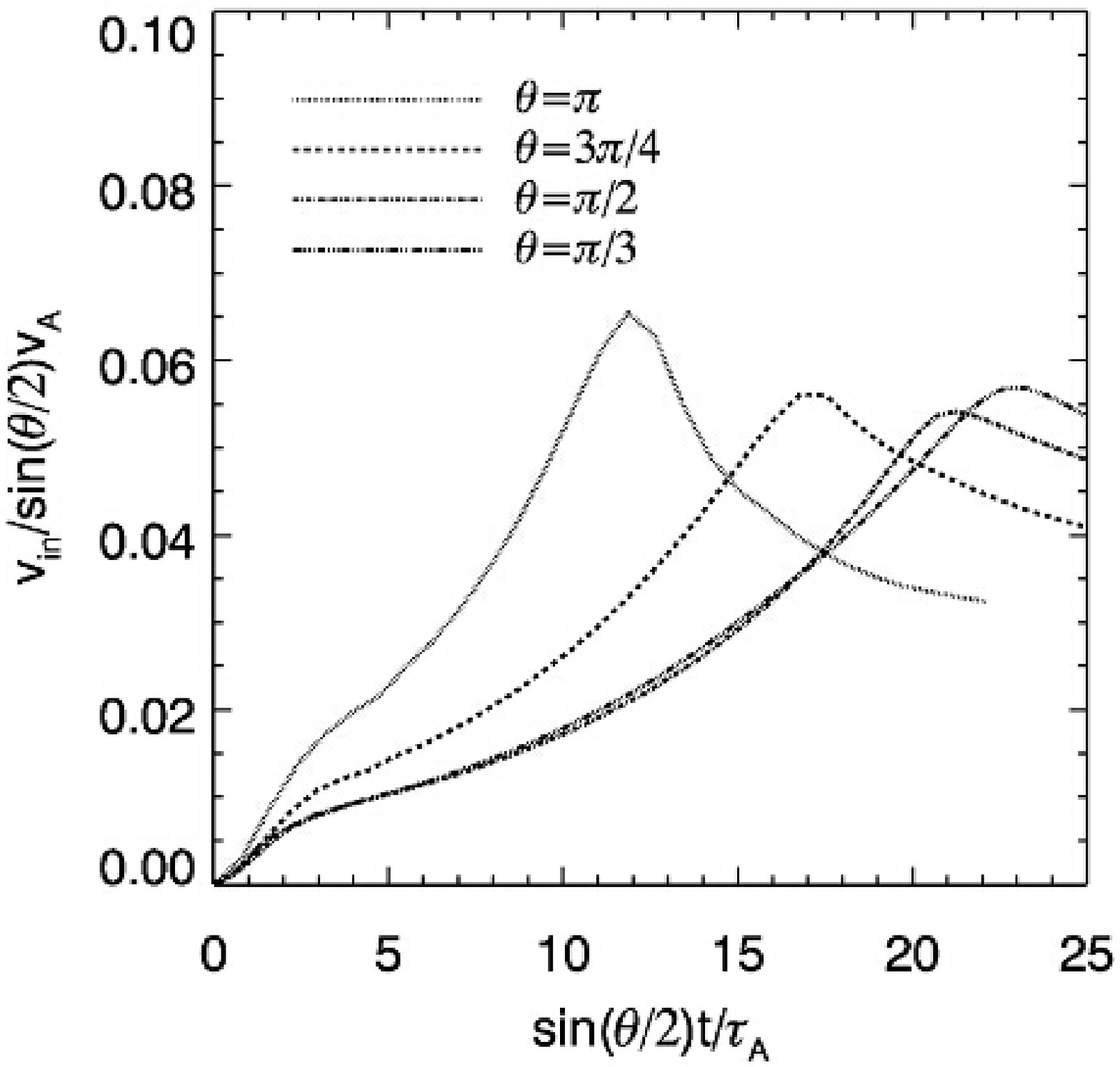}
\caption{Left panel depicts the dependence of maximum velocity indicating the speed of ordinary reconnection jets and the parallel velocity (parallel to the initial magnetic field, it means the velocity of fan-shaped jets) on the reconnection angles. The velocities are measured at the time when the reconnection rate reaches the maximum for different reconnection angles. The dashed-dotted line shows the ratio of two kinds of speed. The right panel gives the reconnection rates (presented by velocity) for different reconnection angles. Note that the normalization of the Alfv\'en speed and the Alfv\'en time scale are different for different reconnection angles.}
\label{fig22}
\end{figure}

Figure~\ref{fig18} depicts 3D visualizations for different reconnection angles. The left panels show gas pressure distributions with a velocity field and in the $z=0$ cross section plane. The right panels also show gas pressure distributions but with a magnetic field and a different view point. The velocity along the guide field is weaker than that in the typical case. Figures~\ref{fig19} and~\ref{fig20} show the 2D gas pressure distributions, which are similar to that in the typical case. Figure~\ref{fig21} depicts the maximum and minimum densities, pressures and temperatures as a function of the reconnection angle. These values are measured at the time when the reconnection rate reaches the maximum. If we decrease the reconnection angle, the difference between the maximum and the minimum values becomes smaller, and finally the maximum value will be equal to the minimum one and being the initial value 1. Linear relation is good in this dependence. The left panel of Figure~\ref{fig22} shows the maximum velocity in the computational box which stands for the velocity of the ordinary reconnection jets, the other is the maximum parallel velocity (parallel to the initial magnetic field) which indicates the speed of fan-shaped jets. From the left panels in Figures~\ref{fig17} and~\ref{fig22}, we found that the speed of fan-shaped jets is around half of the speed of ordinary reconnection jets, and it also depends on the reconnection angle (the reconnecting component of the magnetic field). The right panel of Figure~\ref{fig22} shows the reconnection rates for different angles. The maximum reconnection rate is about 0.06. The smaller angles correspond to a longer evolution time, which is reasonable.

\section{DISCUSSION AND SUMMARY}
\label{discussion}

We perform three dimensional resistive MHD simulations to study the magnetic reconnection using an initially shearing magnetic field configuration, which is similar to the papers by~\cite{Pontin2005} and~\cite{Ugai2010}. It is shown that there are two types of reconnection jets: the ordinary reconnection jets which move along the ambient magnetic field lines, and the fan-shaped jets which move along the guide field lines. In this paper we analyse the fan-shaped jets in detail. This may provide a new understanding of the dynamic phenomena in astrophysics. One of the possibility is that it can be applied to interpret the solar sunsport penumbral jets (\citealt{Katsukawa2007}) or anemone jets (\citealt{Shibata2007}) in the solar chromosphere.

Actually, before the fan-shaped jets, there is a slow mode wave propagating along the magnetic field lines. Because of its fast speed this wave can be seen as the front of the fan-shaped jets. But it is very hard to distinguish which is the slow mode wave and which are the jets. There is no clear boundary between them. The part which has gas pressure increasement but with small velocity is the slow mode wave as shown in Figure~\ref{fig02}. Of course, the ordinary jets connect the fan-shaped jets, the slow mode wave also exists at the front of ordinary reconnection jets.

The driving force for the fan-shaped jets is another interesting point. As there is no gravity in our simulations, the analysis of the Lagrangian fluid elements is easy to understand. Before Lagrangian fluid elements move along the magnetic field lines, both the gas pressure gradient force and the Lorentz force can drive the elements. The key point is that the magnetic tension force dominates the element movement first and then the element is affected by the gas pressure gradient force. In case 1 listed in Table~\ref{cases} with a small total grid points, we incidentally found that a fluid element is only accelerated by the Lorentz force. That is due to the initial position of that element being at the upper edge (outflow side) of the diffusion region. It goes out of the diffusion region soon, so that only the Lorentz force affects this element.

The dependence of the simulation variabes (namely density, pressure, temperature and velocity) on resistivity and reconnection angle is reasonable. The measurement of reconnection rate is simple in our paper and needs to be improved. However, from this method, people can get a general understanding about the 3D magnetic reconnection rate. It is roughly but useful, at least we can know when the magnetic reconnection reaches the maximum. Another disadvantage in our simulation is the limitation of the total grid points. A finer grid mesh or a higher order finite difference scheme or algorithm is under consideration.

If the effect of gravity is considered, the density stratification effect in the solar atmosphere can greatly increase velocity amplitude of the upward fan-shaped jets when it propagates up to the low density region (like in the solar upper chromosphere and the corona) and eventually becomes a shock or jets (\citealt{Shibata1982a, Shibata1982b}). The extended development of this study with gravity considered is in progress.

In summary, we give the conclusions as follows:

1. From 3D MHD simulations we have studied a type of jets which move along the guide magnetic field. Because of the rotation of the initial magnetic field these jets have a fan-shaped structure. Of course, the ordinary reconnection jets ejected by the magnetic tension force also exist.

2. The driving forces of these fan-shaped jets are Lorentz force and gas pressure gradient force. The magnetic pressure gradient force drives the Lagrangian fluid elements first, and then the force on the elements is dominated by the gas pressure gradient in the later stage.

3. The pressure, density, temperature and velocity do not sensitively depend on the initial resistivity value.

4. The ratio between the velocity of the fan-shaped jets and the velocity of the ordinary reconnection jets is almost a constant being about 0.5 for different reconnection angles.

\begin{acknowledgements}
 We thank E. Asano, A. Hillier and K. Nishida for helpful discussions and to K. A. P. Singh for a critical reading of the manuscript. This work was supported by the National Natural Science Foundation of China (NSFC) under the grant numbers 10878002, 10610099, 10933003 and 10673004, as well as the grant from the 973 project 2011CB811402 of China, and in part by the Grant-in-Aid for Creative Scientific Research ``The Basic Study of Space Weather Prediction'' (Head Investigator: K. Shibata) from the Ministry of Education, Culture, Sports, Science and Technology (MEXT) of Japan, and in part by the Grand-in-Aid for the Global COE program ``The Next Generation of Physics, Spun from Universality and Emergence'' from MEXT. Numerical computations were carried out on Cray XT4 at Center for Computational Astrophysics, CfCA, of National Astronomical Observatory of Japan, and in part performed with the support and under the auspices of the NIFS Collaboration Research program (NIFS07KTBL005), and in part performed with the KDK system of Research Institute for Sustainable Humanosphere (RISH) at Kyoto University as a collaborative research project.
\end{acknowledgements}

\normalem

\label{lastpage}

\end{document}